\documentclass[twocolumn]{aastex62}
\turnoffediting
\usepackage{natbib}
\usepackage{amsmath, mathtools}
\usepackage{hyperref}
\usepackage{color}
\usepackage{comment}
\usepackage{textcomp, gensymb}
\usepackage[T1]{fontenc}
\usepackage[caption=false]{subfig}
\usepackage[para,online,flushleft]{threeparttable}
\usepackage{booktabs}
\usepackage{rotating}
\usepackage{graphicx}
\usepackage{subfig}
\usepackage{lineno}
\usepackage[nameinlink]{cleveref}
\usepackage[utf8]{inputenc}
\DeclareUnicodeCharacter{2212}{-}
\graphicspath{{./}}

\shorttitle{Young stellar population in the Galactic Center}
\shortauthors{Jia, Xu et al.}

\begin{document}

\title{Stellar Populations in the Central 0.5 pc of Our Galaxy III: The Dynamical Sub-structures}

\correspondingauthor{Jessica Lu}
\email{jlu.astro@berkeley.edu}

\author[0000-0001-5341-0765]{Siyao Jia}
\affil{Astronomy Department, University of California, Berkeley, CA 94720, USA}

\correspondingauthor{Ningyuan Xu}
\email{alexxny0337@berkeley.edu}

\author[0000-0003-0874-1439]{Ningyuan Xu}
\affil{Astronomy Department, University of California, Berkeley, CA 94720, USA}

\author[0000-0001-9611-0009]{Jessica R. Lu}
\affil{Astronomy Department, University of California, Berkeley, CA 94720, USA}

\author[0000-0003-3765-8001]{D.S. Chu}
\affil{UCLA Department of Physics and Astronomy, Los Angeles, CA 90095-1547, USA}

\author[0000-0003-2400-7322]{K. Kosmo O'Neil}
\affil{UCLA Department of Physics and Astronomy, Los Angeles, CA 90095-1547, USA}

\author{W.B. Drechsler}
\affil{Astronomy Department, University of California, Berkeley, CA 94720, USA}

\author[0000-0003-2874-1196]{M. W. Hosek Jr.}
\altaffiliation{Brinson Prize Fellow}
\affil{UCLA Department of Physics and Astronomy, Los Angeles, CA 90095-1547, USA}

\author[0000-0001-5972-663X]{S. Sakai}
\affil{UCLA Department of Physics and Astronomy, Los Angeles, CA 90095-1547, USA}

\author[0000-0001-9554-6062]{T. Do}
\affil{UCLA Department of Physics and Astronomy, Los Angeles, CA 90095-1547, USA}

\author[0000-0001-5800-3093]{A. Ciurlo}
\affil{UCLA Department of Physics and Astronomy, Los Angeles, CA 90095-1547, USA}

\author[0000-0002-2836-117X]{A. K. Gautam}
\affil{UCLA Department of Physics and Astronomy, Los Angeles, CA 90095-1547, USA}

\author[0000-0003-3230-5055]{A. M. Ghez}
\affil{UCLA Department of Physics and Astronomy, Los Angeles, CA 90095-1547, USA}

\author{E. Becklin}
\affil{UCLA Department of Physics and Astronomy, Los Angeles, CA 90095-1547, USA}

\author[0000-0002-6753-2066]{M. R. Morris}
\affil{UCLA Department of Physics and Astronomy, Los Angeles, CA 90095-1547, USA}

\author[0000-0001-7017-8582]{R. O. Bentley}
\affil{UCLA Department of Physics and Astronomy, Los Angeles, CA 90095-1547, USA}

\begin{abstract}
We measure the 3D kinematic structures of the young stars within the central 0.5 parsec of our Galactic Center using the 10 m telescopes of the W.~M.~Keck Observatory over a time span of 25 years.
Using high-precision measurements of positions on the sky, and proper motions and radial velocities from new observations and the literature, we constrain the orbital parameters for each young star. 
Our results show two statistically significant sub-structures: a clockwise stellar disk with 18 candidate stars, as has been proposed before, but with an improved disk membership; a second, almost edge-on plane of 10 candidate stars oriented East-West on the sky that includes at least one IRS 13 star.
\edit1{We estimate the eccentricity distribution of each sub-structure and find that the clockwise disk has <$e$> = 0.39 and the edge-on plane has <$e$> = 0.68.}
\edit1{We also perform simulations of each disk/plane with incompleteness and spatially-variable extinction to search for asymmetry.}
Our results show that the clockwise stellar disk is consistent with a uniform azimuthal distribution within the disk.
The edge-on plane has an asymmetry that cannot be explained by variable extinction or \edit1{incompleteness} in the field.
The orientation, asymmetric stellar distribution, and high eccentricity of the edge-on plane members suggest that this structure may be a stream associated with the IRS 13 group.  
The complex dynamical structure of the young nuclear cluster indicates that the star formation process involved complex gas structures and dynamics and is inconsistent with a single massive gaseous disk.
\end{abstract}


\section{Introduction} 
\label{sec:intro}
Nuclear Star Clusters (NSCs) and Supermassive Black Holes (SMBHs) are found to coexist within the central parsec (pc) of many different types of galaxies \citep{Graham_2009}.  
Furthermore, there are clear indications that NSCs, SMBHs, and their host galaxies evolve together. 
For example, \citet{Ferrarese_2006} found that the $M - \sigma$ relationship, the empirical correlation between the mass of the SMBH, $M$, and the stellar velocity dispersion $\sigma$ of the galaxy's bulge, applies both for SMBHs and NSCs with similar slopes, although at a certain $\sigma$, NSCs are 10 times more massive than SMBHs.
Additionally, \citet{Kormendy_2013} showed that the combined mass of the SMBH and NSC scales with a galaxy's bulge mass with much less scatter than either the SMBH or the NSC separately, which suggests a strong dependency between their mutual formation and growth.
The formation mechanism for NSCs is still widely debated as the strong tidal force at the Galactic Center will disrupt normal star formation. 

Our own Galactic Center provides a unique test bed for understanding star formation around SMBHs since it harbors a population of around 200 young massive stars within 0.5 pc of the central SMBH, SgrA$^\ast$ \citep{Genzel_2000}, including OB main-sequence stars, Wolf-Rayet (WR) stars, giants, and supergiants \citep{Paumard_2006, Bartko_2010}. 
Because the age of this population \citep[3 - 8 Myr][]{Lu_2013} is much less than the relaxation timescale in the Galactic Center ($\gtrsim$ 1Gyr, \citet{Hopman_2006}),  the origin of these stars can be constrained through studies of their dynamical structures. 
Only in the Galactic Center can we resolve individual stars and measure their motion, photometry, and spectroscopy with sufficient precision to constrain their dynamics and therefore constrain theories of star formation around SMBHs.

%====In situ Formation theory==========
Observations of young stars at the Galactic Center currently favor \textit{in situ} formation models, meaning that young stars are formed roughly where we see them today, within 0.5 pc of the SMBH \citep{Paumard_2006, Lu_2009, Meyer_2015}.  
$\textit{In situ}$ formation is theoretically possible in an accretion disk around the SMBH, if it is massive enough to collapse vertically under its own self-gravity \citep{Kolykhalov_1980,Morris_1996,Sanders_1998,Goodman_2003,Levin_2003,Nayakshin_2005}. 
When a disk reaches a surface density that is just large enough to initiate star formation, the first protostars form. 
Feedback from those stars will then heat the disk up to a point where it stabilizes against collapse and shuts off further disk fragmentation. 
In the meantime, those stars remain embedded in the disk and continue gaining mass at very high rates. 
As a result, an average star created in such a disk may become very massive. 
This process happens throughout the disk, from $\sim$0.01 pc to a few parsecs, with a peak effect at R $\sim$ 0.1 pc \citep{Morris_1996, Vollmer_2001, Nayakshin_2005, Nayakshin_2006_massivestar}. 
This scenario can explain the existence of the disk of stars and the top-heavy Initial Mass Function (IMF) in our Galactic Center \citep{Lu_2013}, although it may not be the only explanation.
%since many formation scenarios still don't uniquely predict the IMF. 

However, if the disk is formed through steady accretion of gas, stars with circular orbits are more likely to form, which may contradict the observed eccentricity distribution that peaks at e=0.27 \citep{Yelda_2014}. Furthermore, only 20\% of the young stars are estimated to be in the disk and it is not clear whether 80\% of the stars could be scattered from the disk in only 4 - 8 Myr.
Modified \textit{in situ} formation scenarios have been proposed, including:  (a) the initial gas is not uniformly distributed, (b) stars form in repeated episodes, (c) after the gas disk collapses, the stars that form in it dynamically evolve off the disk.  
These formation scenarios can be disentangled by comparing the dynamical structures among different dynamical sub-groups.
For example, if the initial gas is not uniformly distributed, we should see asymmetric sub-structures in their stellar systems.

%==========our galaxy=========
Previous studies show that the young stars at the Galactic Center can be divided into three dynamical groups: (1) $\thicksim 12\%$ of the young stars (within 0.03 pc) are in the innermost region with high eccentricities $\bar{e} = 0.8$ and randomly oriented orbits. (2) $\thicksim 20\%$ of the young stars are on a well-defined clockwise (CW) rotating disk (0.03 - 0.5 pc) with moderate eccentricities $\bar{e} = 0.3$. (3) $\thicksim 68\%$ of them are off-disk stars that extend over the same radius but have a more random distribution and eccentricity distribution with higher $\bar{e} = 0.6$ \citep{Lu_2013, Yelda_2014}.
At smaller radii, dynamical effects will randomize the stellar orbits within 4-6 Myr, which is the case for the first group. 
The existence of the CW disk has already been verified to be significant ($\sim 20 \sigma$) \citep{Paumard_2006,Bartko_2010,Lu_2009,Yelda_2014}. More recently, as many as 5 distinct sub-structures, including the CW disk, have been proposed by \citet{von_Fellenberg_2022}. 
\edit1{A detailed comparison between our result and previous work is presented in \S\ref{sec:discuss_previous} and Appendix \ref{app:comp}.}

%============layout=====================
In this paper, we present improved dynamical measurements of the young stars at the Galactic Center derived from adaptive optics observations from the 10m Keck telescopes. 
We conduct novel simulations and comparative analyses of the properties of the different dynamical sub-groups of young stars within the central 0.5 pc of our Galaxy.
The observational setup and data reduction are presented in Section \ref{sec:obs}. 
Orbital parameters are derived in \S\ref{sec:orbitfit} and the disk membership analysis is shown in \S\ref{sec:disk}.
Results of stellar disk properties and cluster simulations are presented in \S\ref{sec:result}. 
We discuss our findings in \S\ref{sec:discussion} and summarize in \S\ref{sec:summary}.

\section{Observations and Data Reduction}
\label{sec:obs}

The kinematic analysis of the young stars at the Galactic Center requires both proper motion and radial velocity (RV) measurements in order to determine their orbital planes and disk membership probability.
Details of the observations, data reduction, and image analysis are presented in \citet{Jia_2019} and \citet{Do_2009}.
Here, we briefly summarize the analysis methods most relevant to this work.

% \textcolor{red}{Devin\&Tuan: a brief description of GCOWS observation. Observation table needed?}

The photometry for those young stars is extracted from a deep, wide mosaic image \citep{Lu_2013}. 
We applied the \edit1{full} extinction map from \cite{Schodel:2010} to correct extinction \edit1{with an average value of} $A_{ks}=2.7$.

\subsection{Sample Selection}
\label{sec:sample}
In this work, we included all spectroscopically identified young (early-type) stars with well measured radial velocities (RV) and proper motions.
To get a young star list \edit1{with well-understood completeness}, we combined new Galactic Center OSIRIS Wide-field Survey (GCOWS) observations (\S\ref{sec:obs_rv}) with previous GCOWS observations \citep{Do_2013} and with other spectral types from the literature.

\begin{figure}[htp]
\centering
\includegraphics[width=0.5\textwidth]{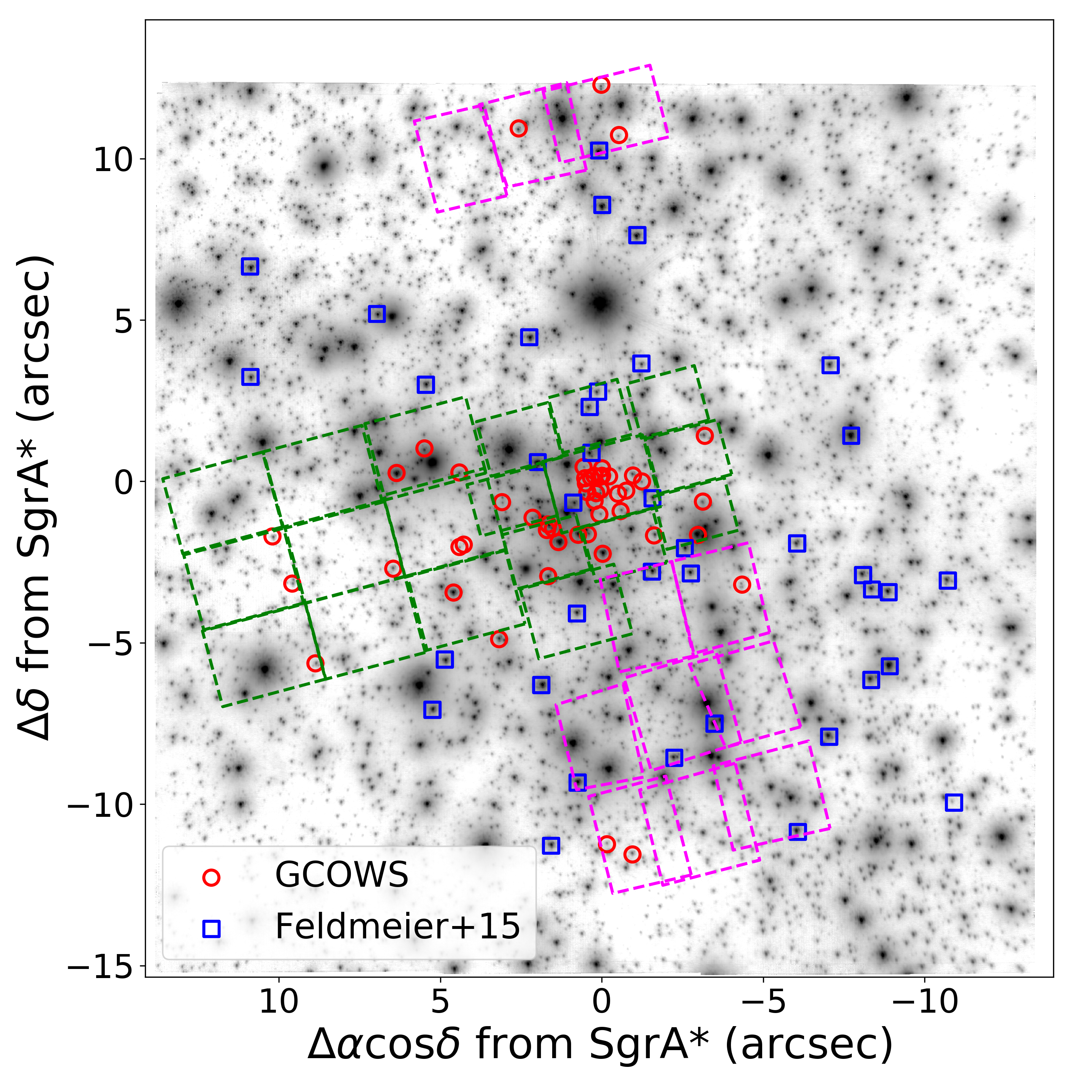}
\caption{The spatial distribution of our final sample of 88 stars. Red circles are the 50 stars from our GCOWS observations and blue squares are the 38 stars from \citet{Feldmeier_2015}. 
The dashed lines show our GCOWS sky coverage, where the green boxes are fields previously published by \citet{Do_2013} and magenta boxes are new areas first published in this paper.}
\label{fig:sample}
\end{figure}

The GCOWS survey consists of observations with the Keck OSIRIS spectrograph behind the laser-guide-star adaptive optics system on the W. M. Keck Observatory \citep{Larkin:2006}.
We obtained diffraction-limited, medium spectral-resolution (R $\sim$ 4000) spectra with the Kn3 filter (2.121-2.220 $\micron$). 
We used two different plate scales: 35 mas in the central fields where the stellar densities are highest and 50 mas for the outer fields having relatively lower stellar density.
Details on the GCOWS survey are presented by \citet{Do_2009} and \citet{Do_2013}, who investigated young stars located in the central region and eastern field in the GC (green boxes in Figure \ref{fig:sample}). 
In this work, we have added new observations in the South and North (magenta boxes in Figure \ref{fig:sample}). 
The new spectroscopic observations are reported in Table \ref{tab:newobs}.
Data were reduced using the latest version of the OSIRIS data reduction pipeline \citep{2017ascl.soft10021L, Lockhart_2019}.
This resulted in 7 more young stars with good quality RVs: S10-261, S10-48, S11-176, S11-21, S11-246, S12-76, S5- 106 (see \S\ref{sec:obs_rv} for reduction details).
The sample of GCOWS stars used in this work includes those that are spectroscopically identified as early-type and that have sufficient Signal-to-Noise ratio (SNR) to measure a radial velocity (RV). 

\begin{deluxetable*}{lccccccc}
\tablecolumns{8} 
\tablewidth{0pc} 
\tablecaption{Summary of New Keck OSIRIS Observations}
\tablehead{ 
    \colhead{Field Name}    &
    \colhead{Field Center\tablenotemark{a}}  &
    \colhead{Date}  &
	\colhead{$N_{\text{frames}} \times t_{\text{int}}$} &
	\colhead{Scale} &
	\colhead{FWHM\tablenotemark{b}} &
	\colhead{Filter} &
	\colhead{PA} \\
	\colhead{}  &
	\colhead{(")} &
    \colhead{(UT)} &
	\colhead{(s)} &
	\colhead{(mas)} &
	\colhead{(mas)} &
	\colhead{} &
	\colhead{($^{\circ}$)}
}
\startdata
N5-3	&  -0.37,11.21 &	2013-05-17 	&	$3\times900$	&	50  &  	75	&	Kn3     &    285     \\
S4-1	&	-0.87,-11.03    &   2014-05-20 	&	$6\times900$	&	50	&	90	&	Kn3	&	195 \\
S4-2	&	-3.19,-10.41    &   2014-05-20 	&	$8\times900$	&	50	&	89	&	Kn3	&	195 \\
S4-3	&	-5.50,-9.79    &   2014-06-05 	&	$9\times900$	&	50	&	140	&	Kn3	&	195 \\
S3-2	&	-2.35,-7.31    &   2014-07-18 	&	$6\times900$	&	50	&	94	&	Kn3	&	195 \\
N5-1	&	4.40,9.93    &   2016-07-21 	&	$5\times900$	&	50	&	77	&	Kn3	&	195 \\
N5-1	&	4.40,9.93    &   2016-07-22 	&	$3\times900$	&	50	&	61	&	Kn3	&	195 \\
N5-2	&	2.01,10.58    &   2016-07-22 	&	$5\times900$	&	50	&	98	&	Kn3	&	195 \\
S2-3	&	-3.80,-3.59    &   2019-05-25 	&	$5\times900$	&	50	&	89	&	Kn3	&	195 \\
S2-2	&	-1.49,-4.21    &   2019-05-27 	&	$5\times900$	&	50	&	    103	&	Kn3	&	195 \\
S3-1	&	-0.03,-7.95    &   2019-05-27 	&	$2\times900$	&	50	&	    135	&	Kn3	&	195 \\
S3-1	&	-0.03,-7.95    &   2019-07-08 	&	$3\times900$	&	50	&	    145	&	Kn3	&	195 \\
S3-3	&	-4.66,-6.67    &   2019-07-08 	&	$4\times900$	&	50	&	    133	&	Kn3	&	195 \\
\enddata
\tablenotetext{a}{R.A. and decl. offset from Sgr A* (R.A. offset is positive to the east).}
\tablenotetext{b}{Average FWHM of a relatively isolated star for the night, found from a two-dimensional Gaussian fit to the source.}
% \tablenotetext{c}{Taken with NIRC2 slit spectrograph}
% \tablecomments{}
\label{tab:newobs}
\end{deluxetable*}

Then, we combined our list of young stars with those identified by Paumard, Bartko, and Feldmeier \citep{Paumard_2006, Bartko_2009, Feldmeier_2015}.
S2-66 was claimed to be young by \citet{Paumard_2006}, but later proven to be old by \citet{Do_2009}, so this star was excluded.
From all the sources combined, the sample consisted of RVs for 149 young stars. 
Unfortunately, \citet{Paumard_2006} and \citet{Bartko_2009} did not report their spectral completeness curve, so we cannot use stars that are found only in their paper. \edit1{This is because we require spectroscopic completeness information for each star, which is crucial in \S\ref{sec:disk_simulation} when comparing observed and simulated stellar distributions.} However, we can still use their RV for stars that are identified in GCOWS or \citet{Feldmeier_2015}, which leaves us 91 stars with both RV and completeness correction curves.

Among those stars, we are able to extract proper motion for 88 stars from a combination of Keck AO observations in the inner region and HST observations in the outer region (see \S\ref{sec:obs_pm} for details).
The spatial distribution of our young star sample is plotted in Figure \ref{fig:sample}.
Although seven new young stars were identified from the new GCOWS observations,  S10-261 does not have a measured proper motion, so only six new stars are included in our sample, as shown in Figure \ref{fig:sample}.
In summary, we have 88 young stars in our sample, extending out to 14\arcsec\ ($\thicksim 0.5$ pc), down to a 90\% limiting magnitude of K$_\mathrm{lim}$= 15.3.

\subsection{Radial Velocities}
\label{sec:obs_rv}

As mentioned in \S\ref{sec:sample}, RVs used in this work come from two sources:
(1) Our GCOWS survey from Keck observations \citep{Do_2009, Do_2013}, and (2) other published RV data for Galactic Center, including \citet{Paumard_2006}, \citet{Bartko_2009}, \citet{Feldmeier_2015} and \citet{Zhu_2020}.

We derive radial velocities for all Keck OSIRIS data (both previously reported and new) using full spectral fitting with a synthetic spectral grid. We use the spectral fitting code StarKit \citep{Kerzendorf_2015} to fit the radial velocity along with physical properties such as effective temperature, surface gravity, metallicity, and rotational velocity. By fitting the physical parameters simultaneously, we can capture the effect of correlations between the parameters. We use the BOSZ spectral grid \citep{Bohlin_2017} to generate the spectra for our Bayesian inference model. Additional discussion of this method is given by \citet{Do_2018,Do_2019}. In general, the statistical uncertainties dominate the radial velocity measurement, but for the brightest sources, the systematic uncertainties dominate at the level of about 11 km s$^{-1}$ mainly due to residuals from the OH line subtraction. See \citet{Do_2019} for a complete discussion of radial velocity systematic uncertainties.

Both \citet{Paumard_2006} and \citet{Bartko_2009} used the AO-assisted, near-infrared integral field spectrometer SPIFFI/SINFONI on ESO VLT. 
Since \citet{Bartko_2009} is claimed to have improved measurements relative to \citet{Paumard_2006}, we will always adopt the RV from \citet{Bartko_2009} if the reported RVs \footnote{We assume stars are not in binary systems.} are different between the two papers.
\citet{Feldmeier_2015} used the integral-field spectrograph, KMOS, on VLT.
For IRS 13E2, IRS 13E3 and IRS 13E4, \citet{Zhu_2020} report the latest RVs with smaller uncertainties, so we adopt their measurements for those three stars.

We match the catalogs from the literature with star-lists from our high-resolution images based on stars' magnitudes and positions.
However, due to different spatial resolutions between our observations and other published observations, not all stars can be matched.
For example, star 3308 from \citet{Feldmeier_2015} is matched to a clump of 3 stars in our image, and it is difficult to determine which one produces the RV signal they report.
All RVs that are successfully matched to our catalogs are reported in Table \ref{tab:rv}.

For stars detected in GCOWS, we always use the GCOWS RV measurements. 
Most stars have only one detection, which is adopted as their final RV measurement.
Some stars in the central region  have multiple detections, which are marked with asterisks in Table \ref{tab:rv}.
For those multiple-detection stars, we will use the weighted mean RV if they are detected less than 5 times or show no significant physical acceleration.
However if stars show significant acceleration in either proper motion or RV (S0-1, S0-2, S0-3, S0-4, S0-5, S0-8, S0-16, S0-19, S0-20, see details in Chu et al. in prep), they will be fit with a full Keplerian orbit in \S\ref{sec:efit}.

For stars not detected in our GCOWS database, we use their literature RVs.
For stars reported multiple times in the literature, we use the weighted mean RV, where the weight, $w$, is:
\begin{equation}
w = \frac{1}{\sigma_\mathrm{RV}^{2}}
\end{equation}
All stars with RVs from the literature agree with each other within 2 sigma, except S7-236, for which we adopt the most recent RV from \cite{Feldmeier_2015}.

\subsection{Spectroscopic Completeness}
\label{sec:completeness}

\begin{figure}[htp]
\centering
\includegraphics[width=0.5\textwidth]{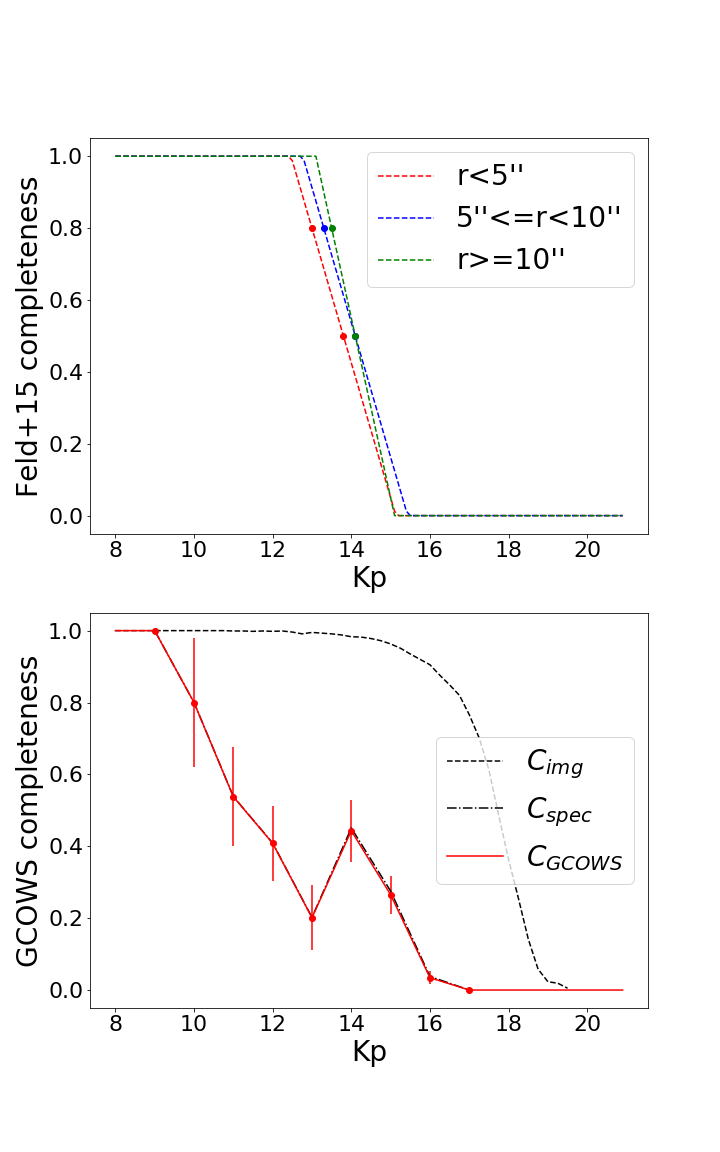}
\caption{Completeness curve as a function of magnitude $K_p$ for data from \cite{Feldmeier_2015} (top) and for our data (bottom).
Completeness for \citet{Feldmeier_2015} is a linear interpolation from the 80\% completeness and 50\% completeness point they reported in their paper.
Completeness also depends on radius; inner regions are less complete because of crowding. 
Completeness from our observations (red line) is a product of imaging completeness and spectral completeness, but is mostly determined by spectral completeness.
The dip in $K_p$ = 13 is probably a result of the fact that stars are transitioning to main sequence at that stage, so most spectra we get in that magnitude bin are featureless and are without usable RV.
}
\label{fig:completeness}
\end{figure}

In order to properly correct for incompleteness in our selection sample, we utilize results from star-planting simulations.
The sample was selected from two sources: \citet{Feldmeier_2015} and GCOWS, and we describe the completeness for each below.

For stars in the GCOWS observations, the completeness $C$ is a product of imaging completeness, $C_{img}$, and spectral completeness, $C_{spec}$. 
Imaging completeness is estimated using star-planting simulations (see details in Appendix C.1 of \citet{Do_2013}), and is 90\% complete down to Kp=16, as shown in Figure \ref{fig:completeness}.
\edit1{Uncertainties in the completeness are shown in the lower panel of Figure \ref{fig:completeness} and are derived from the number of observed stars at given magnitude (see details also in \citet{Do_2013}).}

For spectral completeness $C_{spec}$, we follow a process similar to that in \citet{Do_2013}, where each star from the GCOWS survey is assigned a probability of being young $P_E$. 
So for young stars, $P_E$ = 1; for old stars, $P_E$ = 0; for unknown type stars, $P_E$ is simulated based on the Bayesian evidence for the early-type and late-type hypotheses using the known type stars as a training sample. 
However, not all young stars have spectra with good enough quality to measure RV. 
Therefore, the completeness used in this work is defined as "completeness for young stars with measured RV".

Stars are divided into 8 magnitude bins based on extinction-corrected magnitudes, Kp$_\mathrm{ext}$, from 9 to 17 with 1 magnitude interval, and the spectral completeness curve is a linear interpolation.
In each magnitude bin, the completeness $C_{spec}$ is calculated using the following equation:

\begin{equation}
C_{spec} = \frac{N_{yngRV} + N_{yngWR}}{N_{yng} + \sum_{unk}P_{E}}
\end{equation}
where $N_{yngRV}$ is the number of young stars with well-measured RV, $N_{yngWR}$ is the number of WR young stars, $N_{yng}$ is the total number of young stars (including all spectrally identified young stars, no matter whether they have well-measured RV or not) and $\sum_{unk}P_{E}$ is the sum of the probability of being young for all unknown type stars. 
WR stars are all very bright and have high SNR spectra, but currently we cannot fit their emission lines due to lack of a good model.
So we decided to include WR stars in the numerator, since the missing RVs from them are not because of incompleteness.
The total completeness for young stars with RV is shown with the red line in Figure \ref{fig:completeness}.
Usually completeness will decrease towards fainter magnitudes, but a dip in the completeness curve appears at $Kp$ = 13.
This is because young stars are at the pre-main sequence turn-off point at this magnitude, so they are partially obscured by dust, making them harder to study.

For stars from \citet{Feldmeier_2015}, those authors reported 80\% and 50\% completeness in different radial bins. 
A linear interpolation is derived based on those two data points. 
Notice that their completeness is based on extinction-corrected magnitudes, $Kp_\mathrm{ext}$, but we assume it is a reasonable approximation to the completeness curve in the observed magnitude system.
The completeness curve for \citet{Feldmeier_2015} stars is shown in Figure \ref{fig:completeness}.

\begin{figure}[htp]
\centering
\includegraphics[width=0.45\textwidth]{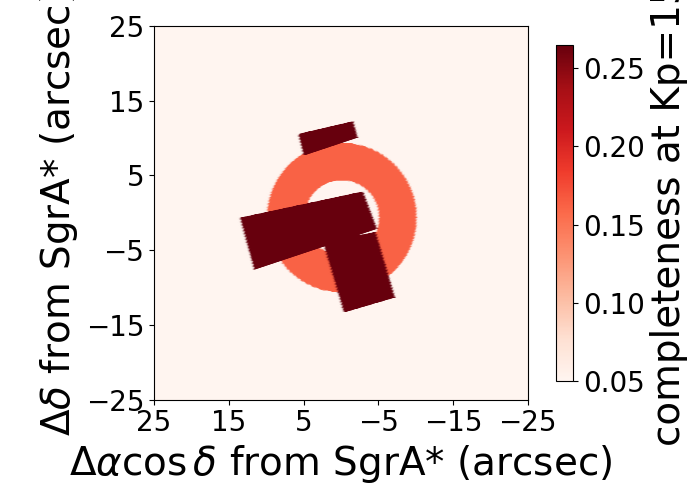}
\caption{An example of completeness map at $k_p$=15 with color showing the completeness.
For stars at this magnitude, completeness is higher in our curve, so GCOWS completeness is adopted for observed GCOWS regions (darker polygons) while completeness from \citet{Feldmeier_2015} is adopted elsewhere (two concentric circles separating regions with radial distances $r < 5''$, $5'' \leq r < 10''$, and $10'' \leq r$).}
\label{fig:completeness_example}
\end{figure}

When determining which completeness to use for each star, we first need to determine whether this star is within our GCOWS field. 
If a star is within our GCOWS field, we will use the higher completeness fraction between GCOWS and  \citet{Feldmeier_2015}.
Otherwise, completeness from \citet{Feldmeier_2015} will be applied.
An example of our completeness map at $K_p$ = 15 is shown in Figure \ref{fig:completeness_example}.

\subsection{Position, Proper motion and Acceleration}
\label{sec:obs_pm}
Projected positions and proper motions on the sky are derived from high-resolution, infrared (IR) images obtained over a 10$-$25 yr time-baseline. Depending on the distance from Sgr A*, we either use observations from the 10 m telescopes at the W. M. Keck Observatory (WMKO) or the Hubble Space Telescope (HST), as described below.

(1) The central 10\arcsec $\times$ 10\arcsec\; region of the GC (approximately centered on Sgr A*) has been monitored with diffraction-limited, near-infrared imaging cameras at WMKO since 1995.
%From 1995 to 2005, Speckle data sets were obtained in the K-band ($\lambda_0$ = 2.2$\micron$) using the Near Infrared Camera (NIRC; \cite{Matthews_1994; Matthews_1996}) on the Keck I telescope with a FOV $\sim$ 5\arcsec $\times$ 5\arcsec and a pixel scale of 20 mas. 
%Since 2005, we have used the Keck II LGS-AO system \citep{vanDam_2006, Wizinowich_2006} with the near infrared camera NIRC2 (PI: K.Matthews) in its narrow-field mode, which has a FOV of $\sim$ 10\arcsec $\times$ 10\arcsec and a plate scale of 9.952 mas/pixel \citep{Yelda_2010}. 
For stars in this region, we have the longest time baseline and the highest spatial resolution, which gives precise proper motions and even significant accelerations on the sky plane.
The complete catalog of measured positions, proper motions, and accelerations and analysis details is presented in \citet{Jia_2019}.

(2) To measure the proper motions of the young stars at larger radii, we use a widely dithered mosaic of shallow Keck IR images covering a 22\arcsec $\times$ 22\arcsec\ FOV as described in \citet{Sakai_2019}.   
%(2) To measure the proper motions of the young stars at larger radii, we used a widely dithered (6\arcsec $\times$ 6\arcsec) nine-point box pattern with multiple images at each of the nine positions.   
%Starting in 2012, additional observations of a four-point box pattern (with a smaller dither of 3\arcsec $\times$ 3\arcsec), with multiple images per dither position, were also obtained.
%The full field of the maser mosaic is 22\arcsec $\times$ 22\arcsec.
The astrometric uncertainties in this mosaicked dataset are typically larger than in the central 10\arcsec\ data, because of the shorter time baseline and lower SNR.

(3) For stars at even larger radii (R $>$ 7\farcs5), we use proper motions measured from the HST WFC3-IR instrument. 
This dataset consists of 10 epochs of observations centered on Sgr A* that were obtained between 2010 -- 2020 in the F153M filter (2010.5: GO 11671/PI Ghez, 2011.6: GO 11671/PI Ghez, 2012.6: GO 12318/PI Ghez, 2014.1: GO 13049/PI Do, 2018.1: GO 15199, PI Do, 2019.2: GO 15498/PI Do, 2019.6: GO 16004/PI Do, 2019.7: GO 16004/PI Do 2019.8: GO 16004/PI Do, 2020.2: GO 15894/PI Do).
While the HST spatial resolution is $\sim$2.5 times lower than that achieved with the Keck observations (FWHM $\sim$ 0.17" versus FWHM $\sim$ 0.06"), HST's FOV of 120" $\times$ 120" is much larger than can be realistically achieved with current AO systems. 
The astrometry from each HST epoch is first transformed into the \emph{Gaia} absolute reference frame \citep{Mignard_2018} and then further transformed into the AO reference frame via 2nd-order bivariate polynomial transformations. 
The resulting HST catalog achieves an average precision of 0.33 mas and 0.07 mas/yr for the positions and proper motions of the stars in the sample.
A detailed description of the HST catalog will be provided in a future paper (Hosek et al., in prep). 

%\textcolor{red}{[mwhosek: Are you planning to include some sort of table for the HST observations? We could use this to specify things like date of each epoch, some estimate of depth, PI/GO number of each observation, etc]}

%\textcolor{red}{[mwhosek: How much detail would you like here? The HST proper motion catalog has never been described in the literature. So, do you want things like how the catalog was constructed (e.g. astrometric transformations to the Gaia reference frame) and how the Gaia reference frame compares to the AO reference frame (and the offsets applied to get them in the same frame)?}

%\textcolor{red}{WAIT FOR Matt for a brief description of HST proper motion}.

In summary, among 91 young stars from our sample described in \S\ref{sec:sample}, we are able to cross-match and measure proper motions for 88 stars. 
The proper motions for the final sample include 54 stars 
%of 54 stars are 
from data set 1, 20 %are 
from data set 2, and 14 %are 
from data set 3, as shown in Table \ref{tab:pm}.

\subsection{Photometry and Extinction}
To ensure that our final sample shares a common photometric system, we adopt the Kp magnitude for each star from the deep wide mosaic image analysis reported in \citet{Lu_2013} which covers all 88 stars in our sample. 
The Kp magnitude for each star is reported in the Kp column in Table \ref{tab:pm}.
Then we applied the latest extinction map from \citet{Lara_2018} to differentially de-redden all stars to a common $A_{Ks} = 2.7$, and the extinction corrected magnitude is reported as $Kp_\mathrm{ext}$ in Table \ref{tab:pm}.

\section{ORBIT ANALYSIS}
\label{sec:orbitfit}

For stars with measured ($x_0$, $y_0$, $v_x$, $v_y$, $v_z$, $a_R$),  the six Keplerian orbit parameters (inclination $i$, angle to the ascending node $\Omega$, time of periapse passage $T_0$, longitude of periapse $\omega$, period $P$, and eccentricity $e$) can be analytically determined if the central potential is known \citep{Lu_2009}. 
\S\ref{sec:mc} describes the Monte Carlo process used to estimate stars' orbital parameters given prior estimates on the central potential.
For stars that show significant acceleration (S0-1, S0-2, S0-3, S0-4, S0-5, S0-8, S0-16, S0-19, S0-20), their orbits are best constrained by simultaneously fitting the astrometry and RV measurements as a function of time \citep{Do_2019}.
\S\ref{sec:efit} describes the computationally expensive orbit fitting procedure used for these 9 stars.

\subsection{MC analysis for stars with ($x_0$, $y_0$, $v_x$, $v_y$, $v_z$, $a_R$) }
\label{sec:mc}

\begin{figure*}[htp]
\centering
\includegraphics[width=\textwidth]{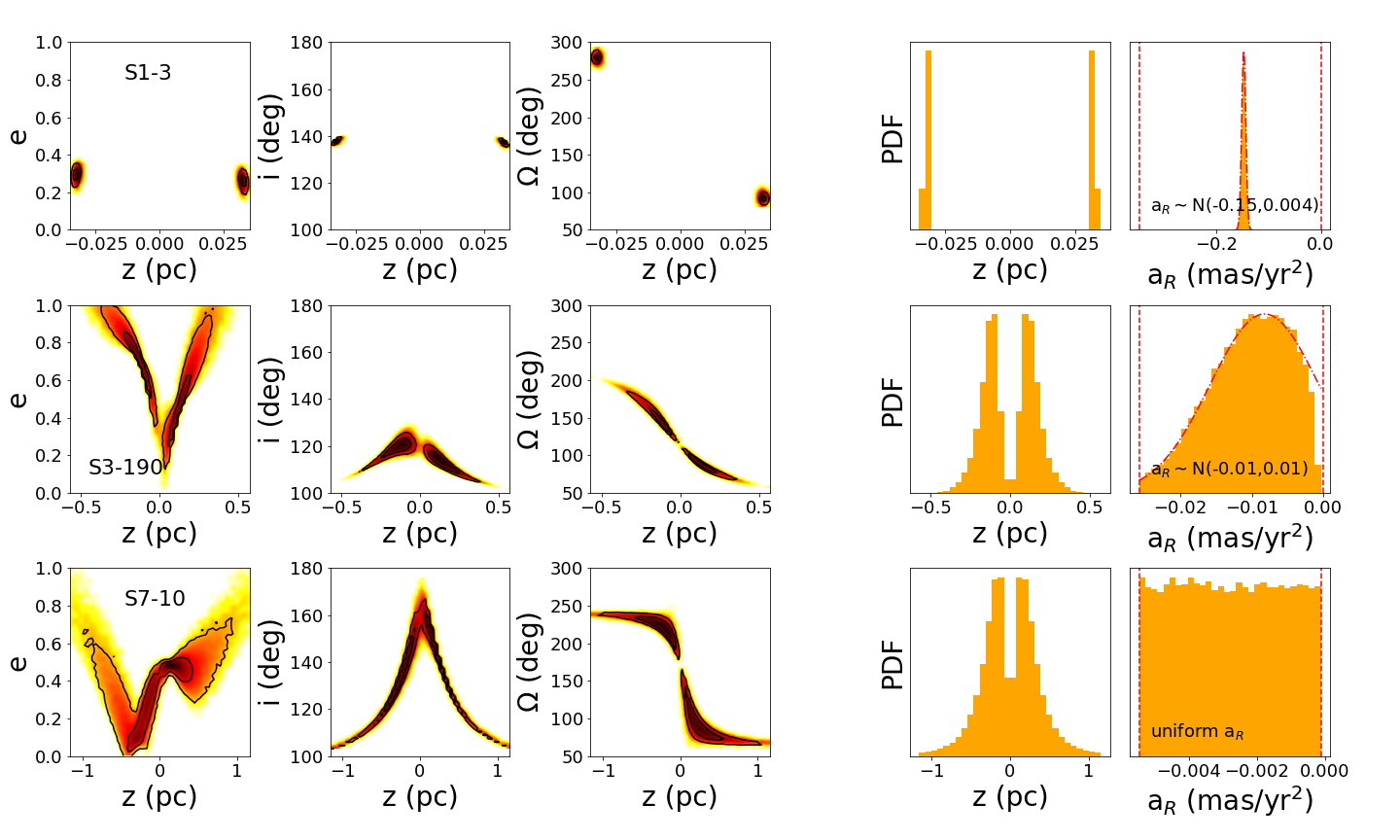}
\caption{Examples of MC analysis for S1-3 (top), S3-190 (middle) and S7-10 (bottom). 
The left three columns show the density map for orbital parameters: $e$, $i$, $\Omega$. 
The right two columns show the MC simulation input prior for $z$ and $a_R$ in orange histograms.
For S1-3 and S3-190, their measured $a_R$ and $\sigma_{a_R}$ (red dashed line) are within the upper and lower limits (vertical red dashed lines), so the simulated $a_R$ are drawn from Gaussian distribution N $\sim$ ($a_R$, $\sigma_{a_R}$). 
For S7-10, no $a_R$ is measured, so it is drawn from a uniform distribution between ${a_R}_{min}$ and ${a_R}_{max}$.
Simulated $z$ is then derived from simulated $a_R$ using Equation \ref{eqn:z}.
}
\label{fig:mc}
\end{figure*}

Orbital parameters are determined for the 79 stars without significant acceleration measurements using a Monte Carlo (MC) analysis as described in detail in \citet{Lu_2009} and \citet{Yelda_2014}.
Each star has measurements of the line-of-sight velocity ($v_z$) as described in \S\ref{sec:obs_rv} and proper motion parameters ($x_0$, $y_0$, $v_x$, $v_y$) as described in \S\ref{sec:obs_pm}. 
The absolute value of the line-of-site distance |$z$| between the star and Sgr A$^*$ can be calculated from the following equation if $a_R$ is known.
\begin{equation}
\label{eqn:z}
a_R = - \frac{G M_{tot}(r) R}{r^3}, r = \sqrt{R^2 + z^2}
\end{equation}
Here, $r$ is the 3D distance and $R$ is the 2D projected sky-plane distance from Sgr A*, where $r^2 = R^2 + z^2$.
We note that there is a sign ambiguity in the line-of-sight distance.

%To convert the measurements to orbital parameters and their uncertainties, we perform a Monte Carlo (MC) simulation 
%as described in \cite{Lu_2009} and \cite{Yelda_2014}.
We sample the 6 measured position, velocity, and acceleration quantities $10^5$ times assuming a Gaussian distribution for each measurement and its uncertainty.
For each sample, the 6 Keplerian orbital parameters are analytically determined assuming an enclosed mass and distance to the Galactic Center as described below. 
The MC simulations produce a joint probability density function (PDFs) for the 6 orbital parameters.

The mass distribution giving rise to the central potential, $M_{tot}$, is a combination of the SMBH mass and an extended mass profile:
\begin{equation}
M_{tot}(r) = M_{BH} + M_{ext}(r)
\end{equation}
The adopted SMBH properties include a mass of $M_{BH}$ = (3.975 $\pm$ 0.058) $\times$ $10^6$ $M_{\odot}$ and a distance of $R_0$ = (7.959 $\pm$ 0.059) kpc, based on the analysis of S0-2's orbit \citep{Do_2019}.
We used the extended mass profiles from \citet{Trippe_2008} and found that adding extended mass has limited impact on the orbit analysis. 
We nonetheless adopt an extended mass profile with
\begin{equation}
M_{ext}(r) = \int_0^r 4\pi  \frac{\rho_0}{1+(r/R_b)^2} r^2 dr 
\end{equation}
where $\rho_0 \simeq 2.1 \times 10^6 M_{\odot}$ and the break radius is $R_b$ = 8.9\arcsec.
We have also tried other extended mass profiles, like that of  \cite{Schodel_2009}, and the results are similar.

\begin{figure*}[htp]
\centering
\includegraphics[width=\textwidth]{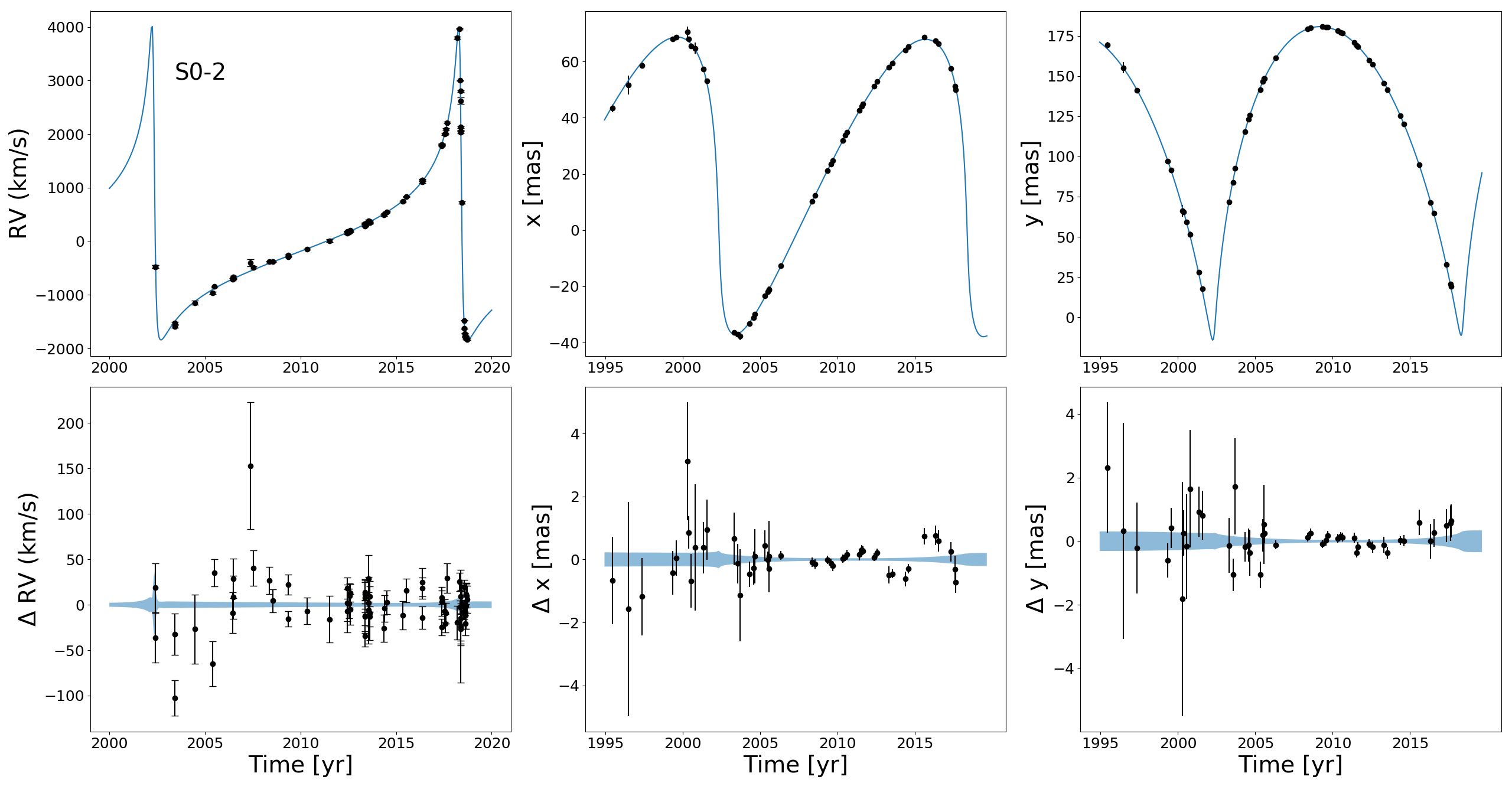}
\caption{Examples of a full Keplerian orbit analysis for S0-2.
The top row shows the line-of-sight velocity $v_z$ and projected position from SgrA*: $x$, $y$ as a function of time, and our model (blue line) fits both very well.
The bottom row shows the offset between data and model as a function of time.}
\label{fig:efit_S0-2}
\end{figure*}

Among the 79 stars, 45 have 
precise and accurate astrometry from \citet{Jia_2019} with well-measured proper motions and constraints on the projected acceleration, $a_R$, as shown in Table \ref{tab:pm}.
However, not all measured $a_R$ values are physically allowed for a single star on a bound orbit around the supermassive black hole and enclosed extended mass.
The maximum allowed $a_R$ is set by the gravitational acceleration when $z$ = 0 pc and we constrain the minimum allowed $a_R$ to be the acceleration at a distance of $z$ = 0.8 pc:
\begin{equation}
\begin{split}
{a_R}_{max} (R) & = - \frac{G M_{tot}(R)}{R^3} \\
{a_R}_{min} (R) & = - \frac{G M_{tot}(r)R}{r^3}, r = \sqrt{R^2 + z^2}, z=0.8pc \\
\end{split}
\end{equation}
The maximum $z$ is chosen to be 0.8 pc because very few young stars are detected outside 0.4 pc and the most distant young star detected in our sample is at 0.6 pc.
If the distribution of young stars is approximately spherically symmetric about the black hole, then the maximum line-of-sight distance should not exceed 0.8 pc.
If a stars' measured $a_R$ ($\pm$ 2$\sigma_{a_R}$) overlaps with the allowed range, we draw its $z$ in each MC trial from a Gaussian distribution with the mean set to the measured $a_R$, the standard deviation set to the measured $\sigma_{a_R}$, and truncated to the allowed $a_R$ range.
This is the case for 35 stars.
For the rest of the stars without a significant or physical $a_R$, we use a uniform distribution in the range of allowed $a_R$.
In Figure \ref{fig:mc}, we use three stars as an example to show how $a_R$ is simulated in the MC analysis. The last column in that plot shows the distribution of $a_R$ and its limits (${a_R}_{max}$ and ${a_R}_{min}$) in red dashed lines, while the orange histogram is the probability density function for simulated $a_R$.  

\edit1{The uniform acceleration prior, in the absence of other constraints, may lead to biased results in angular momentum measurements as shown in \citet{Yelda_2014}.
It produces a deficit of stars close to z=0 as shown in Figure \ref{fig:mc} and \citet{von_Fellenberg_2022}, Fig. 12.
However, as discussed above, we only apply this prior for roughly half of the stars that do not have a physically allowed or significant $a_R$, most of which are located at larger radii where the bias is less apparent. 
Thus the choice of a uniform acceleration prior does not strongly bias the results.}

Even for those 35 stars with measured $a_R$ within the allowed range, the significance of their accelerations varies. 
This is due to many factors, including the location of the star, the brightness of the star, the number of epochs in which the star is detected, etc.   
A more precise acceleration will result in more precise orbital parameters. 
In Figure \ref{fig:mc}, both S1-3 and S3-190 have a measurement-based $a_R$ prior, while S7-10's prior is uniform $a_R$ over the allowed range.
S1-3 has a $\sim$40$\sigma$ significant acceleration, but S3-190 only has 1$\sigma$ significant acceleration.
As a result, S1-3 has much more constrained orbital parameters compared to S3-190.
For S7-10 with a uniform $a_R$ prior, its orbital parameters are even less constrained.
The $z$ distribution for S7-10 decreases with increasing radius even when $a_R$ is evenly distributed, which agrees with observation and validates the uniform $a_R$ prior.

\subsection{Full Keplerain orbit fit for central stars}
\label{sec:efit}

For stars with time-variable RV, we inferred their orbital parameters by simultaneously fitting spectroscopic and astrometric measurements.
We utilized the same orbit-fitting procedure as was used by \citet{Do_2019} to test General Relativity using the orbit of S0-2.
From the orbit-fit posterior distributions, we drew 10$^\mathrm{5}$ samples in order to match the posterior sample size for stars with a single RV from \S\ref{sec:mc}.
S0-1, S0-2, S0-3, S0-4, S0-5, S0-8, S0-16, S0-19, S0-20 in our sample are fitted this way.
We adopted the observable-based prior paradigm from \citet{ONeil_2019} that is based on uniformity in observables to improve orbital solutions for low-phase-coverage orbits.
A full Keplerian orbit fit for S0-2 is shown in Fig \ref{fig:efit_S0-2}.
All 9 stars' fitted results are attached in appendix \ref{app:efit}.

\section{Disk Membership Analysis}
\label{sec:disk}

\subsection{Detecting Stellar Disks}
\label{sec:disk_detect}

Each star's orbital plane can be uniquely described by a unit normal vector $\textbf{\textit{L}}$ that is perpendicular to the orbital plane. 
This normal vector $\textbf{\textit{L}}$ can be expressed in terms of the inclination ($i$) and the angle to the ascending node ($\Omega$) (see Equation 8 in \citet{Lu_2009}). 
Stars moving in a common plane share a common normal vector.
In order to detect a stellar disk or stream, we adopt a nearest-neighbor method similar to that used by \citet{Lu_2009} and \citet{Yelda_2014}. 
In this method, the sky is divided into 49152 pixels with equal solid angle area and for a given  MC simulation the density at each ($i$, $\Omega$) position is calculated using the following equation:
\begin{equation}
    \Sigma = \frac{k}{2\pi (1 - \textrm{cos} \;\theta_k)} \; \textrm{stars} \; \textrm{sr}^{-1}
\end{equation}
where $\theta_k$ is the angle to the $k$th nearest star and we use $k$ = 6. The resulting average density is nearly the same for other choices of $k$ = 4, 5, or 7.
Then the combined density map is an average over all 10$^5$ MC trials from \S\ref{sec:orbitfit}. The resulting density maps are presented in \S\ref{sec:result}.

\begin{figure}[htp]
\centering
\includegraphics[width=0.5\textwidth]{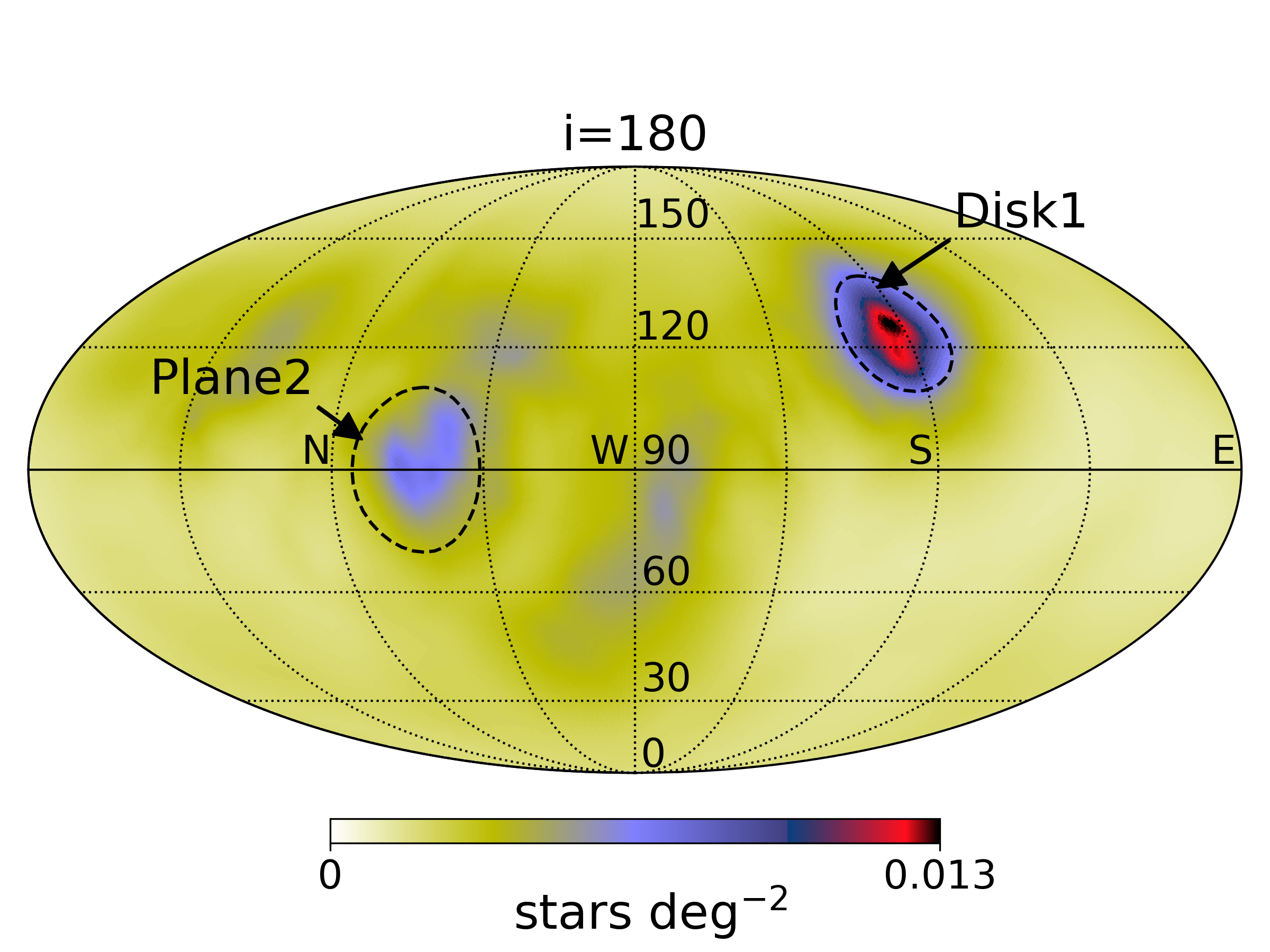}
\caption{Density of normal vectors ($i$, $\Omega$) (in stars deg$^{-2}$) of all 88 stars in our sample. 
Two significant over-dense region are marked as Disk1 and Plane2.
Disk1 has a density of 0.014 stars deg$^{−2}$ located at ($i$, $\Omega$) = (124\degree, 94\degree).
Plane2 has a density of 0.007 stars deg$^{-2}$ located at ($i$, $\Omega$) = (90\degree, 245\degree).
The 1 sigma region is defined as the area inside the contour at which the density drops to half of its peak, and is circled by a black dashed line.}
\label{fig:disk}
\end{figure}

\subsection{Disk/Plane Membership Probabilities}
\label{sec:membership}
With the disk or plane normal-vectors and uncertainties determined from \S\ref{sec:disk_detect}, the membership probability, $L_{\text{disk}}$, can be estimated
for each star following \citet{Lu_2009}.

\begin{equation}
\begin{split}
\label{equ:disk_membership}
L_{\text{non-disk}} &= 1 - L_{\text{disk}} \\
         &= 1 - \left(\frac{\int_{\text{disk}} PDF(i, \Omega) \;d \text{SA}}{\int_{\text{peak}} PDF(i, \Omega) \;d \text{SA}} \right) \\
%         &= \frac{\int_{\text{disk}} PDF(i, \Omega) \;d \text{SA}}{\int_{\text{peak}} PDF(i, \Omega) \;d \text{SA}}
\end{split}
\end{equation}

\begin{equation}
\begin{split}
\int_{\text{disk}} d \text{SA} = \int_{\text{peak}} d \text{SA} 
\end{split}
\end{equation}

Here SA is the solid angle measured at the contour where the disk density drops to half of the peak value; $\int_{disk} PDF(i, \Omega) \;d \text{SA}$ is the integration of each star's density map over the stellar disk region, and  $\int_{peak} PDF(i, \Omega) \;d \text{SA}$ is the integration over its own peak region with the same SA.

In summary, each star's $PDF(i, \Omega)$ is integrated inside the disk or plane area and normalized by the star's peak probability over a similar area. 
Thus, stars with large uncertainties in $i$ and $\Omega$ will only have high plane membership if they are centered on the plane.
The final disk or plane memberships are presented in \S\ref{sec:result}.

\section{Results}
\label{sec:result}

\subsection{Two Stellar Disks}
\label{sec:two_disk}

\begin{figure}[htp]
\centering
\includegraphics[width=0.5\textwidth]{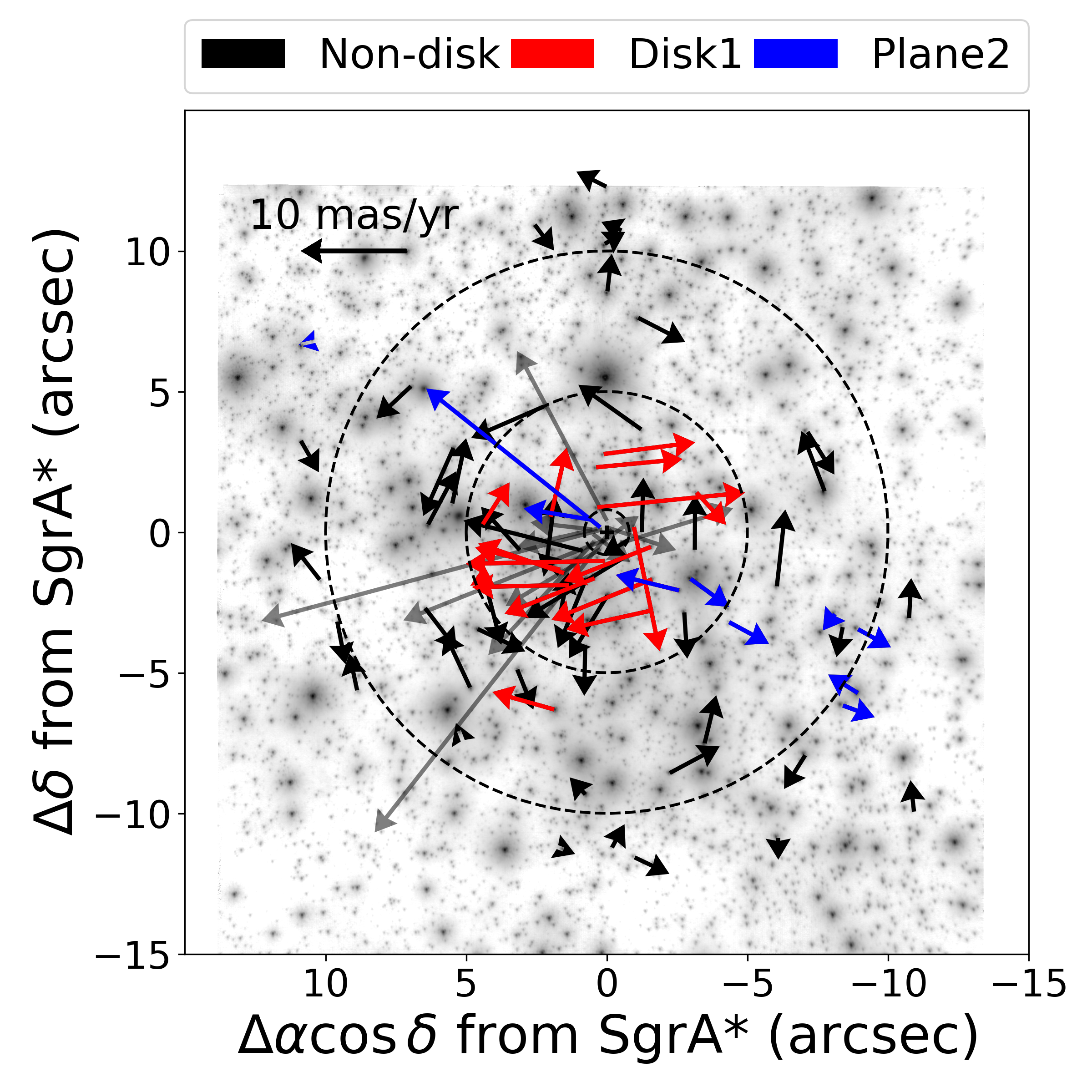}
\caption{The proper motion direction for each star, where red arrows are Disk1 candidates and blue arrows are Plane2 candidates.
The Non-disk stars are shown in black arrows, and the inner region stars are plotted with low opacity due to their large velocity.
}
\label{fig:disk_membership}
\end{figure}

The density map of normal vectors, $\textbf{\textit{L}}$, shows two over-dense regions, indicating the presence of at least two distinct populations, each of which consists of stars that share a common orbital plane (Figure \ref{fig:disk}).
We label the two peaks as Disk1 and Plane2.
The well-known clockwise (CW) disk is the upper right Disk1 located at ($i_\mathrm{Disk1}$, $\Omega_{i, \mathrm{Disk1}}$) = (124\degree, 94\degree), consistent with the past measurements of the disk location \citep{Levin_2003, Genzel_2003, Paumard_2006, Lu_2009, Bartko_2009, Yelda_2014, von_Fellenberg_2022}.
We found another almost edge-on plane which we call Plane2, located at ($i_\mathrm{Plane2}$, $\Omega_{i, \mathrm{Plane2}}$) = (90\degree, 245\degree). 
This may be the same structure identified as F3 in \citet{von_Fellenberg_2022}; \edit1{Plane2 and F3 have similar eccentricity and semi-major axis distributions, along with $\sim 35\%$ of common members. However, the two features are $>$30$\degree$ apart in terms of $\Omega$: \citet{von_Fellenberg_2022} found F3 has ($i$, $\Omega$) = (102\degree, 211\degree) after conversion to our reference frame. A more} detailed comparison is presented in Appendix \ref{app:comp}.
The uncertainty in the location of each of the planar features is defined to be where the density drops to 50\% of its peak value and it is marked with a black dashed line in Figure \ref{fig:disk}.
The uncertainties for Disk1 and Plane2 are ($\sigma_{i,\mathrm{Disk1}}$, $\sigma_{\Omega, \mathrm{Disk1}}$) = (15\degree, 17\degree),  ($\sigma_{i, \mathrm{Plane2}}$, $\sigma_{\Omega,\mathrm{Plane2}}$) = 
(20\degree, 19\degree).
\edit2{We also estimate the fraction of young stars belonging to each structure by calculating the sum of the membership probabilities over the total 88 young stars and found that $\sim 8.4\%$ of our sample belongs to Disk1 while $\sim 6\%$ belongs to Plane2.}

To quantify the significance of both disks, we simulated an isotropic population with synthetic ($x_0$, $y_0$, $v_x$, $v_y$, $v_z$) and extracted orbits in the same way as on the real data.
Each simulated star was first assigned a 2D radius on the sky, $R$, and a 3D velocity, $v_{tot}$, drawn from the observed young stars. Then the orientation of the position and velocity vectors were randomly generated \citep{Yelda_2014}. Any measurements of $a_R$ were kept fixed in amplitude and uncertainty. 
For the 9 stars from \S\ref{sec:efit} with well-measured orbits, we drew 10$^5$ samples from  randomized $i$ and $\Omega$ with their posterior distribution uncertainty.
The significance of Disk1 and Plane2 is defined by:
\begin{equation}
\label{eq:sig}
S = \frac{\rho_{disk} - \rho_{iso}}{\sigma_{iso}}
\end{equation}
where $\rho_{disk}$ is the density of the stellar disk at its peak, $\rho_{iso}$ is the density of the isotropic simulation at the same position, and $\sigma_{iso}$ is the dispersion of the density at the peak among all the isotropic simulations at the same position.
A summary of the peak density, significance, and the simulated isotropic density and its dispersion is presented in Table \ref{tab:disk}. 

The significance for Disk1 is 12.4 and Plane2 has a significance of 6.4. 
The reason we have a slightly lower significance for Disk1 compared to \citet{Yelda_2014} is because we have a smaller sample size due to our requirement that only stars from surveys with published completeness curves can be included. As a result, our uncertainty in the isotropic density is slightly larger. 

The significance of Plane2 is only slightly higher than previously claimed planar structures and disks that were later shown to be statistically insignificant \citep{Paumard_2006,Bartko_2009,Lu_2009}. 
Thus, we must be cautious in claiming a new dynamical structure.
Unlike previous claimed structures, the Plane2 structure is the result of both numerous stars with moderate membership probabilities and at least 5 stars with $P_{plane} > 0.5$.

\begin{table*}
\caption{Disk and Plane Summary}
\label{tab:disk}
\begin{center}
\begin{tabular}{lccccccc}
    \toprule
    Structure & $i$ & $\Omega$ & Solid Angle & $\rho_{disk}$ & $\rho_{iso}$ & Significance & Disk Fraction \tablenotemark{a} \\
     & (\degree) & (\degree) & (sr) & (stars/deg$^{2}$) & (stars/deg$^{2}$) & & \\
    \midrule
    Disk1 & 124 $\pm$ 15 & 94 $\pm$ 17 & 0.20 & 0.013 $\pm$ 0.006 & 0.003 $\pm$ 0.0008 & 12.4 & 0.084\\
    Plane2 & 90 $\pm$ 20 & 245 $\pm$ 19 & 0.37 & 0.007 $\pm$ 0.002 & 0.002 $\pm$ 0.0007 & 6.4 & 0.060\\
    \bottomrule
\end{tabular}
\end{center}
\tablenotetext{a}{This is calculated as sum of membership probabilities of stars belonging to each structure divided by the total 88 young stars.}
\end{table*}

The probability of each star belonging to Disk1 or Plane2 is given  in Table \ref{tab:disk_membership}. 
To view stars in Disk1 and Plane2 in a more direct way, we make a quiver plot showing the proper motion of each star in Figure \ref{fig:disk_membership}.
For illustrative purposes, we identify high-probability disk members as those with $P_{disk} > 0.2$; and Disk1 and Plane2 stars are shown as red and blue arrows, respectively. 
We note that all disk membership probabilities are calculated assuming the existence of only that disk. 
We choose not to determine disk membership using a more complex mixture model based on two simulated disks plus an isotropic population as it would require a prior knowledge of the disk properties such as the eccentricity and semi-major axis distributions. 

\startlongtable 
\begin{deluxetable}{lrrr} 
\tablecaption{Disk Membership\label{tab:disk_membership}}
\tabletypesize{\footnotesize}
\tablehead{\colhead{Name} & \colhead{P$_\mathrm{Disk1}$}  & \colhead{P$_\mathrm{Plane2}$} & \colhead{SA}}
\startdata
S0-1          &   0.00 &   0.00 &  0.001   \\ 
S0-11         &   0.00 &   0.05 &  0.081   \\ 
S0-14         &   0.00 &   0.00 &  0.004   \\ 
S0-15         &   0.76 &   0.00 &  0.004   \\ 
S0-16         &   0.00 &   0.42 &  0.001   \\ 
S0-19         &   0.00 &   0.00 &  0.002   \\ 
S0-2          &   0.00 &   0.00 &  0.000   \\ 
S0-20         &   0.00 &   0.00 &  0.001   \\ 
S0-3          &   0.00 &   0.00 &  0.000   \\ 
S0-30         &   0.00 &   0.00 &  0.106   \\ 
S0-31         &   0.00 &   0.30 &  0.004   \\ 
S0-4          &   0.00 &   0.00 &  0.001   \\ 
S0-5          &   0.00 &   0.00 &  0.001   \\ 
S0-7          &   0.00 &   0.00 &  0.001   \\ 
S0-8          &   0.00 &   0.00 &  0.001   \\ 
S0-9          &   0.00 &   0.00 &  0.003   \\ 
S1-19         &   0.43 &   0.00 &  0.022   \\ 
S1-2          &   0.50 &   0.00 &  0.004   \\ 
S1-22         &   0.54 &   0.00 &  0.189   \\ 
S1-24         &   0.00 &   0.00 &  0.003   \\ 
S1-3          &   0.49 &   0.00 &  0.007   \\ 
S1-33         &   0.00 &   0.00 &  0.122   \\ 
S1-4          &   0.02 &   0.00 &  0.081   \\ 
S1-8          &   0.00 &   0.00 &  0.003   \\ 
S10-185       &   0.00 &   0.73 &  0.132   \\ 
S10-232       &   0.00 &   0.00 &  0.578   \\ 
S10-238       &   0.00 &   0.78 &  0.002   \\ 
S10-32        &   0.18 &   0.00 &  0.119   \\ 
S10-34        &   0.00 &   0.00 &  0.517   \\ 
S10-4         &   0.00 &   0.00 &  0.087   \\ 
S10-48        &   0.00 &   0.00 &  0.037   \\ 
S10-50        &   0.00 &   0.00 &  0.400   \\ 
S11-147       &   0.00 &   0.00 &  1.322   \\ 
S11-176       &   0.00 &   0.00 &  0.237   \\ 
S11-21        &   0.04 &   0.00 &  0.174   \\ 
S11-214       &   0.00 &   0.00 &  0.131   \\ 
S11-246       &   0.00 &   0.01 &  0.209   \\ 
S11-8         &   0.00 &   0.00 &  0.784   \\ 
S12-178       &   0.00 &   0.03 &  0.008   \\ 
S12-5         &   0.00 &   1.00 &  0.003   \\ 
S12-76        &   0.00 &   0.00 &  0.257   \\ 
S14-196       &   0.00 &   0.00 &  0.410   \\ 
S2-17         &   0.37 &   0.00 &  0.009   \\ 
S2-19         &   0.24 &   0.00 &  0.071   \\ 
S2-21         &   0.50 &   0.00 &  0.018   \\ 
S2-4          &   0.22 &   0.00 &  0.007   \\ 
S2-50         &   0.00 &   0.00 &  0.188   \\ 
S2-58         &   0.00 &   0.00 &  0.188   \\ 
S2-6          &   0.50 &   0.00 &  0.009   \\ 
S2-74         &   0.50 &   0.00 &  0.044   \\ 
S3-19         &   0.36 &   0.00 &  0.357   \\ 
S3-190        &   0.35 &   0.00 &  0.084   \\ 
S3-26         &   0.24 &   0.24 &  0.163   \\ 
S3-3          &   0.00 &   0.00 &  0.467   \\ 
S3-30         &   0.00 &   0.00 &  0.012   \\ 
S3-331        &   0.00 &   0.00 &  0.103   \\ 
S3-374        &   0.00 &   0.00 &  0.086   \\ 
S3-96         &   0.00 &   0.00 &  0.351   \\ 
S4-169        &   0.40 &   0.00 &  0.420   \\ 
S4-262        &   0.00 &   0.00 &  0.117   \\ 
S4-314        &   0.47 &   0.00 &  0.231   \\ 
S4-364        &   0.00 &   0.00 &  0.296   \\ 
S4-71         &   0.00 &   0.00 &  0.012   \\ 
S5-106        &   0.00 &   0.34 &  0.029   \\ 
S5-183        &   0.00 &   0.00 &  0.094   \\ 
S5-191        &   0.00 &   0.00 &  0.612   \\ 
S5-237        &   0.00 &   0.00 &  0.247   \\ 
S6-63         &   0.27 &   0.00 &  0.215   \\ 
S6-81         &   0.00 &   0.00 &  0.086   \\ 
S6-89         &   0.00 &   0.00 &  0.398   \\ 
S6-96         &   0.00 &   0.00 &  0.411   \\ 
S7-10         &   0.16 &   0.03 &  0.285   \\ 
S7-216        &   0.00 &   0.00 &  0.153   \\ 
S7-236        &   0.00 &   0.00 &  0.873   \\ 
S7-30         &   0.00 &   0.00 &  0.592   \\ 
S7-5          &   0.05 &   0.00 &  0.499   \\ 
S8-10         &   0.00 &   0.00 &  0.769   \\ 
S8-126        &   0.00 &   0.00 &  1.139   \\ 
S8-196        &   0.00 &   0.22 &  0.142   \\ 
S8-4          &   0.00 &   0.00 &  0.032   \\ 
S8-5          &   0.00 &   0.00 &  0.370   \\ 
S8-8          &   0.01 &   0.00 &  0.271   \\ 
S9-143        &   0.01 &   0.00 &  1.220   \\ 
S9-221        &   0.00 &   0.77 &  0.044   \\ 
S9-6          &   0.04 &   0.00 &  0.267   \\ 
IRS 13E1      &   0.00 &   0.50 &  0.002   \\ 
IRS 16CC      &   0.25 &   0.00 &  0.070   \\ 
IRS 33N       &   0.00 &   0.00 &  0.004   \\ 
\enddata
\end{deluxetable}

\subsection{Disk Properties}
\label{sec:disk_property}

%Using the criteria defined in previous section (p$_\mathrm{prob}$ > 0.2), we divide stars into three groups: disk1, disk2 and non-disk.
In this section, we compare the distribution of eccentricities ($e$), radial distances on the disk plane $R_{plane}$ (derived in \S\ref{sec:radial_profile}) and disk thickness for the different dynamical subgroups. We then use this distribution as the input prior for disk simulations in \S\ref{sec:disk_simulation}.

\subsubsection{Eccentricity Distribution}
\label{sec:ecc_dist}

Stars are divided into Disk1, Plane2 and Non-disk stars using membership probabilities as weights rather than through a hard probability cut.
We assign a weight to each MC trial among the $10^5$ trials for every star (\S\ref{sec:orbitfit}).
For a particular MC trial with a set of Keplerian orbital parameters ($i$, $\Omega$, $e$, $a$), the weight of being on Disk1 (W$_\mathrm{Disk1}$), Plane2 (W$_\mathrm{Plane2}$) and Non-disk (W$_\mathrm{Non-disk}$) structures are calculated as:
\begin{equation}
\begin{split}
W_\mathrm{Disk1}(i,\Omega,e,a) = & N(i|\mu=i_\mathrm{Disk1},\sigma=\sigma_{i,\mathrm{Disk1}}) \times \\
& N(\Omega|\mu=\Omega_\mathrm{disk1},\sigma=\sigma_{\Omega,\mathrm{Disk1}}) \\
W_\mathrm{Plane2}(i,\Omega,e,a) = & N(i|\mu=i_\mathrm{Plane2},\sigma=\sigma_{i, \mathrm{Plane2}}) \times \\ 
& N(\Omega|\mu=\Omega_\mathrm{Plane2},\sigma=\sigma_{\Omega, \mathrm{Plane2}})\\
W_\mathrm{Non-disk}(i,\Omega,e,a) = & 1 - W_\mathrm{Disk1}(i,\Omega,e,a) \\ & - W_\mathrm{Plane2}(i,\Omega,e,a) \\
\end{split}
\end{equation}
where N($\mu$, $\sigma$) is normal distribution with mean of $\mu$ and standard deviation of $\sigma$.

\begin{figure*}[htp]
\centering
\includegraphics[width=0.3\textwidth]{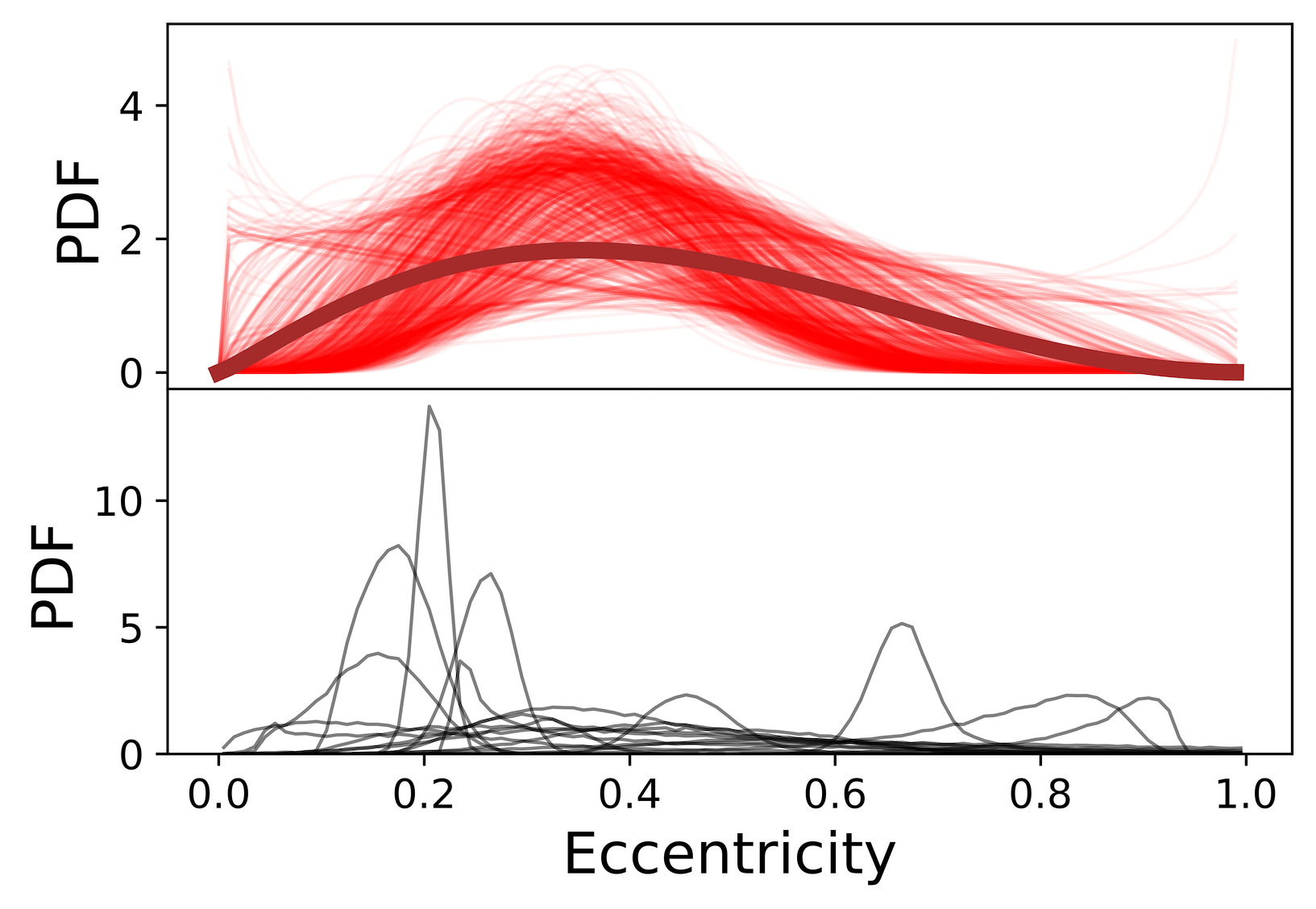}
\includegraphics[width=0.3\textwidth]{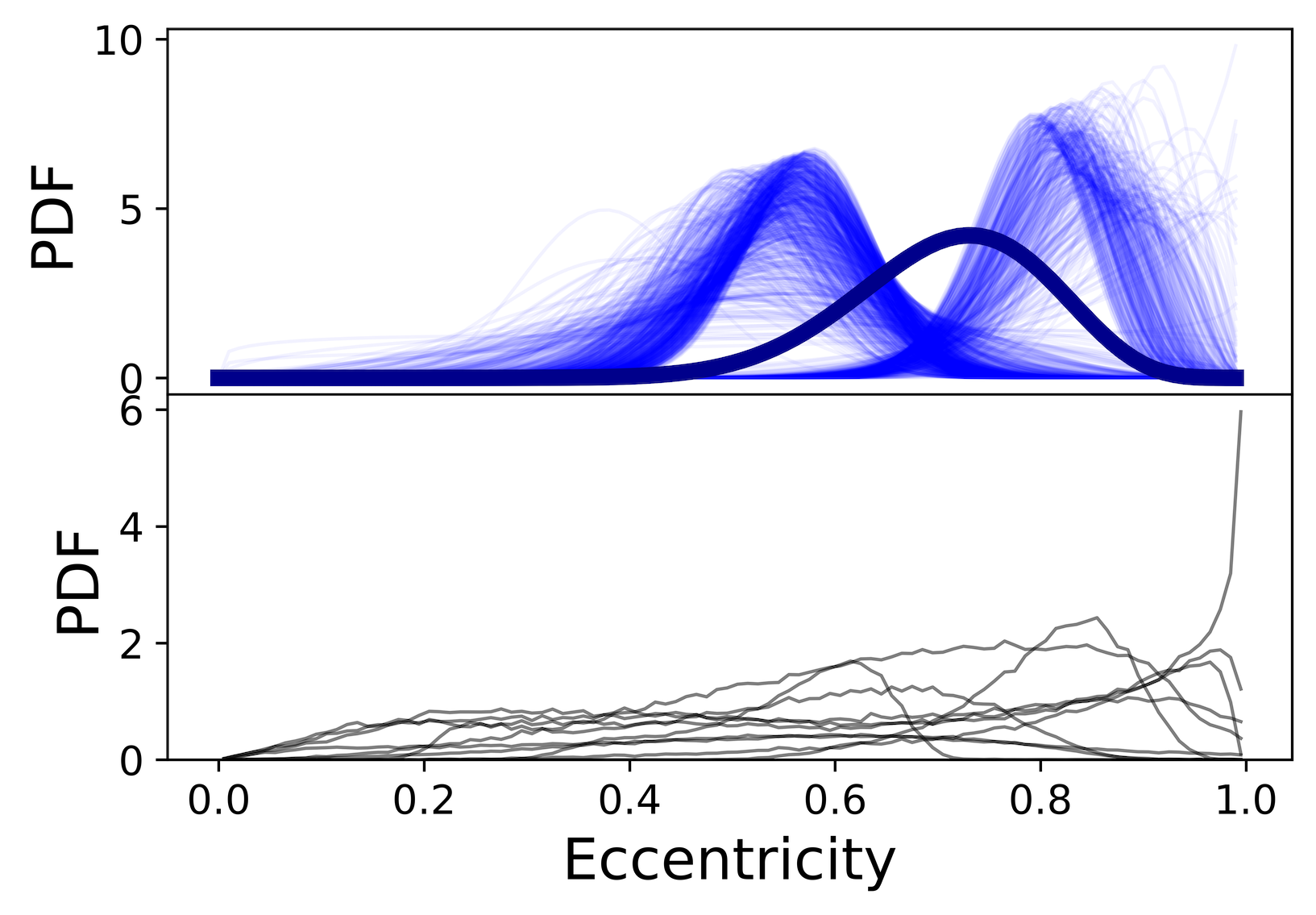}
\includegraphics[width=0.3\textwidth]{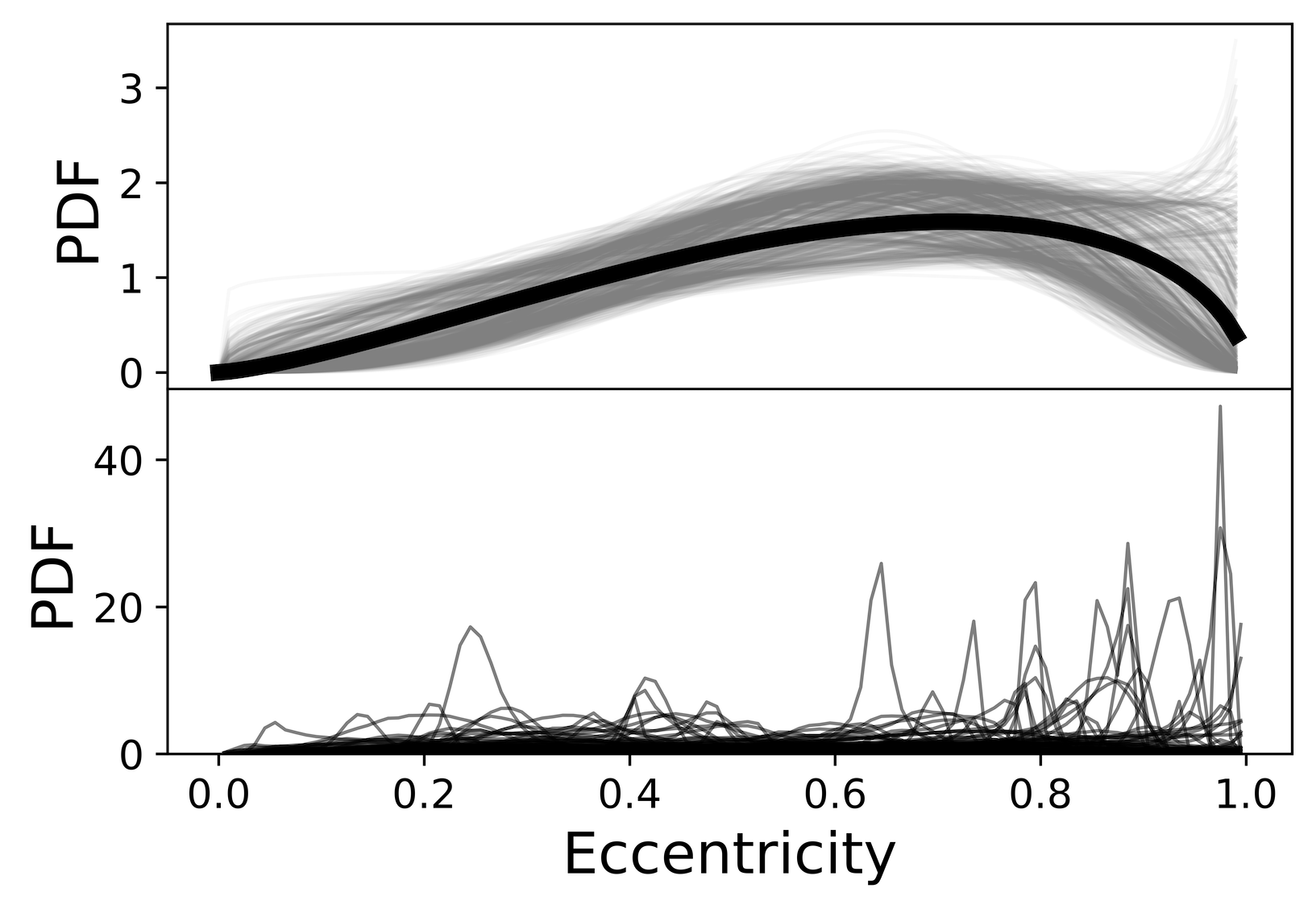}
\caption{
The best-fit (thick line) eccentricity distribution for Disk1 (left), Plane2 (middle), and Non-disk (right) young stars. 
Fit uncertainties are drawn as thin lines in top plots. We randomly draw 1000 parameters from the posterior probability distributions and plot corresponding Beta distributions. 
Bottom plots show individual eccentricity posterior distributions of stars belonging to each structure.
Disk1 stars peak at lower eccentricities while Plane2 and Non-disk young stars have higher eccentricities. Plane2 has large uncertainties and the drawn 1000 
parameters result in distributions that have two prominent peaks. 
This may imply a bi-modal distribution for Plane2; however, given the small sample size, the bi-modality is not significant. 
}
\label{fig:ecc_dist}
\end{figure*}

\begin{figure}
\centering
\includegraphics[width=0.4\textwidth]{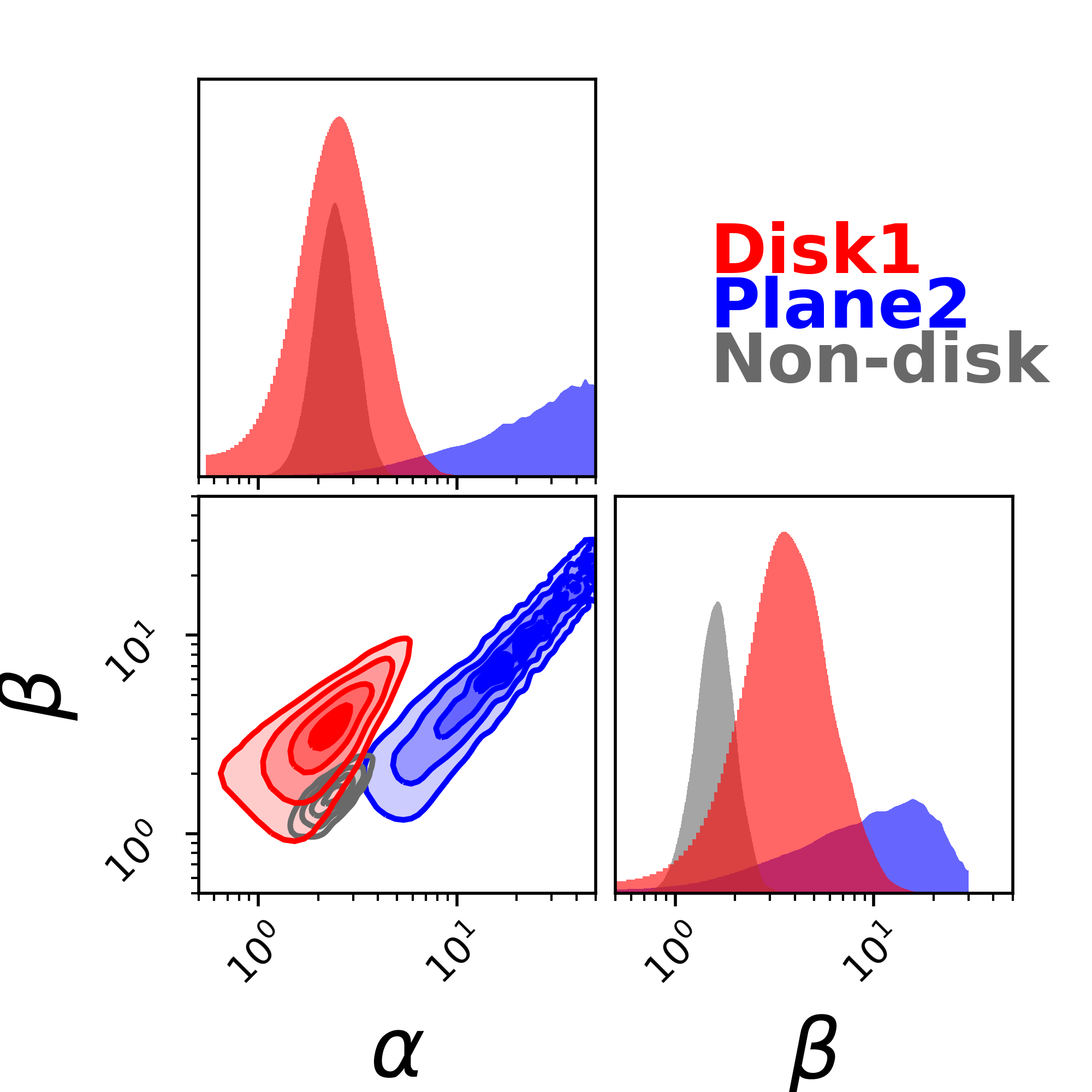}
\caption{
Posterior probability density for the eccentricity distribution parameters, $\alpha$ and $\beta$, for all three groups of young stars. 
Contours show the {0.5, 1, 1.5, and 2 sigma regions}.
}
\label{fig:ecc_posteriors}
\end{figure}

\begin{figure}[htb]
\centering
    \includegraphics[width=0.48\textwidth]{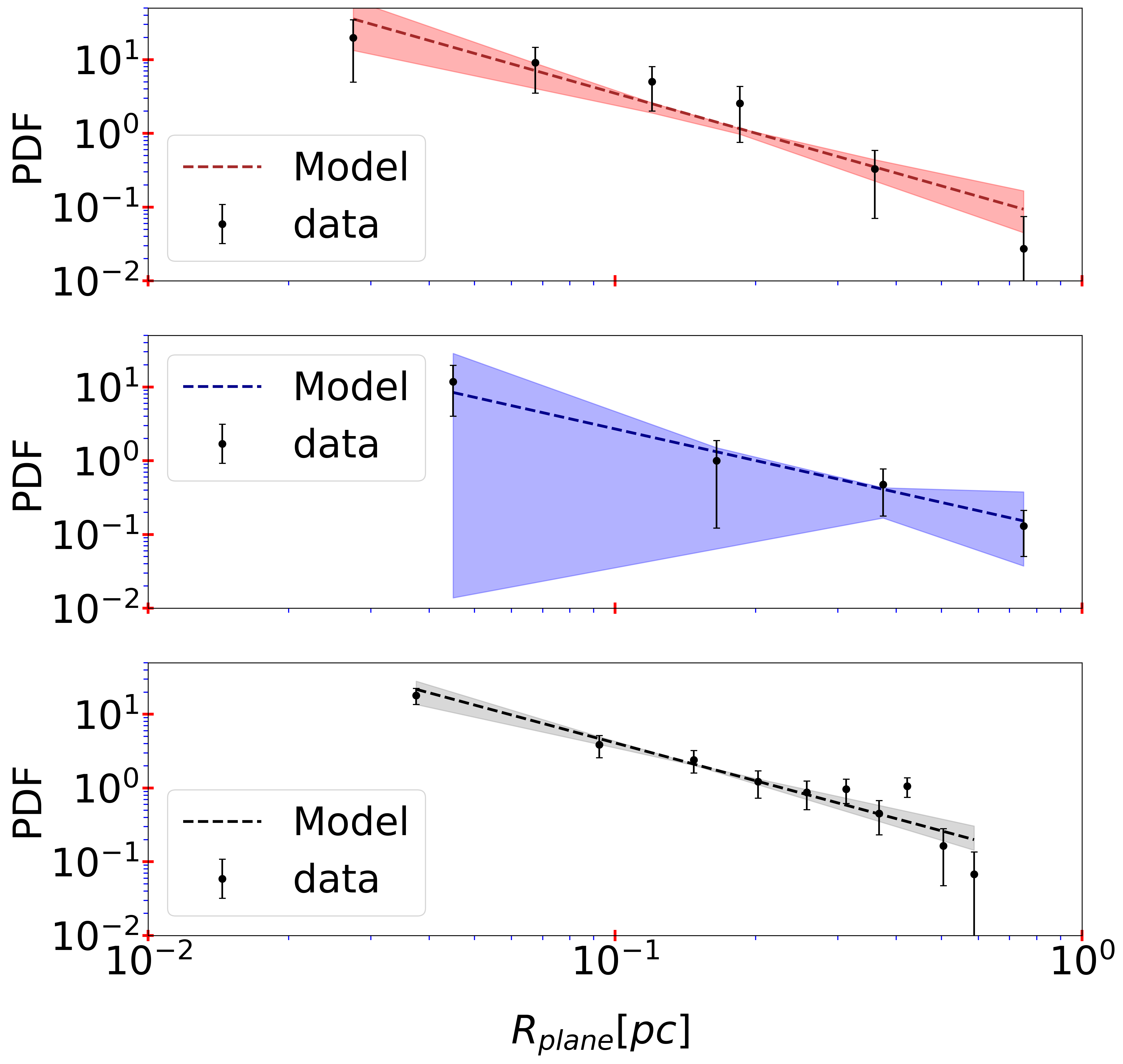}
\caption{
Radial distributions for the three dynamical subgroups and the best-fit model. The top plot is for Disk1 (red), the middle plot is for Plane2 (blue), and the bottom plot is for the Non-disk population (grey). The colored envelope in each plot defines region of uncertainty for the model. We draw 200 samples from the posterior distribution of model parameter and calculate corresponding uncertainty region (plotted as shaded areas).
}
\label{fig:r_dist}
\end{figure}

We modeled the underlying eccentricity distribution for Disk1, Plane2, and the Non-disk populations separately. 
\edit1{We used a hierarchical Bayesian inference (HBI) method similar to \citet{Hogg:2010} and \citet{Bowler:2020} to infer the population-level eccentricity distribution of each sub-structure.}
\edit1{This method was proposed first in constraining the population-level eccentricity distribution of exoplanets, where a standard maximum-likelihood estimator of eccentricity is biased to a high value. 
HBI is especially useful when, as shown in \citet{Bowler:2020}, the measured eccentricity posteriors for individual stars vary significantly with some well constrained, others poorly determined, and most with asymmetric or non-Gaussian probability distributions \citep[see also][]{Wolfgang_2016, Eylen_2019}.
}

\edit1{To apply this framework to our case, we first} adopted a Beta distribution for the eccentricity distribution for each population,
\begin{eqnarray}
P(e|\alpha,\beta) = \frac{\Gamma(\alpha + \beta)}{\Gamma(\alpha)\Gamma(\beta)} e^{\alpha - 1} 
(1 - e)^{\beta - 1}.
\end{eqnarray}
where $\Gamma$ is the usual Gamma function defined as $\Gamma(z) = \int_{0}^{\infty} x^{z-1} e^{x} dx$. Recall that a Beta function can reproduce distributions that are uniform ($\alpha=1, \beta=1$), peaked at low $e$ (small $\alpha$), peaked at high $e$ (small $\beta$), and anything in between, \edit1{which makes the distribution very flexible.}
We then used \edit1{HBI} to find the most probable $\alpha$ and $\beta$ parameters for each population's eccentricity distribution, solving for
\begin{eqnarray}
P(\{\theta_s\}, \mathcal{E} | \{d_s\}) = \frac{P(\{d_s\} | \{\theta_s\}) P(\{\theta_s\} | \mathcal{E}) P(\mathcal{E})}{P(\{d_s\})}
\end{eqnarray}
where $\mathcal{E} = (\alpha, \beta)$ describes the population's eccentricity distribution, $\{\theta_s\}$ is the set of orbital parameters for each star, $s$, and $\{d_s\}$ is the data for each star. This enables us to use posterior samples from an individual star's eccentricity distribution to estimate the population's distribution.
As we already have posterior samples for the individual stars' orbital parameters, $\theta_s$, similar to \citet{Bowler:2020}, we can write the posterior distribution of hyper-parameters $\mathcal{E}$ as: 
\begin{equation}
    P(\mathcal{E} | \{d_s\}) \, \propto \, \mathcal{L}(\{d_s\} | \mathcal{E}) \, \pi (\mathcal{E})
\end{equation}
where $\pi (\mathcal{E})$ is the prior on the hyper-parameters. 
Then we can approximate the likelihood $\mathcal{L}(\{d_s\} | \mathcal{E})$ for the population parameters by importance sampling the existing posteriors with
\begin{eqnarray}
\mathcal{L}(\{d_s\} | \mathcal{E}) \approx \prod_{s=1}^{N_{stars}} \frac{c_s}{K} \sum_{k=1}^K \frac{P(e_{sk} | \mathcal{E})}{\pi (e_{sk})}
\end{eqnarray}
where $e_{sk}$ is the eccentricity of the $k$-th draw for star $s$ as described in \citet{Hogg:2010} and $c_s$ is the membership probability for each star. 
The orbital parameter posteriors for most stars were generated using a Monte-Carlo sampling method and an explicit eccentricity prior was not utilized. 
However, the uniform $a_R$ prior produces a fairly uniform  eccentricity distribution; thus we assume that $\pi (e_{sk})$ is uniform.
Priors on $\alpha$ and $\beta$ were uniform from [0, 50] and [0, 30], respectively. 
The final posterior probability distribution on $\mathcal{E}$ was inferred using the nested sampling package, {\em Dynesty} \citep{Speagle:2020,Skilling:2004,Skilling:2006,Feroz:2009,Higson:2019}.

The best-fit eccentricity distributions are plotted in Figure \ref{fig:ecc_dist}.
%The weighted histogram of eccentricity $e$ is plotted in Figure \ref{fig:histe}.
For the Disk1 population, we see a peak at $e$ = 0.36, which is consistent with the previous determination by \citet{Yelda_2014}, but with a larger spread, implying more high eccentric orbits in Disk1.
For the Plane2 population, the eccentricity distribution is shifted to higher eccentricities, although the uncertainties are large, given the small number of stars. 
We adopt the maximum-likelihood solution for all three populations, including Plane2, even though the intrinsic eccentricity distribution is uncertain and may have a different functional form.
For the Non-disk stars, the distribution is fairly flat, with a slight preference for higher eccentricities. 
Both the Plane2 and Non-disk populations are consistent, within uncertainties, with a relaxed population, which should scale as $P(e) \propto e$; however, a non-relaxed, but high, eccentricity distribution is preferred (Figure \ref{fig:ecc_posteriors}).
To summarize, we find that the eccentricity distributions are described by:

\begin{equation}
\label{equ:e_dist}
\begin{split}
P_\mathrm{Disk1}(e)  =& \ \mathrm{Beta}(\alpha = 3.2 \pm 1.4, \beta = 5.0 \pm 2.4) \\
P_\mathrm{Plane2}(e) =& \ \mathrm{Beta}(\alpha = 31.9 \pm 11.9, \beta = 15.2 \pm 6.8) \\
P_\mathrm{Non-disk}(e)  =& \ \mathrm{Beta}(\alpha = 2.6 \pm 0.6, \beta = 1.7 \pm 0.4).
\end{split}
\end{equation}

\subsubsection{Radial Profile}
\label{sec:radial_profile}

In order to calculate a star's radial position on the disk, we first need to find its \edit1{projected position vector on} the disk plane. 
We have the $x_{sky}$ and $y_{sky}$ positions for each star and thus we \edit1{need to calculate} $z_{plane}$ on the disk plane. 
To do this, we begin by finding the normal vector to the disk, $\textbf{\textit{L}}$, from its $i$ and $\Omega$:

\begin{equation}
\label{equ:Lnormal}
    \textbf{\textit{L}} = 
        \begin{pmatrix}
            $$L_{x}$$ \\
            $$L_{y}$$ \\
            $$L_{z}$$
        \end{pmatrix}
        =
        \begin{pmatrix}
            $$\sin i \cos \Omega$$ \\
            $$-\sin i \cos \Omega$$\\
            $$-\cos i$$
        \end{pmatrix}
\end{equation}

After calculating the normal vector $\hat{L}$, combined with known $x_{sky}$ and $y_{sky}$ positions of stars, we find $z_{plane}$ by projecting stars onto the disk plane:
\begin{equation}
\label{equ:z_plane}
    z_{plane} = \frac{-(L_{x} \, x_{sky} + L_{y} \, y_{sky})}{L_{z}}
\end{equation}
then the radial distances $R_{plane}$ are calculated by Pythagorean Theorem.

%To find $z$, we begin by using the Thiele-Innes constants [CITE]. We will also use the constants $X = \cos E - e$ and $Y = (1-e^2)^{1/2}\sin E$ with the assumption that $e=0$. Combing the equations $x = BX + GY$ and $y=AX+FY$ leads to the equation:

%\begin{equation}\label{1}
%    a\sin(\omega+E) = \left(\frac{x}{\sin\Omega}-\frac{y}{\cos\Omega}\right) \frac{1}{\cos i}[\cot \Omega + \tan\Omega]^{-1}
%\end{equation}

%Combining equation \ref{1} with $z = CX + HY$ gives us the line of sight distance to project at star at the coordinate $(x, y)$ onto a plane defined by $(i, \Omega)$:

%\begin{equation}\label{z}
%    z = \left(\frac{x}{\sin\Omega} - \frac{y}{\cos\Omega}\right)\tan i[\cot\Omega+\tan\Omega]^{-1}
%\end{equation}

%To find the projected radial position in the orbital plane, we start by rearranging equation \ref {1}:

%\begin{equation} \label{3}
%    a = \left(\frac{x}{\cos\Omega}+\frac{y}{\sin\Omega}\right)[\tan\Omega+\cot\Omega]^{-1}\frac{1}{\cos(\omega+E)}
%\end{equation}

%In addition, we can simplify $z = CX + HY$ to get:

%\begin{equation} \label{4}
%    z = a\sin (i) \sin (\omega + E)
%\end{equation}

%combining equations \ref{3} and \ref{4} gives:

%\begin{equation}\label{tan}
%    \tan(\omega + E) = z \left(\frac{x}{\cos\Omega}+ \frac{y}{\sin\Omega}\right)^{-1} [\tan\Omega + \cot\Omega]\frac{1}{\sin(i)}
%\end{equation}

%Equation \ref{tan} can now be plugged into equation \ref{4} to solve for $a$ in terms of $x,y,z,i,$ and $\Omega$, where $a$ here is equivalent to the projected radial position in the plane $R_{plane}$.

To create the PDFs of the $R_{plane}$ distributions for both Disk1 and Plane2, we sample both stars and structures \edit1{simultaneously}. 
First, a sample of $i_{star}$ and $\Omega_{star}$ is drawn for each star from its MC result $5 \cdot 10^4$ times. 
Second, another sample of each structure's $i_{disk}$ and $\Omega_{disk}$ is created by drawing $5 \cdot 10^4$ times from a Gaussian distribution defined by the means and uncertainties \edit1{of $i$ and $\Omega$ shown in Table \ref{tab:disk}}. 
We randomly match elements of these two samples to create a new data set, where each \edit1{data set contains one combination of} a star's $i_{star}$ and $\Omega_{star}$ and a disk's $i_{disk}$ and $\Omega_{disk}$. 
From \edit1{this new data set}, we calculate $R_{plane}$ values for each combination using Eqs. \ref{equ:Lnormal} and \ref{equ:z_plane}.  
Additionally, we imposed a cut of 20 parsecs on our values of $R_{plane}$ \edit1{because these dynamical sub-structures, especially Plane2, are close to edge-on.}. 
Without this cut, we would obtain values of $R_{plane}$ approaching infinity.

We also calculate a weight associated with each data in the following way, assuming no change in uncertainty of disk parameters and a normal distribution. 
\begin{equation}
\begin{split}
    W(i_{star}, \Omega_{star} | i_{disk}, \Omega_{disk}, \sigma_{i, disk}, \sigma_{\Omega, disk}) = \\ N(i_{star}|\mu=i_{disk},\sigma=\sigma_{i,disk}) \times \\
    N(\Omega_{star}|\mu=\Omega_{disk},\sigma=\sigma_{\Omega,disk})
\end{split}
\end{equation}
This is the probability of each sampled star being located on the sampled disk.
We do not use membership probability because it is an integrated result and could not reflect details of sampled data. 
Then a \edit1{weighted} histogram of 88 stars is created for each sampled disk's $i_{disk}$ and $\Omega_{disk}$. 
Each histogram is normalized by 2D bin widths (i.e. annular area between each bin) to account for the 2D geometry of disk structure. 
The final PDF of $R_{plane}$ for each structure is obtained by taking the mean across these histograms and the uncertainties are estimated by the standard deviation. 
For the off-disk population, there is no disk to project the stars onto, so the stars' radial positions on the plane of the sky are considered. 
An $R_{sky}$ distribution is then created \edit1{through} a normalized distribution of the data. 
The uncertainties of these data points are estimated by the Poisson error on each data point.

The data were fit using Bayesian inference with multi-nested sampling by the Dynesty python package. From inspection, we decided to fit the $R_{plane}$ distribution by a truncated single power-law for all three structures. Thus we only need to fit for the slope parameter $\alpha$. 

%\begin{equation}
%\label{equ:chi-sq}
%    \chi^2 = \sum_{i=1}^{n} \frac{(y_{i} - y_{i, %model})^2}{\sigma_{i}^2}
%\end{equation}

To estimate the uncertainty of the model, we sampled the posterior distributions of the model parameters 200 times, each time recalculating the model. We estimate the uncertainty by taking the standard deviation of all these calculated models. The radial distributions and best-fit models for the three subgroups are shown in Figure \ref{fig:r_dist}.

We find that that the radial position distribution in the disk and the sky are described by:

\begin{equation}
\label{equ:r_dist}
\begin{split}
P_\mathrm{Disk1}(R_{plane})  =& \ \mathrm{Plaw}(\alpha = -1.80 \pm 0.17) \\
P_\mathrm{Plane2}(R_{plane}) =& \ \mathrm{Plaw}(\alpha = -1.43 \pm 0.47) \\
P_\mathrm{Non-disk}(R_{sky})  =& \ \mathrm{Plaw}(\alpha = -1.71 \pm 0.10).
\end{split}
\end{equation}

We initially fit the $R_{plane}$ distribution of Disk1 by the truncated broken power law with model parameters $\alpha_{1}$, $\alpha_{2}$, and $r_{break}$. 
However, this results in huge uncertainties on model parameters. 
To justify our choice of a single power law, we compare values of both the Bayesian Information Criterion (BIC) and Akaike Information Criterion with small sample modification (AICc) for these two models.
The results are shown in Table \ref{tab:criterion}. 
From this table, the difference between BIC values of Disk1 for different models is $\sim$ 2, which suggests moderate evidence against single power law. 
However, the AICc strongly prefers a single power law with difference of about 10. 
Thus we conclude that there is no preference between these two models for Disk1 and we choose to present the single power-law result because it has well constrained parameters. 
Similar reasoning applies to the off-disk population. 
For the Plane2 population, we prefer a single power law due to its few data points compared to the number of parameters.

\startlongtable
\begin{deluxetable}{lrrr}
\tablecaption{BIC and AICc \label{tab:criterion}}
\tablehead{\colhead{Model} & \colhead{Structure}  & \colhead{BIC}  & \colhead{AICc}}
\startdata
Single Power Law      &   Disk 1   &  20.72  &  21.93    \\ 
Broken Power Law      &   Disk 1   &  19.85  &  32.48    \\ 
Single Power Law      &   Plane 2  &  5.45   &  8.07     \\ 
Broken Power Law      &   Plane 2  &  8.08   &  N/A  \tablenotemark{a} \\ 
Single Power Law      &   Non-disk    &  15.78  &  15.98    \\ 
Broken Power Law      &   Non-disk    &  12.99  &  16.08    \\ 
\enddata
\tablenotetext{a}{AICc does not apply to Plane2 because the number of data points equals to the number of parameters plus one, which rejects Broken Power Law model (overfitting).}
\end{deluxetable}

%For Disk1, the only population fit a broken powerlaw, the two slopes for the best fit model are $\alpha_1 = 1.12$ and $\alpha_2 = -2.65$. For plane two we obtained $\alpha = -0.43$ and for the off disk population, the sky projected radial profile had a slope of $\alpha=-0.75$.%

% The weighted histogram of semi-major axis $a$ is plotted in Figure \ref{fig:hista}.
% At larger radius ($a$ > 0.3 pc), the distribution of $a$ is consistent between \textit{disk1}, \textit{plane2} and \textit{other} within uncertainties.
% They all follow a power law function with a slope of -2 at larger $a$, which is consistent with the observed surface-density profile \citep{Berukoff_2006, Paumard_2006}.
% However at inner radius region, \textit{disk1} population density drops a lot faster compared to \textit{plane2} and \textit{other} population. 
% See details in Equation \ref{equ:a_dist}.

% \begin{equation}
% \label{equ:a_dist}
% \begin{split}
% P_{\mathrm{disk1}}(a) \propto & 
% \begin{cases} a^{2.3}  \qquad 0.005 \;\mathrm{pc} <= a < 0.120 \;\mathrm{pc}  \\ a^{-2} \qquad  0.120 \;\mathrm{pc} <= a < 300 \;\mathrm{pc} \end{cases} \\
% P_{\mathrm{plane2}}(a) \propto  & 
% \begin{cases} a^{-0.2}  \qquad 0.005 \;\mathrm{pc} <= a < 0.38 \;\mathrm{pc}  \\ a^{-2} \qquad  0.380 \;\mathrm{pc} <= a < 300 \;\mathrm{pc} \end{cases} \\
% P_{\mathrm{other}}(a) \propto  & 
% \begin{cases} a^{-0.35}  \qquad 0.005 \;\mathrm{pc} <= a < 0.330 \;\mathrm{pc}  \\ a^{-2} \qquad  0.330 \;\mathrm{pc} <= a < 300 \;\mathrm{pc} \end{cases} \\
% \end{split}
% \end{equation}
The thickness of the disk can be estimated using the velocity dispersion perpendicular to the disk plane.
Here we follow the process described by \citet{Lu_2009}.
First, each potential disk candidate's three-dimensional velocity, $\vec{v}$, is projected into the direction of $\hat{L}$.
The uncertainty of both $\vec{v}$ and $\hat{L}$ are taken into consideration when calculating intrinsic velocity dispersion $\sigma_{v_n}$, where $v_n$ = $\vec{v}$ $\cdot$ $\hat{L}$. 
Then the disk's scale height ($h/r$) can be derived from the ratio of $\sigma_{v_n}$ and <$\vec{v}$> , where $\sigma_{v_n}$ is the intrinsic velocity dispersion, and <$\vec{v}$> is the average magnitude of the 3D velocity of disk candidates, weighted their by disk membership probability.
Finally, this scale height can be related to disk thickness described in terms of disk-opening angle $h/r \propto \sqrt{1/2} \Delta{\theta}$.
For Disk1, we find $\sigma_{v_n}$ = 33 km s$^\mathrm{-1}$, giving a scale height of 0.09 $\pm$ 0.01 and a disk-opening angle of 7.0\degree $\pm$ 0.9\degree, consistent with previous results \citep{Lu_2009, Yelda_2014}.
For Plane2, we find $\sigma_{v_n}$ = 58 km s$^\mathrm{-1}$, giving a scale height of 0.23 $\pm$ 0.07 and a disk-opening angle of 18.6\degree $\pm$ 6.2\degree.

\subsection{Simulations}
\label{sec:disk_simulation}

\begin{table*}[h!]
\caption{Characteristics of the Simulated Cluster Subgroups}
\label{tab:simulation}
%\centering
\begin{threeparttable}
\resizebox{\textwidth}{!}{
\begin{tabular}{cccc}
    \toprule
    \textbf{Parameters} & \textbf{Disk1} & \textbf{Plane2} & \textbf{Non-disk} \\
    \midrule
    $i$ & N $\propto$ ($i_\mathrm{Disk1} = 124\degree$, $\sigma_\mathrm{i, Disk1} = 15\degree$) &  N $\propto$ ($i_\mathrm{Plane2} = 90\degree$, $\sigma_\mathrm{i, Plane2} = 20\degree$) & P($i$) $\propto$ cos($i$) \\
    $\Omega$ & N $\propto$ ($\Omega_\mathrm{Disk1} = 94\degree$, $\sigma_\mathrm{\Omega, Disk1} = 17\degree$) &  N $\propto$ ($\Omega_\mathrm{Plane2} = 245\degree$, $\sigma_\mathrm{\Omega, Plane2} = 19\degree$) & Uniform(0\degree,360\degree)\\
    $\omega$ & Uniform(0\degree,360\degree) & Uniform(0\degree,360\degree) & Uniform(0\degree,360\degree) \\
    $e$ & $\mathrm{Beta}(\alpha=3.2 \pm 1.4, \beta=5.0 \pm 2.4)$ \tablenotemark{b} & $\mathrm{Beta}(\alpha=31.9 \pm 11.9, \beta=15.2 \pm 6.8)$ & $\mathrm{Beta}(\alpha=2.6 \pm 0.6, \beta=1.7 \pm 0.4)$ \\
    $a$ \tablenotemark{c} & $\mathrm{Plaw}(\alpha = -1.80 \pm 0.17)$ \tablenotemark{b}  & $\mathrm{Plaw}(\alpha = -1.43 \pm 0.47)$ & $\mathrm{Plaw}(\alpha = -1.71 \pm 0.10)$ \\
    $t_0$ & Uniform(1995, 1995+period) & Uniform(1995, 1995+period) & Uniform(1995, 1995+period) \\
    $M_{min}$ &  1$M_{\odot}$ & 1$M_{\odot}$ & 1$M_{\odot}$ \\
    $M_{max}$ &  150$M_{\odot}$ & 150$M_{\odot}$ & 150$M_{\odot}$ \\
    $M_{cluster}$ & 10$^\mathrm{5}$ $M_{\odot}$ & 10$^\mathrm{5}$ $M_{\odot}$ & 10$^\mathrm{5}$ $M_{\odot}$ \\
    Age & 6Myr & 6Myr & 6Myr \\
    IMF & $\xi(m) \propto m ^\mathrm{-2.35}$ &  $\xi(m) \propto m ^\mathrm{-2.35}$  &  $\xi(m) \propto m ^\mathrm{-2.35}$ \\
    distance & 8kpc & 8kpc & 8kpc \\
    \bottomrule
    \end{tabular}}
    \tablenotetext{a}{Equ \ref{equ:e_dist}.}
    \tablenotetext{b}{For simulated cluster, we choose to only use peak parameters corresponding to max likelihood.}
    \tablenotetext{c}{Equ \ref{equ:r_dist}. Here we use $a$ to represent the radial distance $r_{\mathrm{plane}}$. For each broken power law, we choose $a_{\mathrm{min}} = 0.01$pc and $a_{\mathrm{max}} = 1.0$pc.}
\end{threeparttable}
\end{table*}

From Figure \ref{fig:disk_membership}, we can see the spatial distribution within each disk plane does not appear symmetric, especially for Plane2 stars. 
\edit1{Intrinsic asymmetry is important in understanding and constraining the subgroup's dynamical history.
However, there are many reasons that can lead to this observed asymmetry: extinction, incomplete observation, and intrinsic asymmetry.
In order to characterize this asymmetry, we simulate the whole cluster and each dynamical sub-structure using properties determined in \S\ref{sec:disk_property}. 
We also take inverse-completeness (see \S\ref{sec:completeness}) and the full extinction map from \citep{Schodel:2010} into account when doing the simulation in order to diminish the effects from these two factors.
Then we compare the simulated stellar distribution to the observed one to explore whether the structure is intrinsically asymmetric or not.}

Table \ref{tab:simulation} summarizes the input simulation parameters.
First, we use an open-source python package \textit{SPISEA} \citep{Hosek:2020lz} to generate a single age (6 Myr) star cluster with solar metallicity, located 8 kpc away. 
The total mass of the simulated cluster is 10$^5 M_{\odot}$, with minimum mass of 1 M$_{\odot}$,  maximum mass of 100 M$_{\odot}$, and a power-law IMF with a slope of -2.35 \citep{Salpeter_1955}.
While the IMF for the YNC has been shown to be top-heavy \citep{Bartko_2009,Lu_2013}, the analysis of the kinematic and spatial sub-structure is relatively insensitive to the choice of IMF and total cluster mass as we re-scale the simulated clusters to match observed stellar densities. 
We generate the synthetic photometry for each star in the NIRC2 Kp filter, assuming a fixed extinction value of A$_{Ks}$ = 2.7 mag, for easy comparison to the observed  Kp$_\mathrm{ext}$ shown in Table \ref{tab:pm}.
In this step, we assume the three dynamical subgroups have the same age, IMF, and extinction.

\begin{figure*}
\hspace*{-1in}
\subfloat[]{
\includegraphics[width=0.8\textwidth]{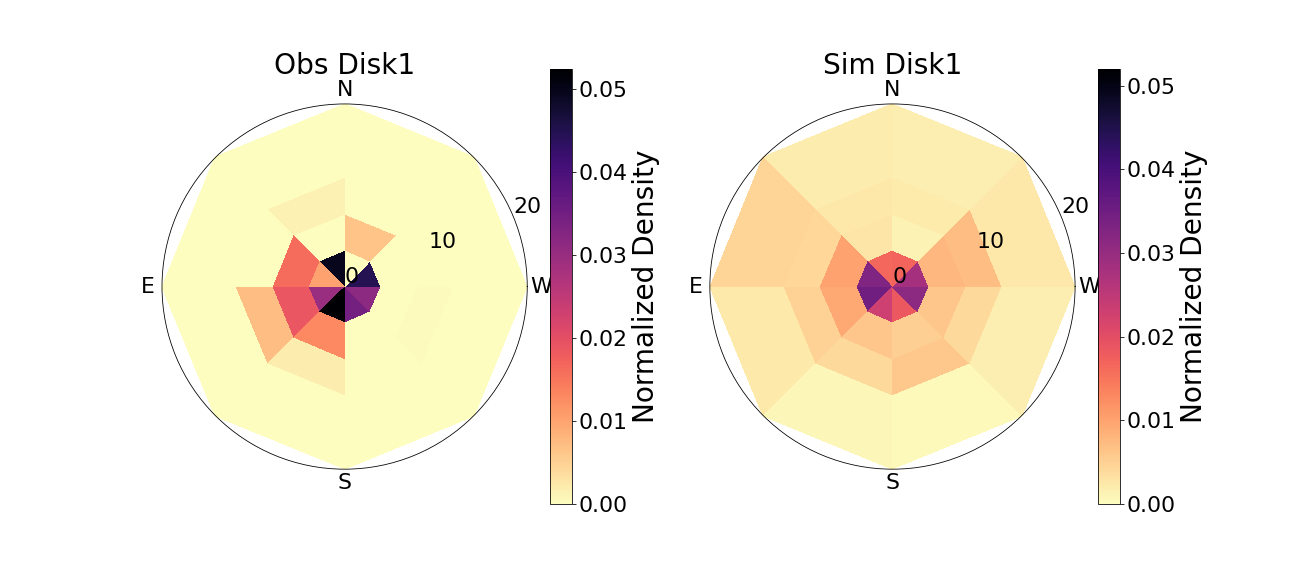}
} 
\hspace*{-0.1in}
\subfloat[]{
\includegraphics[width=0.3\textwidth]{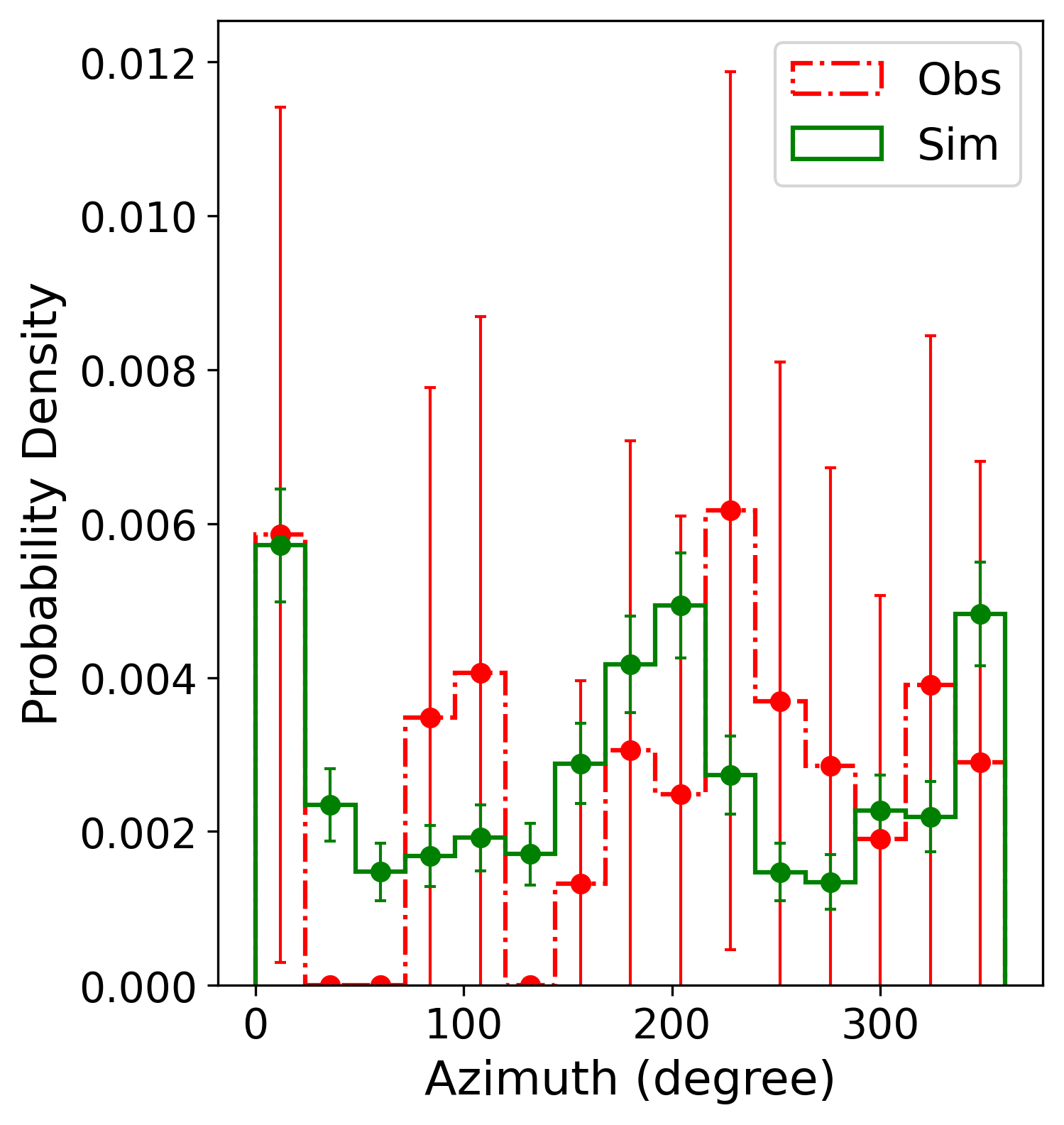} 
} \\

\hspace*{-1in}
\subfloat[]{
\includegraphics[width=0.8\textwidth]{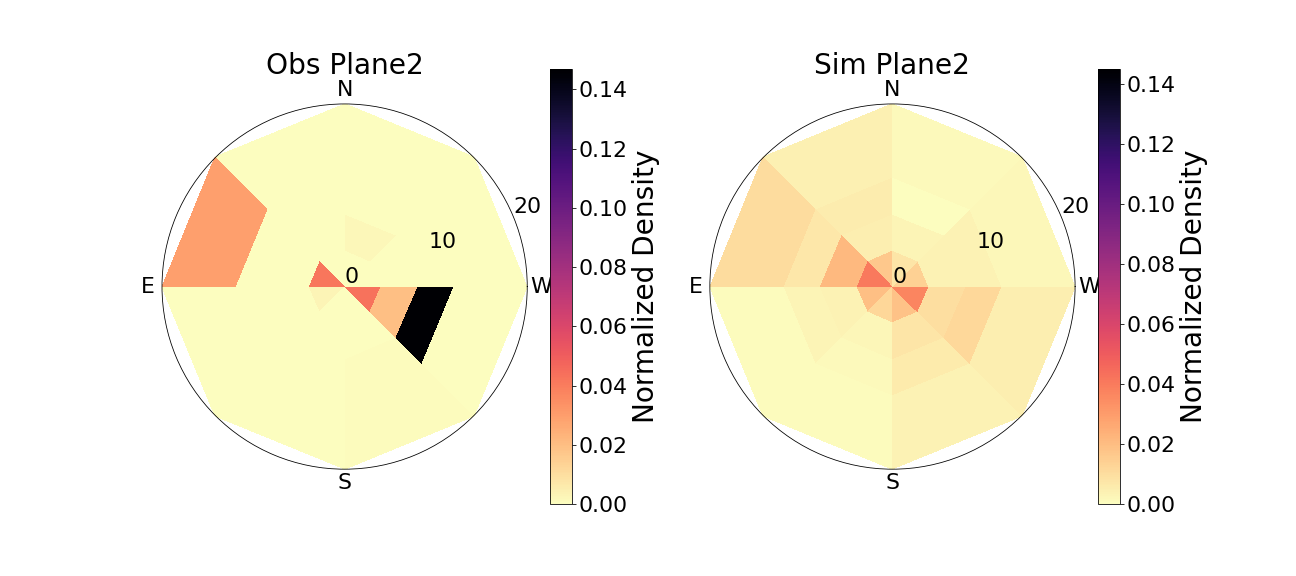}
} 
\hspace*{-0.1in}
\subfloat[]{
\includegraphics[width=0.3\textwidth]{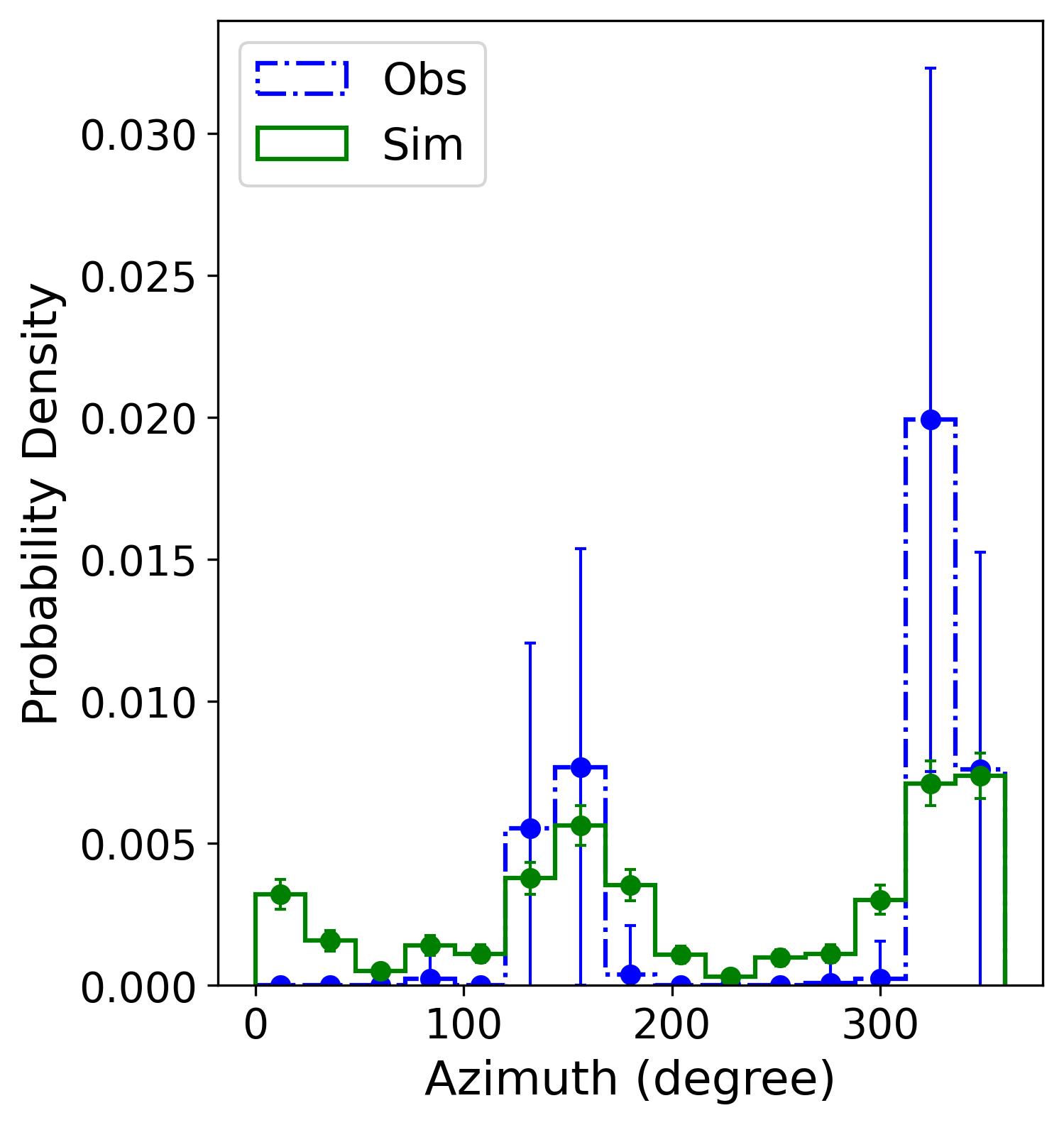} 
} \\

\hspace*{-1in}
\subfloat[]{
\includegraphics[width=0.8\textwidth]{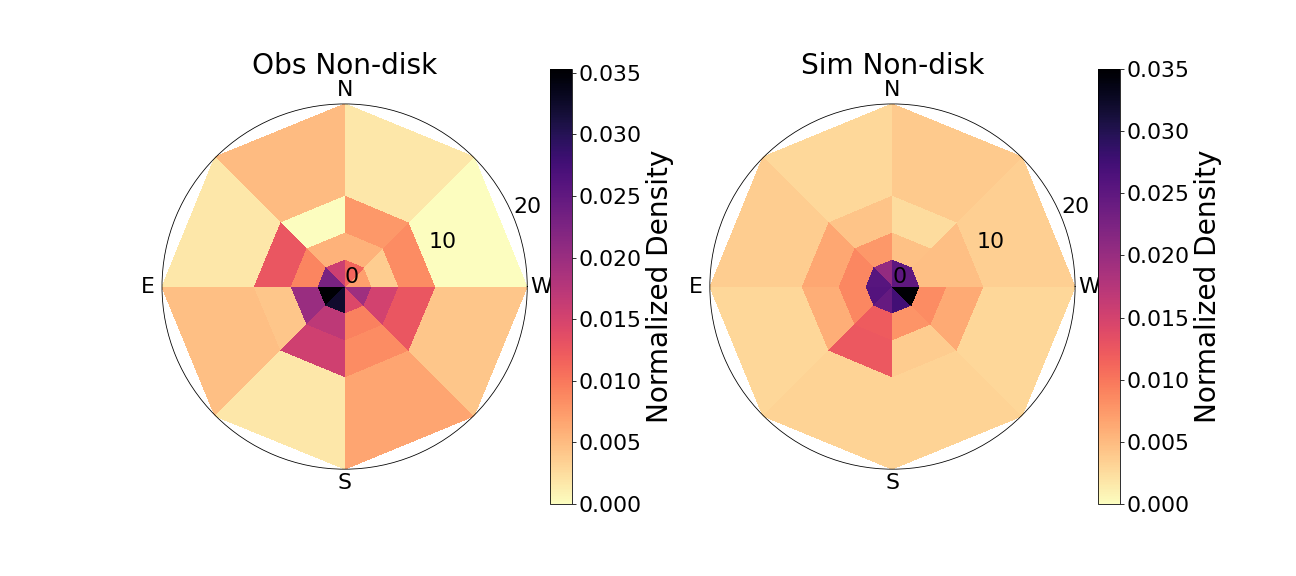}
} 
\hspace*{-0.1in}
\subfloat[]{
\includegraphics[width=0.3\textwidth]{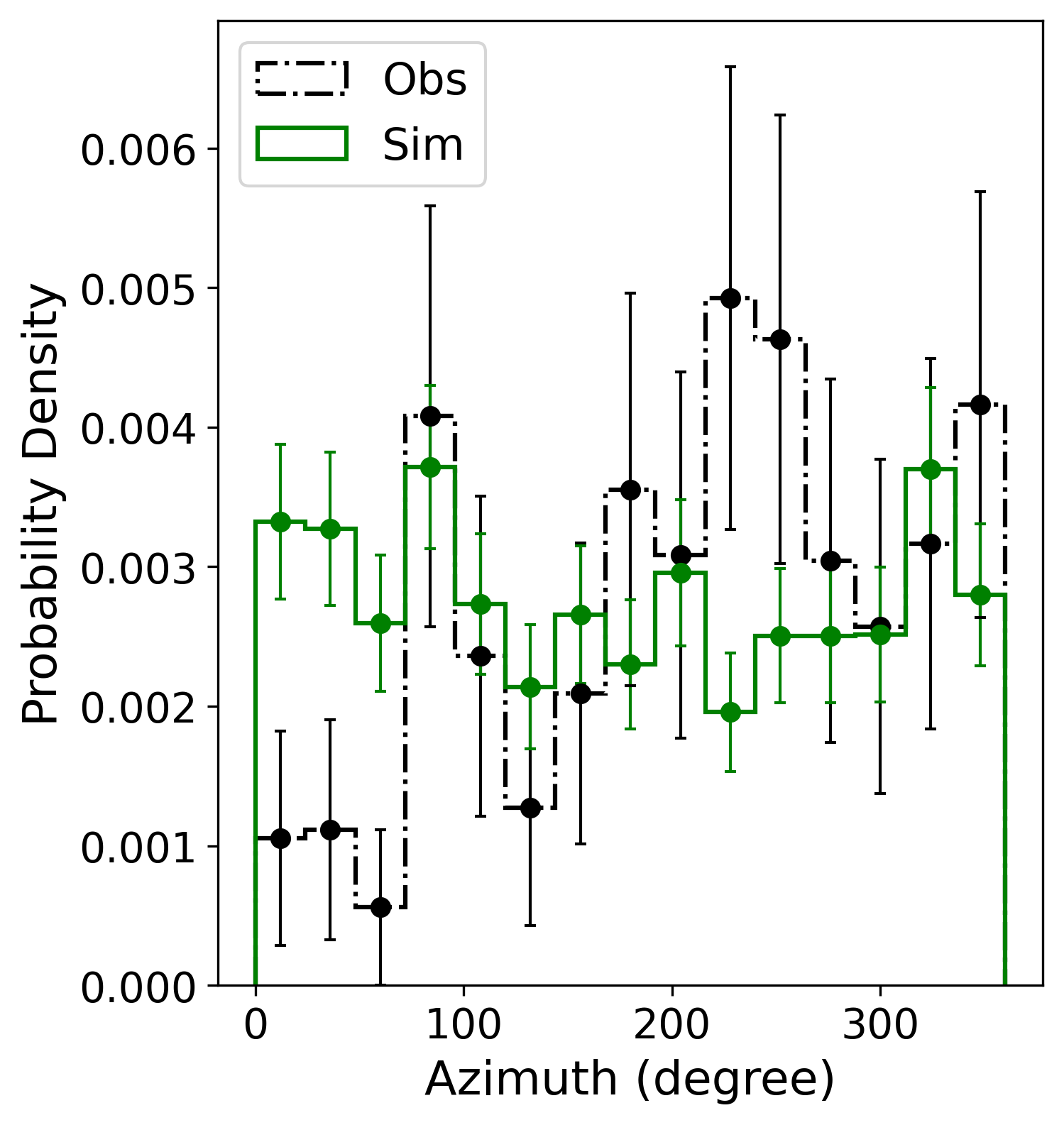} 
} \\

\caption{Stellar density map comparison between observation and simulation for Disk1 (top), Plane2 (middle) and Non-disk  (bottom) subgroups.
The first two columns are the stellar density map in polar system for observation (left) and simulation (middle), in which North is at 90\degree and West is 0\degree.
We choose this coordinate system to be consistent with how $\Omega$ is defined.
The right column is the 1D histogram of azimuth.
For the Disk1 and Plane2 groups, the over-dense regions are mostly attributable to the orientation of the disk, while for the Non-disk group, the asymmetry is less significant, resulting from the differential reddening map. 
Within uncertainties, the 1D azimuthal distribution is consistent between the observation and simulation for Disk1 and Non-disk group.
However for Plane2, the paucity of stars in the northeast sector of the observed map cannot be explained by the simulation itself.
}
\label{fig:simulation}
\end{figure*}

The cluster generated by \textit{SPISEA} gives mass and Kp$_\mathrm{ext}$ for every star in the system. However, our analysis excludes all WR stars because the generated values of mass and Kp$_\mathrm{ext}$ are less trustworthy compared to non-WR stars.
The next step is to assign a position for each star based on their dynamical subgroup.
We use the disk properties from \S\ref{sec:disk_property} to simulate positions for Disk1, Plane2 and Non-disk stars.
To account for differential extinction over the field of view, which introduces asymmetric features in the observed distributions, we redden the Kp$_{ext}$ back to observed Kp using the \edit1{full} extinction map from \citet{Schodel:2010} and then apply an inverse completeness map from \S\ref{sec:completeness} to account for stars that would not be observable. This approximates how the simulated cluster would appear in observations.
The comparison of the observed stellar density profile with the simulated density profile is shown in Figure \ref{fig:simulation}.
We present the density profile in polar coordinates, where North is at 90\degree and West is at 0\degree, so that it is easier to see the azimuthal structure in each dynamical subgroup.
These plots have the same orientation as they appear on the sky.
%This system has the same convention as the Keplerian orbital element, $\Omega$.

For the Non-disk group, the sub-structure is mainly caused by the differential extinction, and our simulation reproduces the observation well within uncertainties.
This indicates that the Non-disk group is nearly isotropic.
For Disk1 and Plane2, we expect an over-density in their disk plane, which can be seen in both the observed and simulated maps.
However, for Plane2, the observed density on the Southwest side is significantly more dense than the observed density on the Northeast side \edit1{(left plot of (c) in Figure \ref{fig:simulation})}, which \edit1{is less significant in the simulated stellar distribution (right plot of (c) in Figure \ref{fig:simulation}). 
Because we have taken both completeness and extinction into account when doing the simulation, this difference between observed and simulated stellar distribution should be explained by reasons other than these two factors.
Thus, we think this difference implies that Plane2 is intrinsically asymmetric and may be a stream rather than a plane.}

\section{Discussion}
\label{sec:discussion}

\edit1{The presence of multiple sub-structures in the young nuclear cluster implies that star formation occurred in dynamically complex gas structures around the SMBH. 
We can infer that the structures that are observed today are related, but not identical, to the initial gas structure since the two-body relaxation timescale is longer ($\geq$ 30 Myr $-$ 1 Gyr \citet{Kocsis_2011}) than the age of the YNC (3 $-$ 8 Myr \citet{Lu_2013}). However, some dynamical evolution is expected due to vector resonant relaxation at radii $< 1''$ (0.04 pc) and resonant friction at larger radii. In particular, \citet{Levin_2022} notes that resonant friction can take a single disk and smear it into transient streams with different orientations in only a few Myrs. Note, the asymmetric gravitational potential due to the surrounding circum-nuclear disk at a few pc may dampen these resonant dynamical processes. More precise theoretical simulations are needed to run back the clock from the sub-structures observed today to the initial configuration at the time of star formation. More detailed discussion is presented below.}

\subsection{Comparisons to Previous Work}
\label{sec:discuss_previous}

While the CW disk has been verified many times in previous work \citep{Levin_2003, Genzel_2003, Paumard_2006, Lu_2009, Yelda_2014, von_Fellenberg_2022}, other kinematic structures in the young nuclear cluster have been controversial.
A counter-clockwise (CCW) disk was originally reported in \citet{Genzel_2003} and confirmed in \citet{Paumard_2006}.
Later work \citep{Bartko_2009, von_Fellenberg_2022} found that the CCW disk was highly extended, anisotropic, and showed evidence for a warped disk on large scales.
However, this CCW disk was not detected by \citet{Lu_2009} and \citet{Yelda_2014} in an independent analysis with different observations for RVs and proper motions.
In the work presented here, we confirm the existence of the CW disk, with properties consistent with previous analyses, and we also do not detect the CCW disk.
Instead, we detect a second edge-on disk called Plane2 (c.f., \S\ref{sec:discuss_disk2}), which might be the same as the F3 structure reported in \cite{von_Fellenberg_2022} (c.f., \S\ref{sec:two_disk}, Appendix \ref{subsec:sig_comp}). 

\begin{figure}
    \centering
    \includegraphics[width=0.5\textwidth]{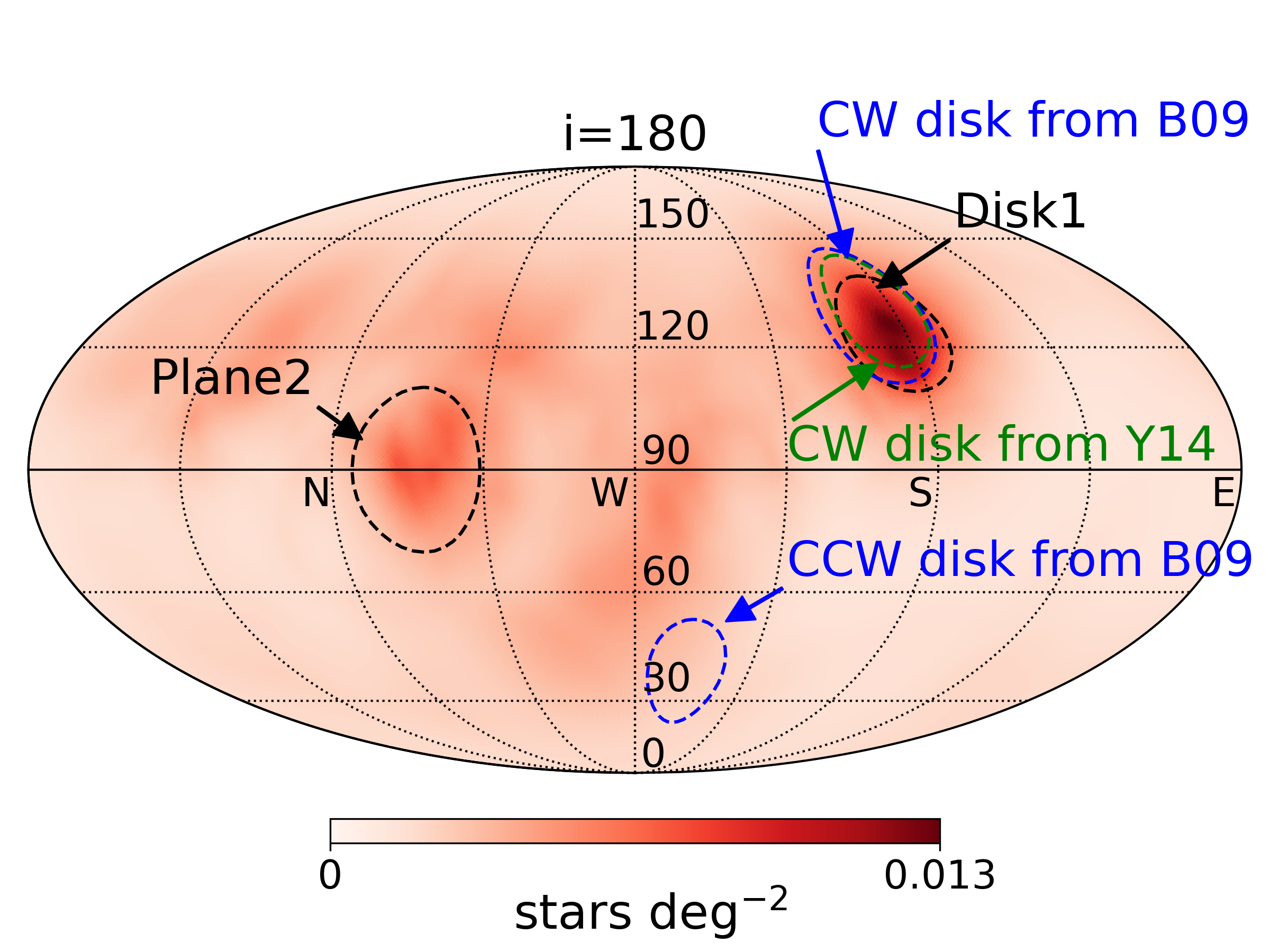}
    \caption{Proposed stellar disk from previous work and our work.
    Here we plot the CW disk and CCW disk from \citet{Bartko_2009} (shortened as B09) in blue and the CW disk from \citet{Yelda_2014} in green.
    The stellar disk reported in our work in shown in black.
    The CW disk, on the other hand, is very consistent between different analyses.}
    \label{fig:disk_old}
\end{figure}

A comparison of the locations of the CW disk, CCW disk and Plane2 on the density map is presented in Figure \ref{fig:disk_old}, from which we conclude that the position of Plane2 is clearly very different from that of the previously claimed CCW disk.

\edit1{There are small discrepancies between our work and that of vF22 for the significance and locations of Disk1 and Plane2. These are likely due to significant differences in proper motions at small radii and large, but not statistically significant, differences in radial velocities.
In addition, our requirement of spectroscopic completeness information for stars to be included in our sample yields an overall smaller number of young stars. Lastly, our use of a uniform acceleration prior (c.f., \S\ref{sec:mc}) on $\sim 50$\% of the stars at large radii may also influence structure detection.}

The stars within 0.8\arcsec are more randomized by dynamical effects like vector resonant relaxation \citep{Rauch_1996, Hopman_2006, Alexander_2007}, so the inner edge of the CW disk is roughly at 0.8\arcsec as reported in \citep{Schodel_2003, Ghez_2005, Gillessen_2009_stars}.
This agrees with our analysis.
In our sample, we have 14 young stars within 0.8\arcsec and all of them have zero probability of being on Disk1. 

\subsection{CW Disk1 Properties}
\label{sec:discuss_disk1}

\begin{figure}
    \centering
    \includegraphics[width=0.4\textwidth]{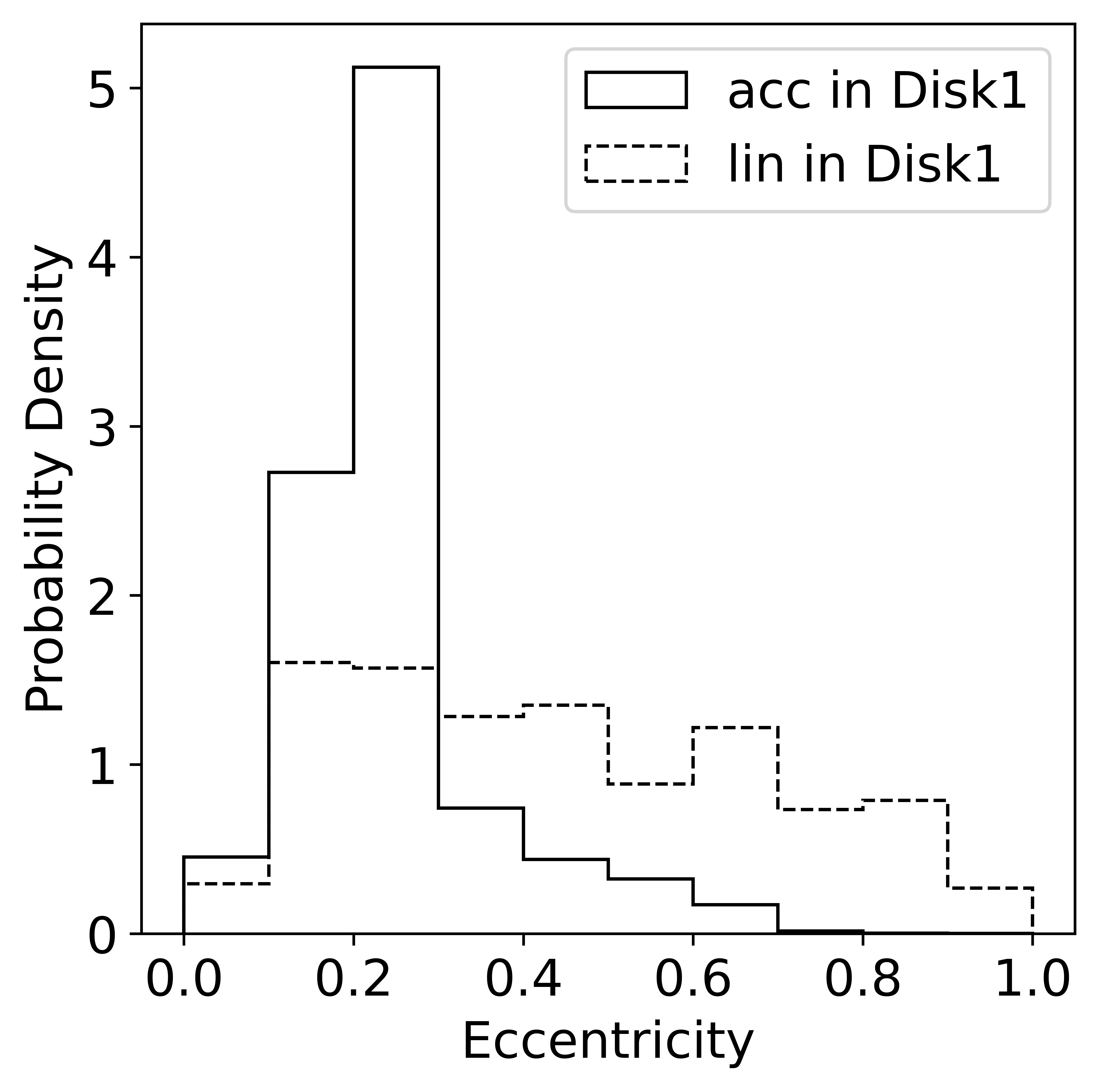}
    \caption{The eccentricity distribution for accelerating sources and non-accelerating sources on Disk1.}
    \label{fig:disk1_ecc}
\end{figure}

The stars in Disk1 are found to have non-circular orbits. \citet{Bartko_2009} combined their results with those of \citet{Gillessen_2009_stars} and reported an eccentricity distribution with <$e$> = 0.36 $\pm$ 0.06. \citet{Yelda_2014} divided stars into accelerating sources and non-accelerating \footnote{These stars also have accelerations but are insignificant that we cannot detect.} sources, having <$e$> = 0.27 $\pm$ 0.09 and <$e$> = 0.43 $\pm$ 0.24, respectively. 
Our eccentricity distribution for Disk1 is shown in Figure \ref{fig:ecc_dist} and has a peak at $e=0.36$ and <$e$> = 0.39 $\pm$ 0.16. Our result has a much larger uncertainty due to our requirement of only using data with completeness information, which results in a smaller overall sample.
We performed an analysis similar to that of \citet{Yelda_2014}, dividing stars into accelerating and non-accelerating moving stars.
The accelerating sources are defined in \citet{Jia_2019}, including 4 stars in our sample: S1-2, S1-3, S2-6 and S4-169.
The comparison of $e$ distribution between accelerating and non-accelerating sources in Disk1 are plotted in Figure \ref{fig:disk1_ecc}, which agrees with the conclusion from \citet{Yelda_2014} - accelerating sources have a well constrained $e$ peaking at 0.2, while the $e$ distribution of non-accelerating sources has a much larger dispersion. Specifically, accelerating sources have <$e$> = 0.25 $\pm$ 0.12 while non-accelerating sources have <$e$> = 0.45 $\pm$ 0.24.
The difference is likely due to accelerating sources having better constraints on their orbital parameters (see Figure \ref{fig:mc}).

Disk1 has a significant intrinsic thickness as was shown by \citet{Paumard_2006, Lu_2009} and \citet{Yelda_2014}. 
For example, \cite{Paumard_2006} reported a disk opening angle of $\Delta_\theta$ = 14\degree $\pm$ 4\degree\ and \citet{Lu_2009} reported a disk thickness of $\Delta_\theta$ = 7\degree $\pm$ 2\degree.
Our intrinsic disk thickness of $\Delta_\theta$ = 7\degree $\pm$ 1\degree\ is consistent with that found in previous work, only with smaller uncertainty.

The surface density profile in the plane of Disk1 is predicted to be $\Sigma$(r) $\propto$ r$^{-2}$ for \textit{in-situ} formation scenarios \citep{Lin_1987, Levin_2007} and has been verified in many observations \citep{Paumard_2006, Bartko_2009, Lu_2009, Yelda_2014}. 
%In our paper, the semi-major axis $a$ also follows a power law function with a slope of -2. \textcolor{red}{Do we have an uncertainty on this slope?}
In our paper, the semi-major axis $a$ distribution was initially fit with methods similar to the eccentricity distribution as discussed in \S\ref{sec:ecc_dist}. However, the fit results were extremely uncertainty and no significant constraint could be placed on the $a$ distribution in the disk plane. We instead explore the distributions of projected radial distances on the disk plane. The slope of the radial profile ($-1.80 \pm 0.17$ from Eq.\ref{equ:r_dist}) is in agreement with the predicted $-2$.

\subsection{Plane2 Properties}
\label{subsec:discuss_plane2}

\edit2{In addition to the CW Disk1, we found a new CCW structure Plane2, which has 10 stars out of the total 88 young stars in our sample.
We estimated that Plane2 has $(i, \Omega) = (90 \pm 20\degree, 245 \pm 19\degree)$, which is similar to the F3 structure reported in \citet{von_Fellenberg_2022}.
A more detailed comparison between these two structures is discussed in Appendix \ref{subsec:sig_comp} and \ref{subsec:disk_comp}.
As presented in \S\ref{sec:ecc_dist}, stars in Plane2 have relatively high eccentricities, with $<e> = 0.68$.
We also estimated the intrinsic disk thickness of Plane2 to be $\Delta_{\theta} = 18.6\degree \pm 6.2\degree$.
The radial profile of Plane2 has a power law index of $-1.43 \pm 0.47$ from Eq.\ref{equ:r_dist}, which is inconsistent with the predicted -2.
This might be caused by the complex dynamical evolution experienced by Plane2 stars such that they have high eccentricities with less steep radial profile.
One of the most prominent feature of Plane2 is its spatial asymmetry.
As discussed in \S\ref{sec:disk_simulation}, this asymmetric feature cannot be explained by differential extinction or incompleteness and is likely intrinsic to the structure.}

\subsection{Is Plane2 Related to the IRS 13 Group or G sources?}
\label{sec:discuss_disk2}
IRS 13 is a group of nearly co-moving massive young stars, clustered together at 3.5\arcsec to the West of the SMBH \citep{Maillard_2004, Paumard_2006, Martins_2007}.
It has been proposed to lie within the previously claimed CCW disk \citep{Maillard_2004, Schodel_2005}, but later works did not detect the CCW disk \citep{Lu_2009, Yelda_2014}, nor do we detect it in this work.
Interestingly, one of IRS 13 group stars, IRS 13E1, is a potential Plane2 star, with a 50\% probability being on Plane2.
The remaining IRS 13 sources E2, E3, and E4, are currently not in our sample as they were excluded either due to their Wolf-Rayet star nature or the lack of published completeness information. 
Here we relax our requirement for a complete sample in order to identify other candidate Plane2 members; although we note that we cannot infer structural properties of Plane2 from this analysis.

\begin{figure}
    \centering
    \includegraphics[width=0.5\textwidth]{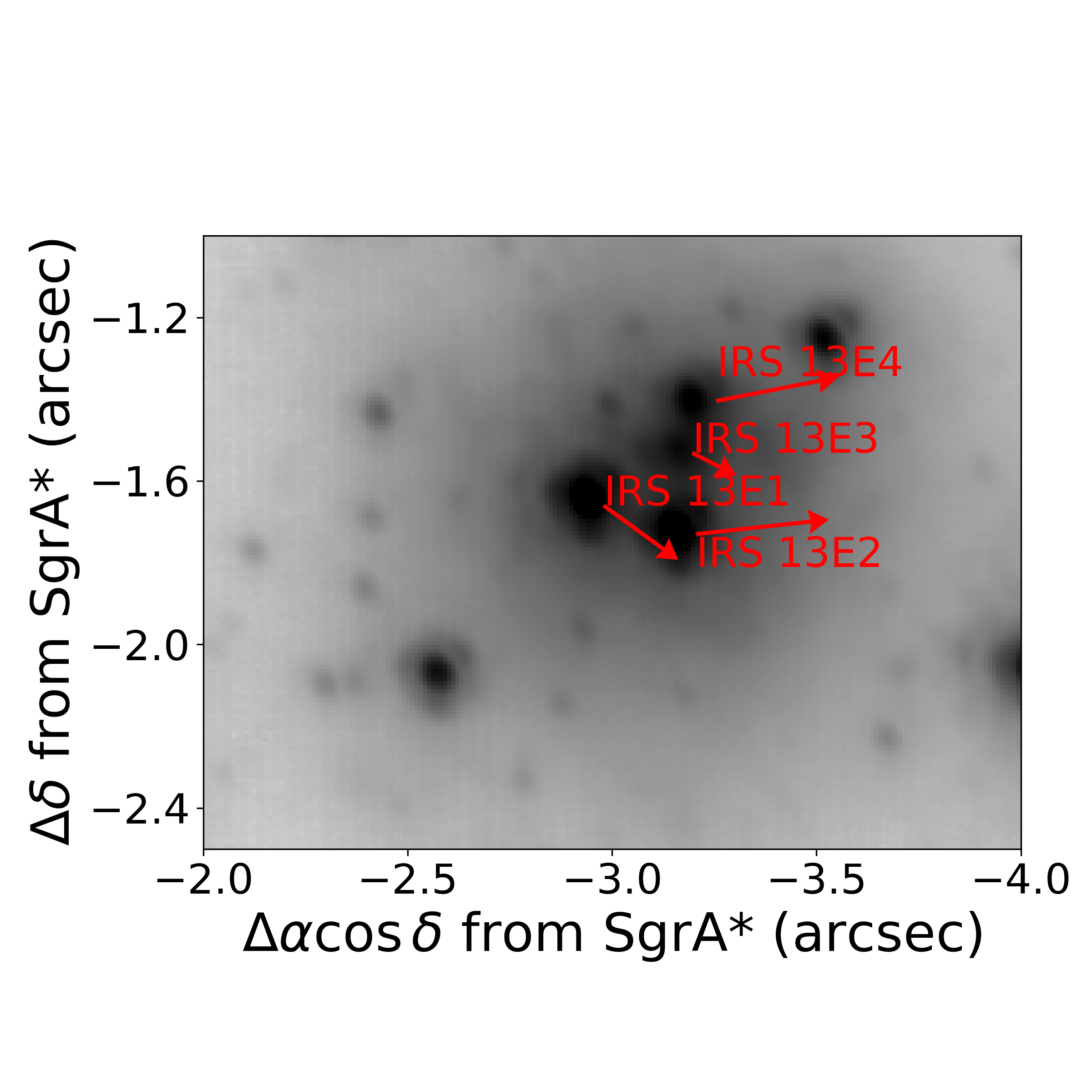}
    \caption{The proper motion for IRS 13 sources, where IRS 13E1 and IRS 13E3 are likely to be on Plane2 while IRS 13E2 and IRS 13E4 are not.}
    \label{fig:irs13_quiver}
\end{figure}

\begin{figure}
    \centering
    \includegraphics[width=0.45\textwidth]{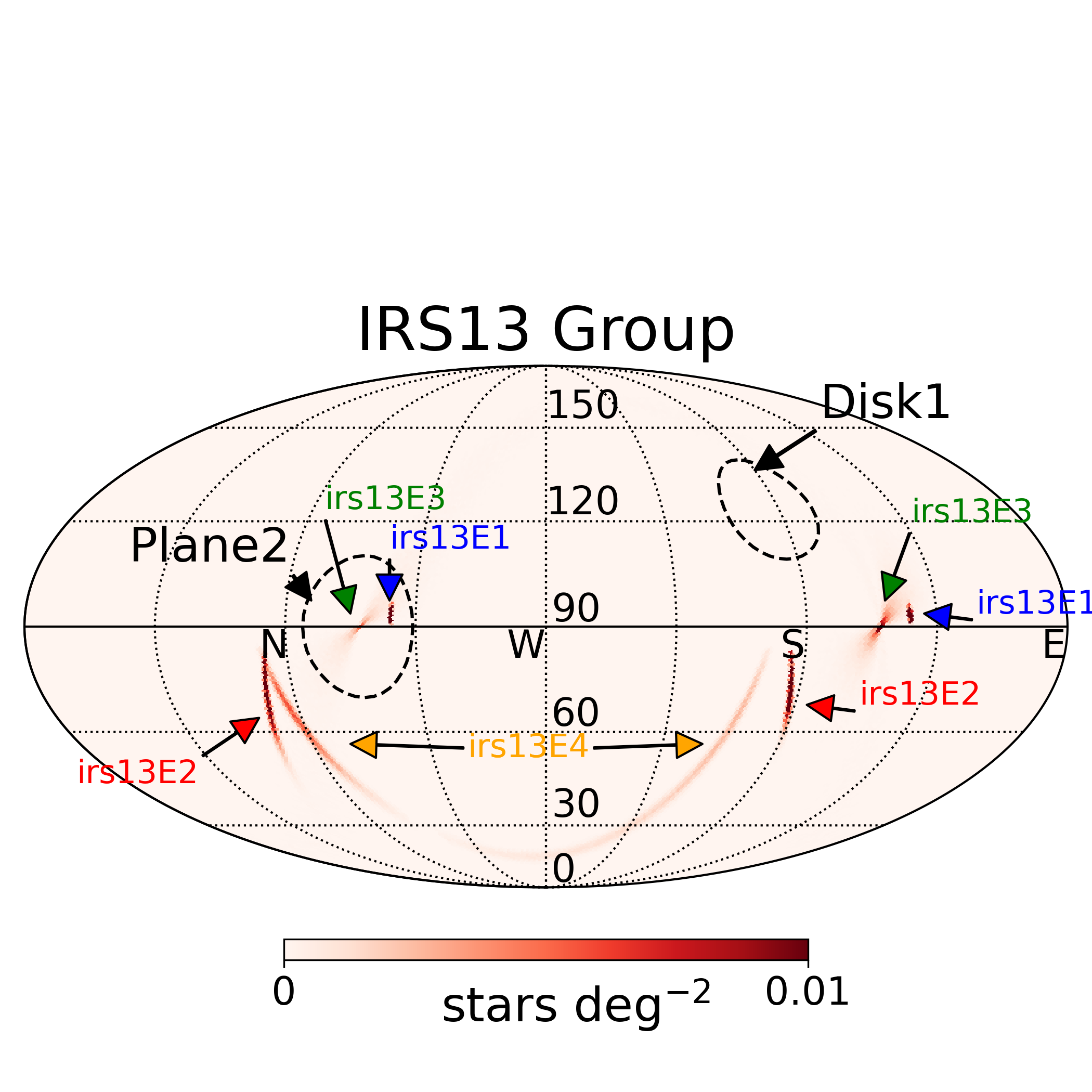}
    \caption{The ($i$, $\Omega$) density map for IRS 13E1, IRS 13E2, IRS 13E3, IRS 13E4.}
    \label{fig:irs13}
\end{figure}

Since all the IRS 13 stars are moving in approximately the similar direction, we run the same orbital and disk membership analysis for three other IRS 13 stars in addition to IRS 13E1. They are IRS 13E2, IRS 13E3, IRS 13E4.
We use the most up-to-date radial velocities reported in \citet{Zhu_2020}.  Note that using RV measurements from \citet{Paumard_2006} and \citet{Bartko_2009} generates similar results. 
The proper motions for IRS 13 stars are plotted in Figure \ref{fig:irs13_quiver} and 
the ($i$, $\Omega$) density map is shown in Figure \ref{fig:irs13}.
% We can see that IRS 13E1 and IRS 13E3b are moving in almost the same direction, implying that both of them are likely to be on plane2.
% IRS 13E2 and IRS 13E4 are moving in a similar direction on the sky, but have a different orbital from IRS 13E1 and IRS 13E3b (more then 20\degree difference in $\Omega$). 
Note that the nature of IRS 13E3 is not entirely clear -- it may be a dusty star or a gas clump at the intersection of colliding winds \citep{Zhu_2020, Wang_2019, Fritz_2010} and the proper motion is quite uncertain \citep{Tsuboi_2022, Fritz_2010}. 
Nevertheless, we find that IRS 13E3 is a potential Plane2 star, with a disk membership probability P$_\mathrm{Plane2}$ = 0.34, but IRS 13E2 and IRS 13E4 are not on Plane2 with P$_\mathrm{Plane2}$ < $10^{-2}$. 
\edit1{Based on the posterior contours of IRS 13E2 and IRS13 E4 in the ($i$, $\Omega$) plane, we think they are close to F1/CCW feature reported in \citet{von_Fellenberg_2022}; however, they did not classify these two IRS13 stars as belonging to F1/CCW feature.}

Even though all four IRS 13 stars are approximately moving in a similar direction on the sky, the dispersion in the proper motions of the stars is significant. 
Recent studies of E1's proper motion propose that this star may not be bound to the IRS13 group, while E2 and E4 are most likely bound to the group \citep{Wang_2019, Mu_i__2008}.
% Besides, IRS 13E3 also has significant differences in proper motion direction and magnitude, which may indicate its unbound relation to the IRS13 group.
Additionally, studies of the spectrum of IRS 13E1 and E2 do not show signs of binarity \citep{Fritz_2010}.
The two different pairs (E1 \& E3 and E2 \& E4) are discrepant enough that they may not be associated with each other.
Thus it is unclear whether Plane2 is related to the potentially bound IRS13 group given the low Plane2 membership probability for E2 \& E4. Further study, including higher-resolution images and continued astrometric and RV monitoring will be needed to resolve the relationship between the apparently bound IRS 13 group and Plane2. 

\begin{figure}
    \centering
    \includegraphics[width=0.5\textwidth]{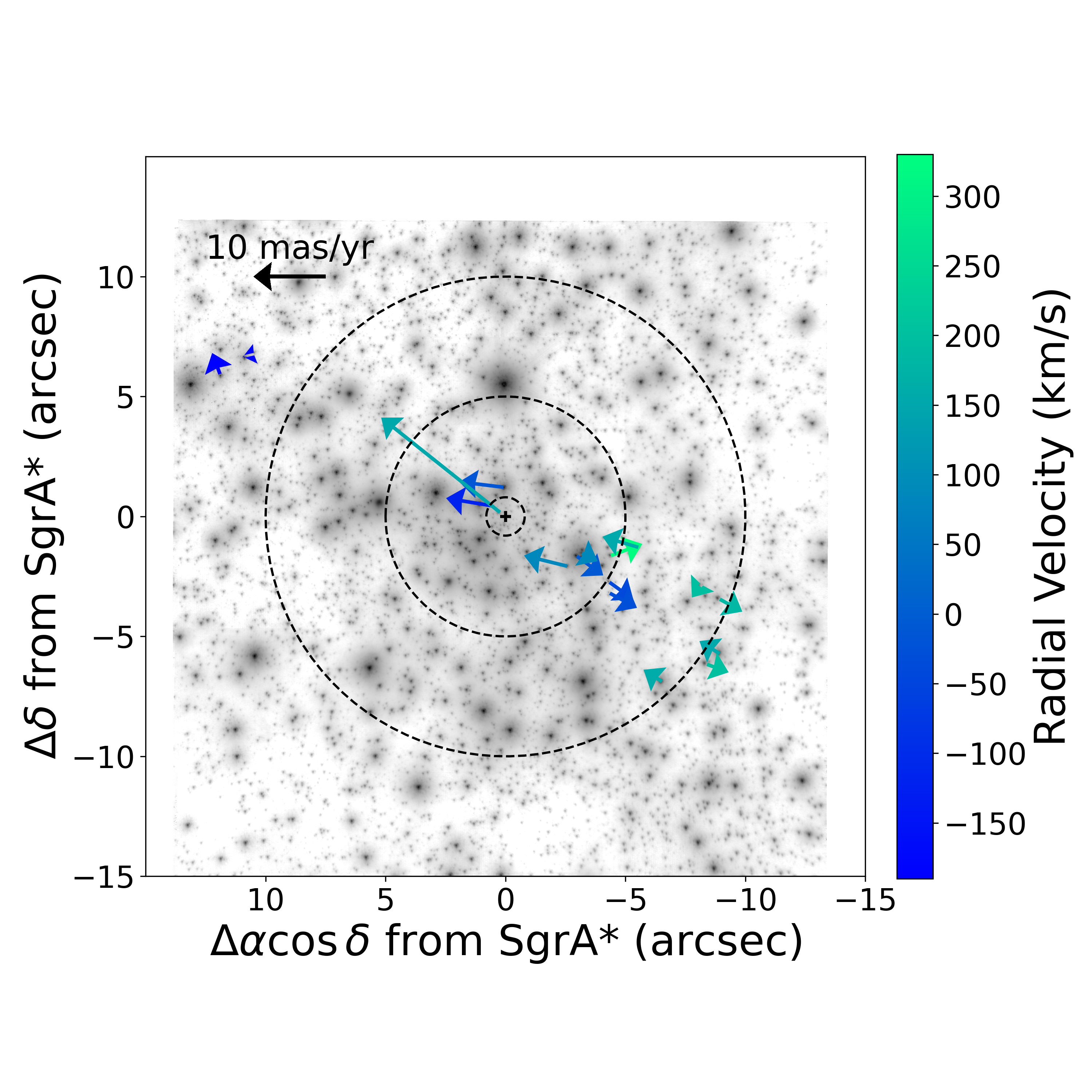}
    \caption{The quiver plot of all potential Plane2 stars, with color showing the RV.}
    \label{fig:disk2}
\end{figure}

In order to include as many Plane2 stars as possible, we report all potential Plane2 stars based on our analysis of all 146 stars with reported RVs and proper motions, listed in Table \ref{tab:rv}.
The potential Plane2 stars and their disk membership is reported in Table \ref{tab:disk2}.
A quiver plot showing the proper motion and RV for those potential Plane2 stars (P$_\mathrm{Plane2}$ > 0.2) is shown in Figure \ref{fig:disk2}. 

\startlongtable 
\begin{deluxetable}{lrr}
\tablecaption{Plane2 Summary\label{tab:disk2}}
\tablehead{\colhead{Name} & \colhead{P$_\mathrm{Plane2}$}  & \colhead{SA}}
\startdata
S0-31         &   0.30 &  0.004   \\ 
IRS 16NW      &   0.38 &  0.046   \\ 
IRS 13E1      &   0.50 &  0.002   \\ 
IRS 13E3      &   0.34 &  0.265   \\ 
S4-258        &   0.40 &  0.093   \\ 
S5-34         &   0.36 &  0.034   \\ 
S5-106        &   0.34 &  0.029   \\ 
S5-236        &   0.71 &  0.024   \\ 
S9-114        &   0.52 &  0.063   \\ 
S9-221        &   0.77 &  0.044   \\ 
S10-185       &   0.73 &  0.132   \\ 
S10-238       &   0.78 &  0.002   \\ 
S12-5         &   1.00 &  0.003   \\ 
S13-3         &   0.27 &  0.133   \\ 
S0-16         &   0.42 &  0.001   \\ 
S3-26         &   0.24 &  0.163   \\ 
S8-196        &   0.22 &  0.142   \\ 
\enddata
\end{deluxetable}

%\begin{figure*}
%    \centering
%    \subfloat[]{
%    \includegraphics[width=0.5\textwidth]{irs13E1.png}
%    }
%    \subfloat[]{
%    \includegraphics[width=0.5\textwidth]{irs13E2.png}
%    } \\
%    \subfloat[]{
%    \includegraphics[width=0.5\textwidth]{irs13E3b.png}
%    }
%    \subfloat[]{
%    \includegraphics[width=0.5\textwidth]{irs13E4.png}
%    }
%    \caption{The ($i$, $\Omega$) density map for IRS 13E1, IRS 13E2, IRS 13E3b, IRS 13E4.}
%    \label{fig:irs13}
%\end{figure*}

\begin{figure}
    \centering
    \includegraphics[width=0.45\textwidth]{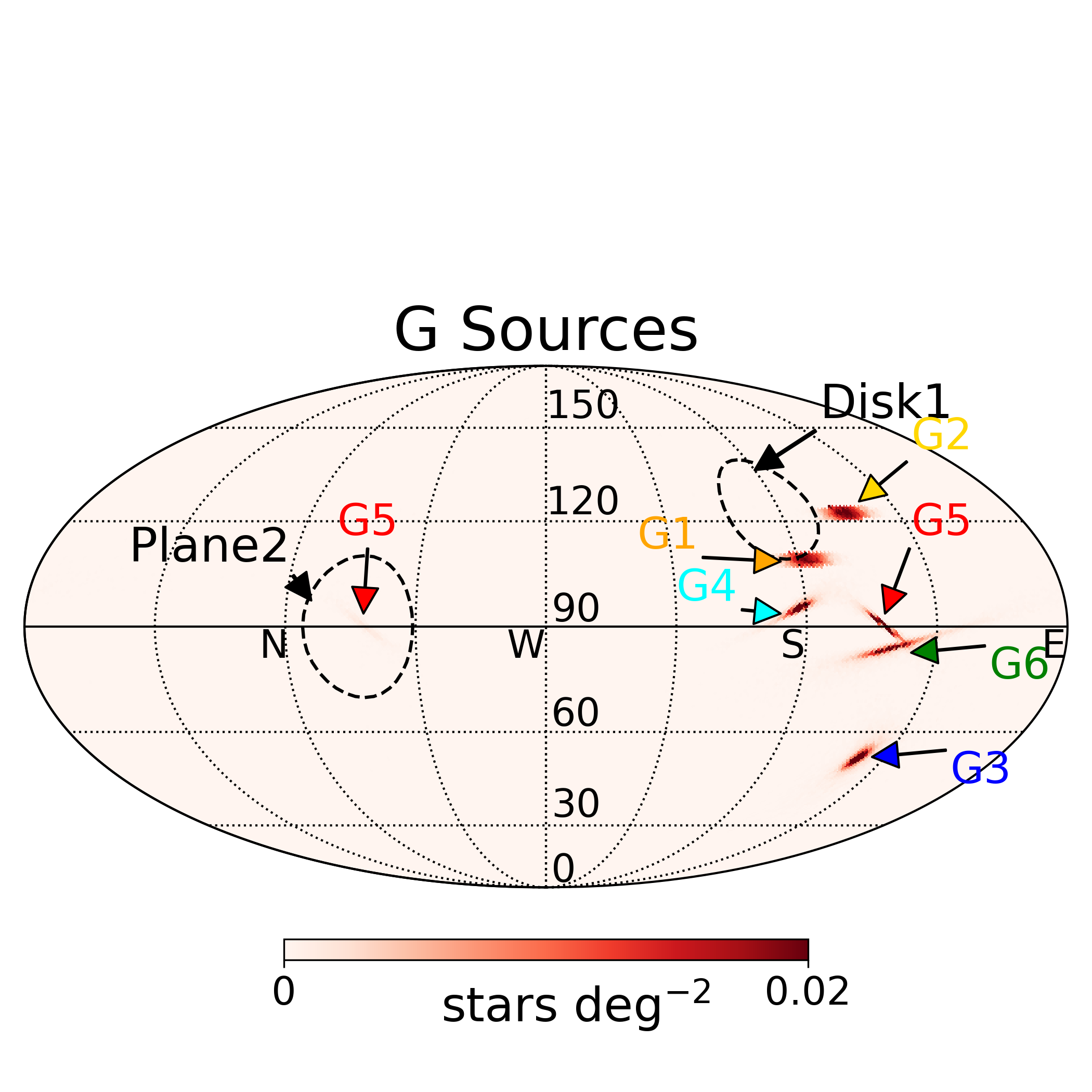}
    \caption{The ($i$, $\Omega$) density map for G1, G2, G3, G4, G5, and G6. We don not have full posterior samples for G1 and G2 so we use their $i$, $\Omega$, and associated uncertainty from \citep{Witzel_2017, Gillessen_2019} to generate the 2D Gaussian distribution.}
    \label{fig:Gsource}
\end{figure}

A population of dust enshrouded objects are found to orbit around the SMBH: the so-called G sources. 
The most famous G source is G2, which first looked like a pure gas cloud \citep{Gillessen_2013_G2, Eckart_2013,Phifer_2013}, yet survived through closest approach to SgrA$^\ast$ in early 2014 \citep{Gillessen_2013_G2,  Abarca_2014, Shcherbakov_2014, Witzel_2014, Valencia_2015}. This implies that G2 must contain a stellar-like object and is perhaps a binary merge product.
However, there is still no broad consensus as to the origin and nature of the G sources.
Using the near-infrared (NIR) spectro-imaging data obtained over 13 years at the W. M. Keck Observatory with the OSIRIS integral field spectrometer, \citet{Ciurlo_2020} reported four more additional G sources, making the total number of G sources to six. 
\citet{Ciurlo_2020} found the six G sources (G1, G2, G3, G4, G5 and G6) have widely varying orbits, suggesting G sources are formed separately. 
We compare the orbital plane direction for G sources with stellar disk plane and the individual G source ($i$, $\Omega$) density map is plotted in Figure \ref{fig:Gsource}.
The conclusion is that none of the G sources are likely to be on either Disk1 or Plane2.
However G5 is also edge-on, and has almost exactly 180\degree difference in $\Omega$ compared to Plane2.
In fact, there is a small probability of P$_\mathrm{Plane2}$ = 0.08 for G5 for its degenerate solution.
In other words, G5 lies in the edge-on Plane2; but is counter-rotating in the plane.

%\begin{figure*}
%    \centering
%    \subfloat[]{
%    \includegraphics[width=0.5\textwidth]{G3.png}
%    }
%    \subfloat[]{
%    \includegraphics[width=0.5\textwidth]{G4.png}
%    } \\
%    \subfloat[]{
%    \includegraphics[width=0.5\textwidth]{G5.png}
%    }
%    \subfloat[]{
%    \includegraphics[width=0.5\textwidth]{G6.png}
%    }
%    \caption{The ($i$, $\Omega$) density map for G3, G4, G5, G6.\textcolor{red}{AC: are you going to add G1 and G2 here?}}
%    \label{fig:Gsource}
%\end{figure*}

\subsection{The Star Formation Process}
Besides the widely accepted \textit{in-situ} formation theory, it has been suggested that an inspiraling star cluster to explain the formation of young stars in our GC, where a massive young cluster migrate towards the center of the galaxy under dynamical friction \citep{Gerhard_2001}.
However, this scenario will deposit stars with a profile of r$^{-0.75}$, which is inconsistent with the observed density profile.
So the fact that radial profile from \S\ref{sec:disk_property} follows P(r) $\propto$ r$^{-1.80}$ supports \textit{in-situ} formation scenario.

Although we did not detect the counter-clockwise disk claimed in \citep{Genzel_2003, Paumard_2006, Bartko_2009}, we found another almost edge-on disk (Plane2 in this paper) with highly asymmetric stellar distribution. 
This edge-on disk is mostly determined by stars on the southwest side and very few stars on the northeast side are found in this disk.
While the uneven distribution of stars on Plane2 is not fully understood, possible explanations include:
(1) The young stellar population is not uniformly distributed  within Plane2, with more stars in the southwest region compared to northeast region (see \S\ref{sec:disk_simulation}).
\edit1{One possible situation that could explain this asymmetry is the non-uniform initial gas distribution, leading to uneven star formation.}
(2) If Plane2 is indeed related to the IRS13 group, the possible explanations for the formation of IRS13 group can also be used to explain Plane2.  
For example, Plane2 might be a remnant stream from the disruption of an IRS 13 cluster that may or may not contain an intermediate-mass black hole (IMBH)  \citep{Maillard_2004}. 
%Another possible explanation is the colliding winds from a massive binary stars and the stellar winds will significantly affect the motion of ionized gas \citep{Coker_2000, Zhao_2009, Zhu_2020}.

%\textcolor{red}{
%\begin{itemize}
%    \item Comparison to prior work.
%    \item Comparison to G-sources.
%    \item Comparison to structures in formation (Perets, et al.).
%    \item Discussion of binarity impact.
%\end{itemize}
%}

\subsection{Biases Induced by Binaries}

In our analysis of disk membership and disk properties, we assume that stars are not in binary systems. 
However, this may lead to biases in our results as presented by \citet{Naoz_2018}: if we ignore binaries, the disk memberships and disk fractions are likely to be biased to lower values; observed eccentricity and dispersion angle are likely to be biased to higher values.

Here we present an analysis of the influence of ignoring binaries on disk membership on each sub-structure, Disk1 and Plane2. 
We simulate a sample of stars drawn from each structure and generate 3D positions and velocities for each star based on the radial profile and eccentricity distribution presented in \S\ref{sec:ecc_dist} and \S\ref{sec:radial_profile}.
Then we randomly assign each star as a binary or not based on the binary fraction of $70\%$ as is appropriate for massive stars \citep{Raghavan_2010, Stephan_2016}. 
If a star is assigned to be in a binary system, we assign its binary properties, such as binary eccentricity, mass ratio, and inclination, drawn from distributions of massive star binary properties described in \citet{Sana_2012}. 
The primary stellar masses are sampled based on the IMF described in \citet{Lu_2013}. 

\begin{figure}[htp]
\centering
\includegraphics[width=0.4\textwidth]{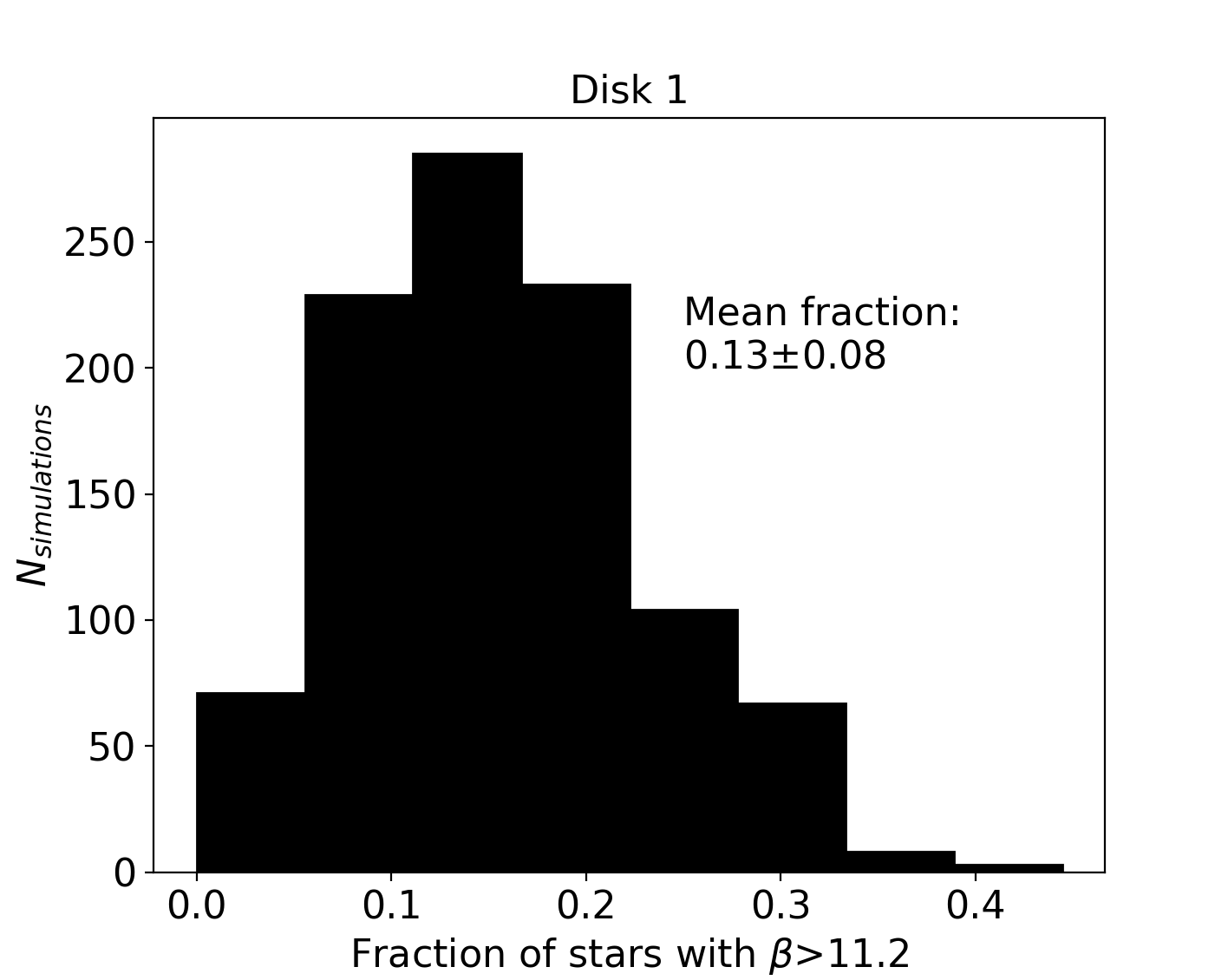}
\includegraphics[width=0.4\textwidth]{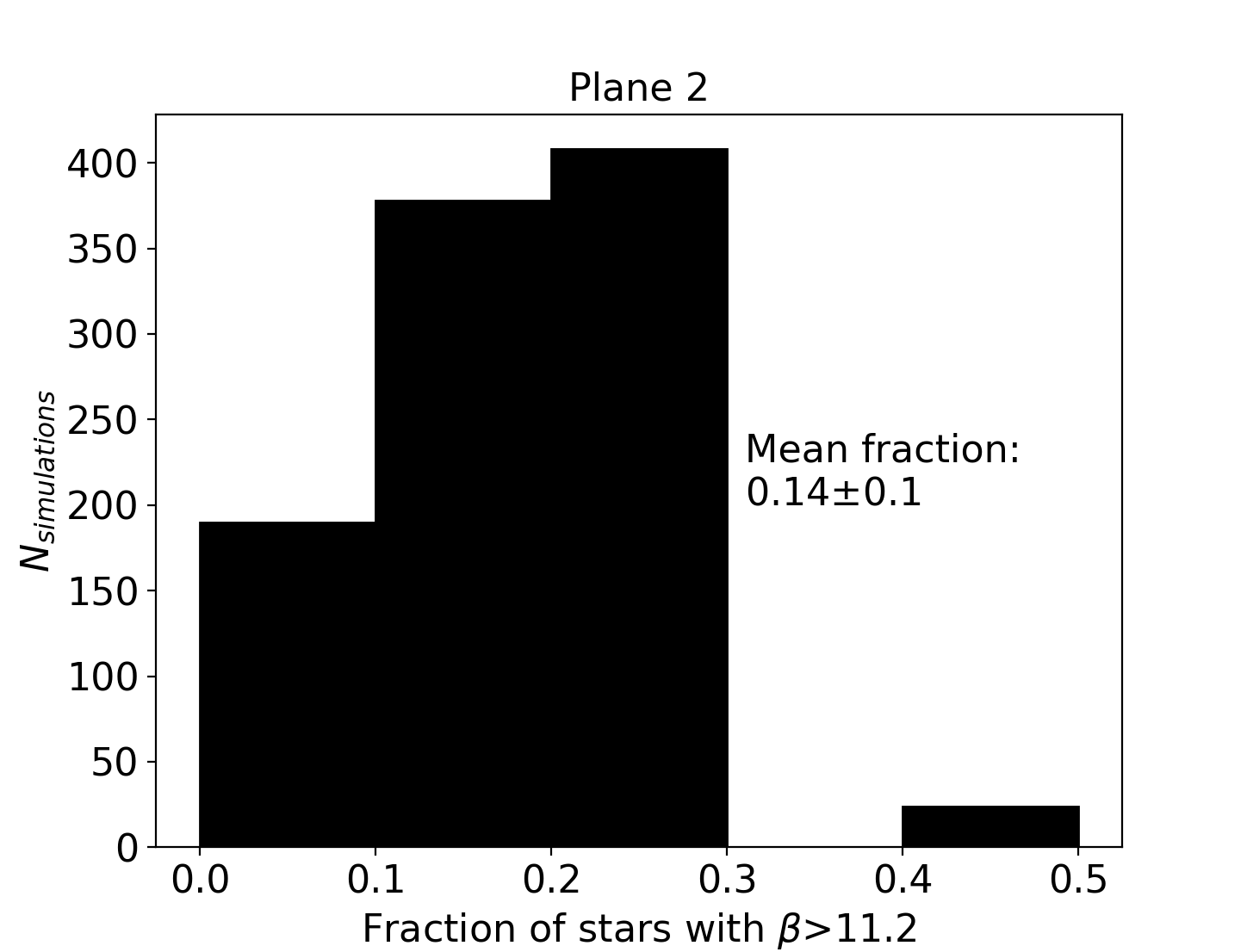}
\caption{The fraction of misidentified off-disk stars when binaries are neglected. The histogram shows the fraction of stars with an angular difference in the input vs. output orbital plane ($\beta$) exceeding expected measurement errors, averaged over 1000 simulations, for Disk1 (top) and Plane2 (bottom).
10-20\% of stars are likely misclassified as off-disk due to their binary nature.}
\label{fig:beta_dist}
\end{figure}

These binary properties, combined with position and velocity data of stars, are used to calculate a proxy parameter $\beta$ defined by \citet{Naoz_2018}, Eq 5. 
$\beta$ is the apparent deflection angle from the true orbital plane around the SMBH if the binary RV is ignored. We follow a similar criterion used in \citet{Naoz_2018} \S3, where a star with $\beta > 11.2\degree$ is considered as an on-disk star misidentified as an off-disk star.
We simulate each structure 1000 times and calculate the mean fraction for $\beta > 11.2\degree$, as shown in Figure \ref{fig:beta_dist}.

We find that for Disk1, the misclassified disk fraction is $0.13 \pm 0.08$ and for Plane2 is $0.14 \pm 0.1$. 
This implies that the presence of binaries should make $\sim 14\%$ of the on-disk stars appear as off-disk stars. 
Because we are ignoring binaries in our analysis, taking them into account would increase the Disk1 sample size from 18 stars to $\sim$ 21 stars and Plane2 sample size from 10 stars to $\sim$ 11 stars. 
Given this relatively small change in sample size, we think the bias induced by ignoring binaries is not significant if the binary fraction is 70\%. However, binary RV surveys are needed to fully quantify the true binary fraction and the size of the bias.

In this section, we only calculate the bias on disk membership. 
Further analysis could be done to quantifying biases on eccentricity and disk dispersion angle, which can provide a more realistic set of disk properties.

\section{Summary}
\label{sec:summary}
We analysed the dynamical structure of 88 young stars at the Galactic Center with projected radii from $0\arcsec$ to $15\arcsec$ with well measured proper motions, radial velocities and spectroscopic completeness information.
%The proper motion comes from a long time baseline observation with 10m Keck telescope since 1995 in the inner region and HST WFC-IR observation in the outer region.
%The radial velocity comes from our GCOWS survey using diffraction-limited Keck OSIRIS spectra, combined with RVs published in \citet{Feldmeier_2015}.
This is the largest sample of GC young stars published with proper motion, radial velocity and completeness correction. 
We also simulate star clusters with different dynamical sub-structure to directly compare with what is actually observed on sky.
We detect the well known clockwise disk (Disk1 in this paper), consistent with what has been previously published.
We also find a second, almost edge-on, counter-clockwise disk  (Plane2 in this paper).
Plane2 is asymmetric with stars concentrated in the southwest direction, and two IRS 13 stars are found to potentially reside in this plane. 
\edit1{By applying a hierarchical Bayesian inference framework, we calculate that} Disk1 has an eccentricity peaking at 0.2 \edit1{for sources with well-measured accelerations}, consistent with previous analysis, but Plane2 is even more eccentric than Disk1, with P(e) $\propto$ e.
By simulating Disk1, Plane2 and Non-disk stars using the observed disk properties, we are able to reproduce observed density maps for Disk1 and the Non-disk stars, but the highly asymmetric structure in Plane2 can not be explained with a uniform disk and extinction alone.
In the future, we will use this sample with known completeness correction to constrain the IMF for different dynamical subgroups, which will greatly help us understand their formation history.

%%%%%%%%%%%%%%
% Acknowledgements
%%%%%%%%%%%%%%
\section{Acknowledgements}

\edit1{We thank the referee for a very quick, yet thorough report which helped to improve the paper.} We also thank the staff of the W.M.~Keck Observatory for all their help in obtaining observations.
We acknowledge support from the W. M. Keck Foundation, the Heising Simons Foundation, and the National Science Foundation (AST-1412615, AST-1518273).
M. W. Hosek Jr. also acknowledges support by Brinson Prize Fellowship. 
The W.M. Keck Observatory is operated as a scientific partnership among the California Institute of Technology, the University of California and the National Aeronautics and Space Administration.  
The Observatory was made possible by the generous financial support of the W. M. Keck Foundation.  
The authors wish to recognize and acknowledge the very significant cultural role and reverence that the summit of Maunakea has always had within the indigenous Hawaiian community.  
We are most fortunate to have the opportunity to conduct observations from this mountain.

\vspace{5mm}

\facilities{Keck Observatory}

\software{
AstroPy \citep{astropy_2013},
Matplotlib \citep{matplotlib_2007},
SciPy \citep{scipy_2001}
%KS2 (Anderson et al. 2008)
}


\clearpage
\appendix

\section{Radial Velocities \& Proper Motion Data}
\label{tabs}

\startlongtable 
\begin{deluxetable*}{lrrcrrcrrccc}

\tabletypesize{\footnotesize}
\tablewidth{0pt}

\tablecaption{RVs Summary\label{tab:rv}}
\tablehead{\colhead{Name} & \colhead{RV$_\mathrm{GCOWS}$ \tablenotemark{a}}  & \colhead{t0$_\mathrm{GCOWS}$} & 
            \colhead{ID$_\mathrm{P06}$}  & \colhead{RV$_\mathrm{P06}$ \tablenotemark{b}} & 
            \colhead{RV$_\mathrm{B09}$ \tablenotemark{c}} &  
            \colhead{ID$_\mathrm{F15}$}  & \colhead{RV$_\mathrm{F15}$ \tablenotemark{d}} & 
            \colhead{RV$_\mathrm{use}$}  & \colhead{ref \tablenotemark{e}} & 
            \colhead{WR} & \colhead{Data Available \tablenotemark{f}} \\ 
            \colhead{} & \colhead{(km s$^{-1}$)}  & \colhead{} & 
            \colhead{}  & \colhead{(km s$^{-1}$)} & 
            \colhead{(km s$^{-1}$)}  &  
           \colhead{}  & \colhead{(km s$^{-1}$)} & 
           \colhead{(km s$^{-1}$)} & \colhead{} & \colhead{} & \colhead{} } 
\startdata
S0-20         &    261 $\pm$    25$^{\ast}$  &  2008.4 & E3    &   -280 $\pm$     50 &       --          &  --   &       --          &   multiple RVs    & 1      &  -  & RV + C + PM  \\ 
S0-2          &   -473 $\pm$    27$^{\ast}$  &  2002.4 & E1    &  -1060 $\pm$     25 &       --          &  --   &       --          &   multiple RVs    & 1      &  -  & RV + C + PM  \\ 
S0-1          &  -1054 $\pm$    19$^{\ast}$  &  2005.5 & E4    &  -1033 $\pm$     25 &       --          &  --   &       --          &   multiple RVs    & 1      &  -  & RV + C + PM  \\ 
S0-8          &   -389 $\pm$    40$^{\ast}$  &  2006.5 & E7    &   -390 $\pm$     70 &       --          &  --   &       --          &   multiple RVs    & 1      &  -  & RV + C + PM  \\ 
S0-16         &    139 $\pm$    28$^{\ast}$  &  2011.5 & E2    &    300 $\pm$     80 &       --          &  --   &       --          &   multiple RVs    & 1      &  -  & RV + C + PM  \\ 
S0-3          &   -709 $\pm$    15$^{\ast}$  &  2006.5 & E6    &   -570 $\pm$     40 &       --          &  --   &       --          &   multiple RVs    & 1      &  -  & RV + C + PM  \\ 
S0-5          &    631 $\pm$    14$^{\ast}$  &  2005.5 & E9    &    610 $\pm$     40 &       --          &  --   &       --          &   multiple RVs    & 1      &  -  & RV + C + PM  \\ 
S0-19         &    164 $\pm$    46$^{\ast}$  &  2006.5 & E5    &    280 $\pm$     50 &       --          &  --   &       --          &   multiple RVs    & 1      &  -  & RV + C + PM  \\
\enddata
\tablenotetext{a}{RV measurements from our database. \\Stars with multiple measurements in our database are marked with asterisks and only one of their measurements is shown here.}
\tablenotetext{b}{RV measurements from \cite{Paumard_2006}}
\tablenotetext{c}{RV measurements from \cite{Bartko_2009}}
\tablenotetext{d}{RV measurements from \cite{Feldmeier_2015}}
\tablenotetext{e}{Reference for final $RV_{use}$: 1 - database, 2 - P06\citep{Paumard_2006}, 3 - B09\citep{Bartko_2009}, 4 - F15\citep{Feldmeier_2015}, 5 - Z20\citep{Zhu_2020}}
\tablenotetext{f}{The available data for each star: RV - Radial Velocity, C - spectral completeness, PM - proper motion. See details in \ref{sec:sample}. Our final sample of 88 stars has measurements of RV, proper motion and spectral completeness.} 
\end{deluxetable*}

\startlongtable 
\begin{deluxetable*}{lrrrrrrrrl}
\tabletypesize{\footnotesize}
\tablewidth{0pt}
\tablecaption{Proper Motions Summary\label{tab:pm}}
\tablehead{\colhead{Name} & \colhead{Kp} & 
            \colhead{Kp$_\mathrm{ext}$ \tablenotemark{a}} & 
            \colhead{ar} & \colhead{x0}  & \colhead{y0} & 
            \colhead{vx} & \colhead{vy} & 
            \colhead{t0} & \colhead{source \tablenotemark{b}} \\ 
            \colhead{} & \colhead{} & \colhead{} & \colhead{(km s$^\mathrm{-1}$ yr$^\mathrm{-1}$)} &
            \colhead{($\arcsec$)}  & \colhead{($\arcsec$)} & 
            \colhead{(mas yr$^\mathrm{-1}$)} & \colhead{(mas yr$^\mathrm{-1}$)} & 
            \colhead{} & \colhead{}} 
\startdata
S0-20         &   15.7 &   16.1 &  -167.68  $\pm$  17.25 &   0.162 $\pm$ 0.00665 &   0.096 $\pm$ 0.00824 &    32.32 $\pm$    2.17 &    -8.66 $\pm$    2.22 &  2011.3 &  1  \\ 
S0-2          &   14.1 &   14.5 &  -111.17  $\pm$  12.27 &  -0.029 $\pm$ 0.00213 &   0.160 $\pm$ 0.00390 &    -6.50 $\pm$    0.91 &    -2.13 $\pm$    1.54 &  2010.2 &  1  \\ 
S0-1          &   14.7 &   15.2 &  -103.22  $\pm$   2.09 &   0.111 $\pm$ 0.00076 &  -0.294 $\pm$ 0.00296 &    19.00 $\pm$    0.32 &    -7.58 $\pm$    1.39 &  2009.9 &  1  \\ 
S0-8          &   15.8 &   16.1 &   -77.39  $\pm$   2.75 &  -0.179 $\pm$ 0.00153 &   0.099 $\pm$ 0.00065 &    22.49 $\pm$    1.02 &   -28.73 $\pm$    0.34 &  2010.3 &  1  \\ 
S0-16         &   15.3 &   15.8 &   -41.50  $\pm$  15.75 &   0.230 $\pm$ 0.00586 &   0.171 $\pm$ 0.00234 &    16.51 $\pm$    2.12 &    13.17 $\pm$    1.63 &  2008.0 &  1  \\ 
S0-3          &   14.5 &   14.9 &   -43.15  $\pm$   1.09 &   0.340 $\pm$ 0.00115 &   0.114 $\pm$ 0.00034 &     7.62 $\pm$    0.58 &    -1.71 $\pm$    0.16 &  2009.8 &  1  \\ 
S0-5          &   15.1 &   15.6 &   -62.51  $\pm$   0.82 &   0.170 $\pm$ 0.00083 &  -0.359 $\pm$ 0.00175 &    -3.48 $\pm$    0.39 &     2.49 $\pm$    0.80 &  2009.8 &  1  \\ 
S0-19         &   15.5 &   15.9 &   -59.35  $\pm$   1.77 &  -0.007 $\pm$ 0.00037 &   0.400 $\pm$ 0.00201 &     8.52 $\pm$    0.10 &    16.10 $\pm$    0.98 &  2009.7 &  1  \\
\enddata
\tablenotetext{a}{Extinction corrected Kp to Aks=2.7}
\tablenotetext{b}{Reference for final proper motion: 1 - central 10 \arcsec, 2 - maser mosaic, 3 - HST}
\end{deluxetable*}

\section{Comparison to The von Fellenberg et al. 2022 Work}
\label{app:comp}

In following sections, we compare our data set with the one used in \citet{von_Fellenberg_2022}, where they reported a total of 5 dynamical structures, including the CCW feature that we consider insignificant. 
In total, their work includes a sample of 195 stars from a ${\sim 30'' \times 30''}$ spectroscopic survey of the GC, and they report 95 stars belonging to 5 features: Inner Clockwise Disk (CW1), Outer Clockwise Disk (CW2), Counter-clockwise Disk/Filament (CCW/F1), Outer Filament 2 (F2), and Outer Filament 3 (F3). 
Our data set has a total of 88 stars extending to $\sim 14''$, where we find 28 stars belonging to Disk1 and Plane2. 

While the two works agree on the clockwise disk (CW1$_{F22}$ = Disk1) and nearly agree on a new filament (F3$_{F22}$ = Plane2), we do not find the other structural features rise to significance in our data set.
\edit2{As explained in Appendix \ref{subsec:sig_comp}, the discrepancy is likely caused by several factors such as differences in the measured proper motions, radial velocities, and sample of stars used.}

In Appendix \ref{subsec:vel_comp} we compare proper motions, in Appendix \ref{subsec:sig_comp} we discuss differences between how the significance of dynamical features are calculated, and finally in Appendix \ref{subsec:disk_comp} we analyze membership to different structures of individual star.

\subsection{Velocity Comparison}
\label{subsec:vel_comp}

For the common 48 stars that have reported proper motions and radial velocities from both data sets, we analyze the data consistency for each star. 
We first calculate the combined uncertainty for each star's velocity data as the quadratic sum of uncertainties from both data sets (eg: $\sigma_{x} = \sqrt{(\sigma^2_{x, curr} + \sigma^2_{x, F22})}$, where $\sigma_{x, curr}$ is the uncertainty in x direction in our data set while $\sigma_{x, F22}$ is the uncertainty in von Felleberg's data set). 
Velocity data are considered to be inconsistent with one another if $|\Delta_{v_x}| = |v_{x, curr} -v_{x, F22}| > 3 \sigma_{x}$. 
The comparison of $v_{x}$, $v_{y}$, and $v_{z}$ for each star is shown in Table \ref{tab:comp_vel}.
We compare velocities (km s$^{-1}$) rather than direct measurements of proper motion (mas yr$^{-1}$) since \citet{von_Fellenberg_2022} does not report the distance used for conversion.
Even though $\Delta v_x$ and $\Delta v_y$ are usually small compared to $\Delta v_z$, the uncertainties on the proper motions are significantly smaller than that of the radial velocities. Thus, $\sim 73\%$ of stars classified as having inconsistent velocity data in either x or y direction while only one star, namely S4-169, has inconsistent $v_{z}$ with opposite signs. 

\begin{figure*}[!ht]
  \centering
  \subfloat[][PM Difference in X and Y Directions] {\includegraphics[width=.4\textwidth]{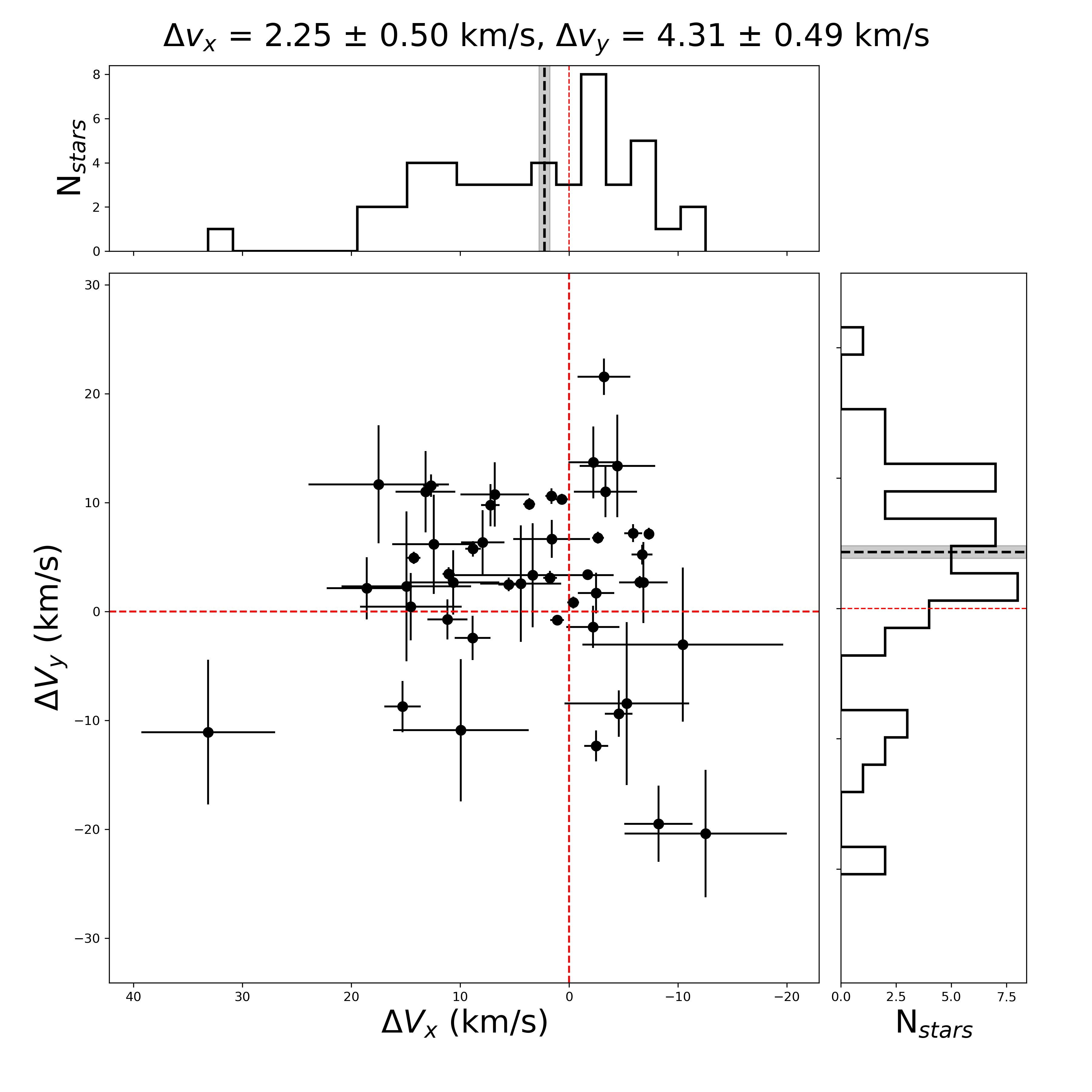}}\quad
  \subfloat[][PM Difference in X and Y Directions in units of $\sigma$] {\includegraphics[width=.4\textwidth]{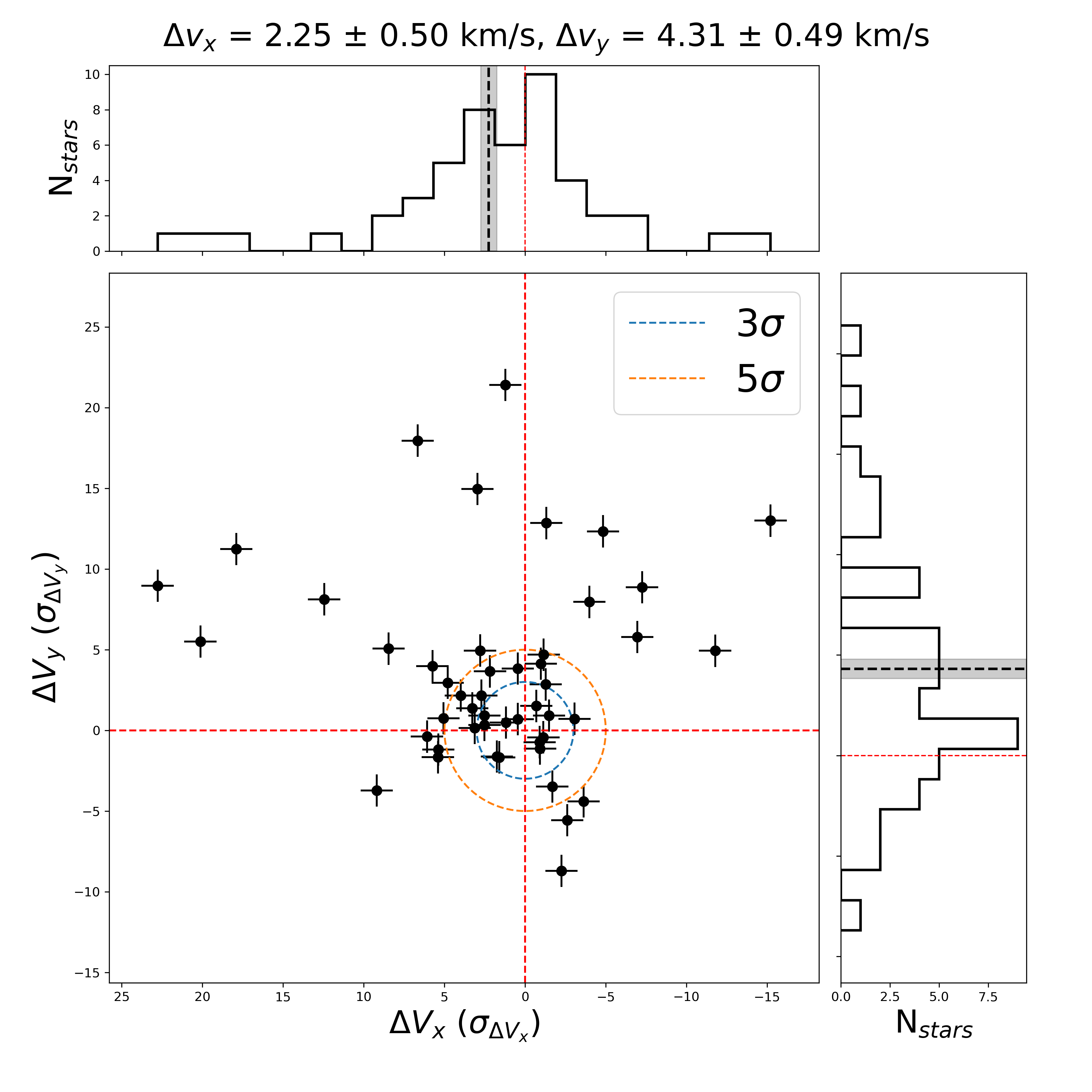}}\\
  \subfloat[][PM Difference in Tangential and Radial Directions] {\includegraphics[width=.4\textwidth]{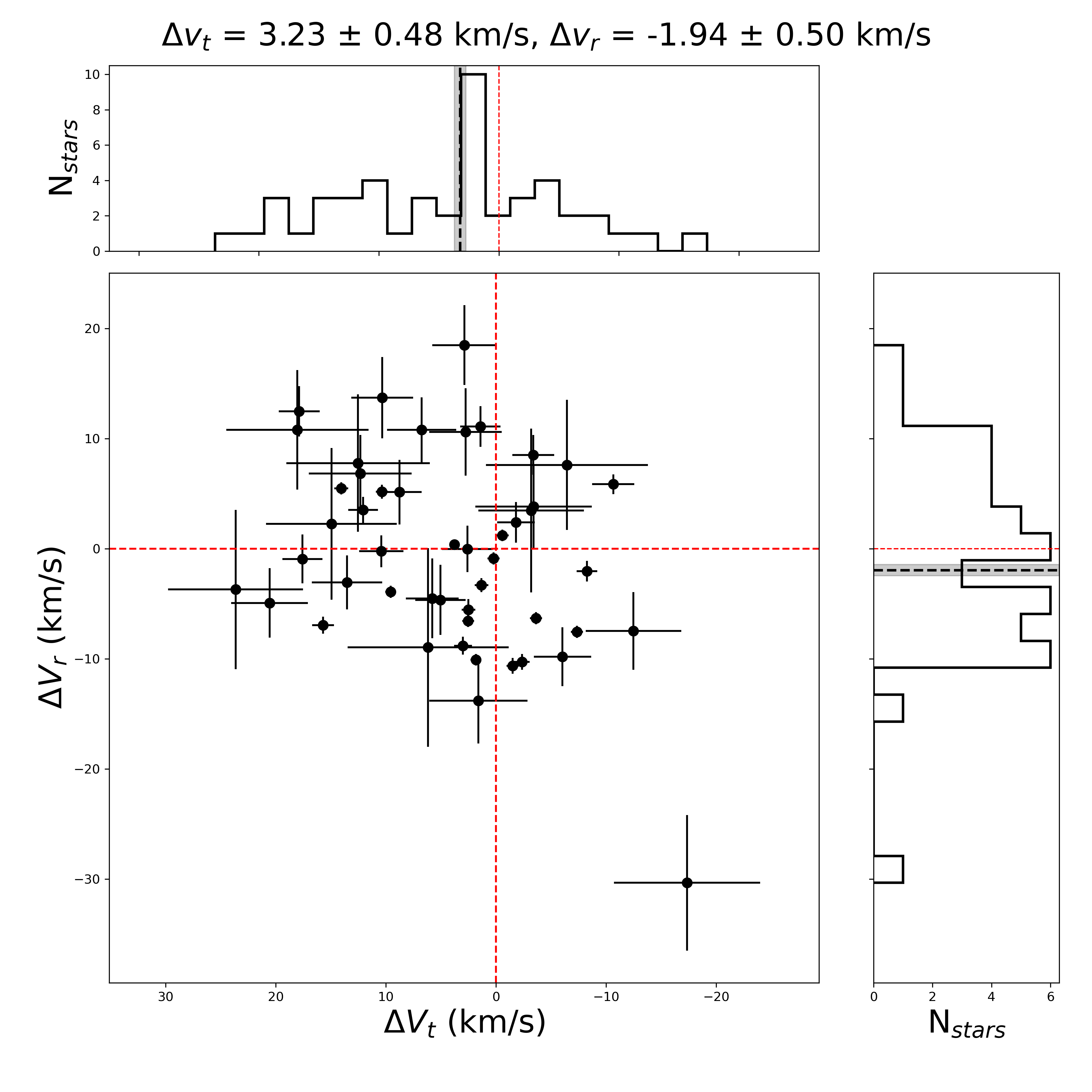}}\quad
  \subfloat[][Same as (c) But in units of $\sigma$] {\includegraphics[width=.4\textwidth]{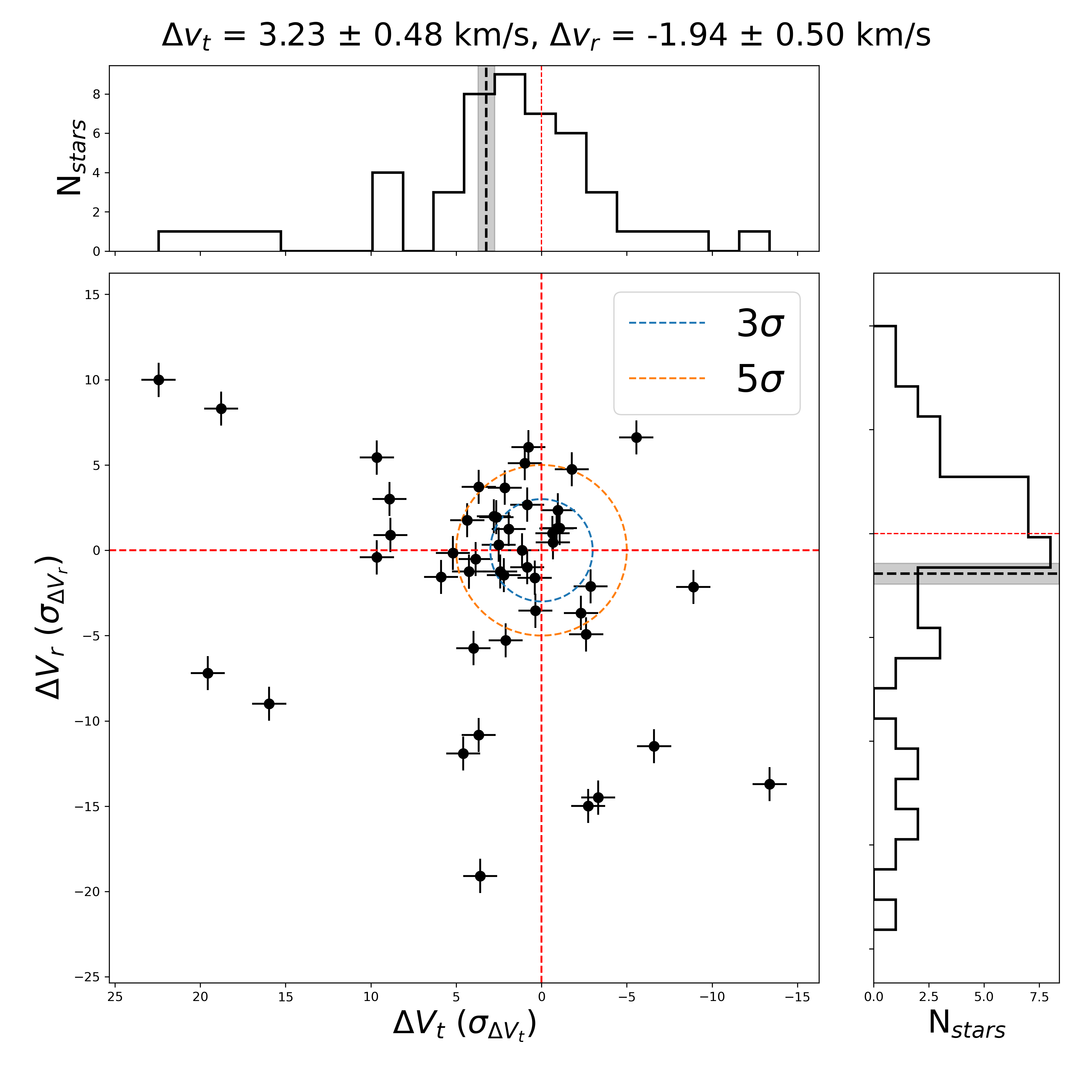}}
  \caption{Comparison of proper motions from the current work with that of \citet{von_Fellenberg_2022}. The upper left panel shows differences in x and y directions, plotted in unit of km s$^{-1}$. The upper right panel shows the same differences, plotted in units of star's corresponding $\sigma$. Similar comparisons in the radial and tangential directions are shown in the lower two panels. For the two $\sigma$-plots (right), we also draw 2 circles showing 3$\sigma$ (blue) and 5$\sigma$ (orange) regions to illustrate our criterion of consistency.}
  \label{fig:pm_diff}
\end{figure*}

\begin{figure*}[htp]
\centering
    \includegraphics[width=\textwidth]{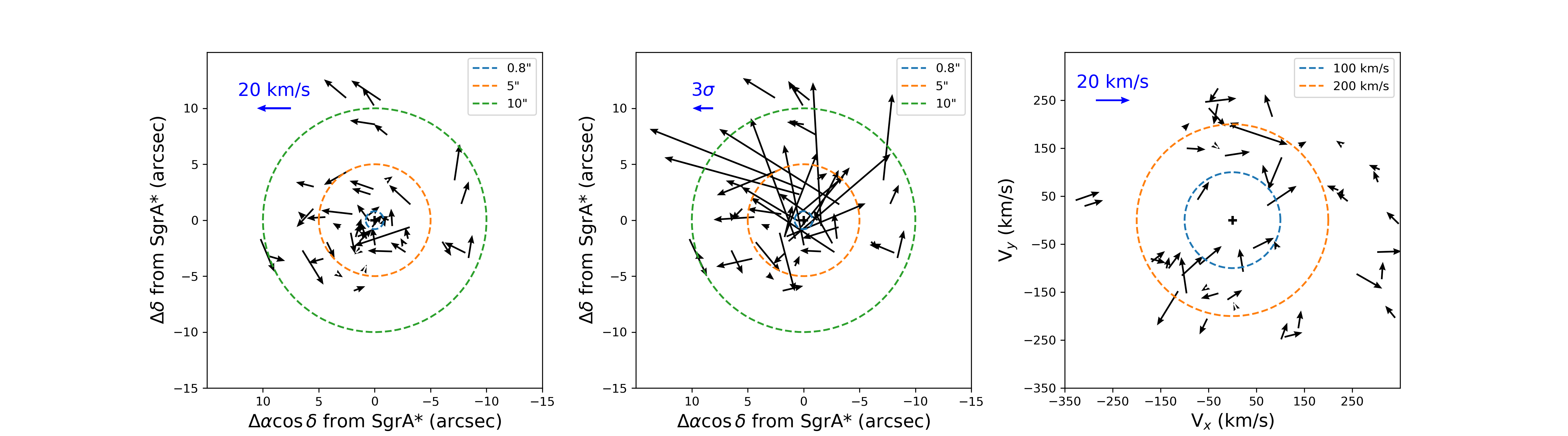}
\caption{A sky map showing the orientation and significance of the proper motion differences for the 45 common stars with reported radial velocities \edit2{in this work and in \citep{von_Fellenberg_2022}.} The arrow indicates \edit2{the proper motion difference} in units of km s$^{-1}$ (left) and in units of each star's corresponding $\sigma$ (\edit2{middle}). The \edit2{left and middle plots have} $(x, y)$ position of each star from our data set; the three circles show $0.8''$ (blue), $5''$ (orange), and $10''$ (green) regions from SgrA*. \edit2{The right figure plots the differential proper motion vectors over $(v_x, v_y)$ from our data set; the two circles represent contours of 100 km s$^{-1}$ (blue) and 200 km s$^{-1}$(orange).} The upper right blue arrow is plotted as reference, indicating a length of 20 km/s (left \edit2{and right}) and
3$\sigma$ (\edit2{middle}). \edit2{Notice that for illustration purposes, arrows in the right figure do not have the same scale as the x and y axes.}}
\label{fig:qvr_pm}
\end{figure*}

To assess whether there are differences in the astrometric reference frames of the two papers, we analyze the mean proper motion offsets.
We calculate the weighted mean velocity and error on weighted mean velocity to find that $<\Delta v_x> = 2.25 \pm 0.50$ km $\mathrm{s}^{-1}$, $<\Delta v_y> = 4.31 \pm 0.49$ km $\mathrm{s}^{-1}$.
To search for scale or rotation differences in the reference frames, we also convert the proper motions into their projected radial and tangential components, $v_r$ and $v_t$. 
Following the same procedures we find that $<\Delta v_r> = -1.94 \pm 0.50$ km $\mathrm{s}^{-1}$ and $<\Delta v_t> = 3.23 \pm 0.48$ km $\mathrm{s}^{-1}$. This suggests there are linear, scale, and rotation differences in the reference frames (Figure \ref{fig:pm_diff}). 
Figure \ref{fig:qvr_pm} shows a map on the sky of the orientation and significance of the proper motion difference vectors.
We find that there is not an obvious relation between $\sigma$ and stars' angular position in the sky, but stars within the inner $5''$ (orange line) are more likely to have inconsistent proper motion data. 
The proper motion measurements in \citet{von_Fellenberg_2022} come from several previous papers and the astrometric reference frame is likely not consistent between them. 
The inconsistencies we find may originate from systematic reference frame errors. \citet{Sakai_2019} reports typical reference frame errors in proper motion of $\sim$1.2 km s$^{-1}$ and rotation of $\sim$2$\degree$ per decade, which largely explain the discrepancies above.

\startlongtable 
\begin{deluxetable*}{lrrrrrrrrr}
\tablecaption{Comparison of Star Velocity\label{tab:comp_vel}}
\tablehead{\colhead{Name} & \colhead{$v_{x, curr}$}  & \colhead{$v_{x, F22}$} & \colhead{$\Delta v_{x} / \sigma v_{x}$} & \colhead{$v_{y, curr}$} & \colhead{$v_{y, F22}$} & \colhead{$\Delta v_{y} / \sigma v_{y}$} & \colhead{$v_{z, curr}$} & \colhead{$v_{z, F22}$} & \colhead{$\Delta v_{z} / \sigma v_{z}$} \\
\colhead{} & \colhead{(km s$^{-1}$)} & \colhead{(km s$^{-1}$)} & \colhead{} & 
\colhead{(km s$^{-1}$)} & \colhead{(km s$^{-1}$)} & \colhead{} & 
\colhead{(km s$^{-1}$)} & \colhead{(km s$^{-1}$)} & \colhead{}}
\startdata
S0-11         & -133.27 & -140.50 &   8.46 & -100.43 & -110.20 &   5.06 & -21.50 & -41.60 &   0.30 \\ 
S0-9          & 339.41 & 345.30 &  -7.25 & -203.50 & -210.70 &   8.87 &  95.30 & 156.70 &  -1.13 \\ 
S0-31         & 240.50 & 247.20 &  -6.95 &  39.64 &  34.40 &   5.79 & -119.60 & -262.70 &   1.42 \\ 
S0-14         &  92.50 &  94.20 &  -3.98 & -55.50 & -58.90 &   7.96 & -31.50 & -57.00 &   1.14 \\ 
S1-33         &  -5.66 & -11.20 &   5.74 & 195.19 & 192.70 &   3.97 &  26.20 &   3.20 &   1.28 \\ 
S1-22         & 311.47 & 310.80 &   1.22 & -123.08 & -133.40 &  21.40 & -229.70 & -293.60 &   0.63 \\ 
S1-24         & 101.56 &  97.90 &   6.65 & -248.42 & -258.30 &  17.96 & 125.20 & 153.50 &  -0.68 \\ 
S2-4          & 307.32 & 313.80 & -11.78 & 105.71 & 103.00 &   4.93 & 217.60 & 208.20 &   0.33 \\ 
S2-6          & 303.55 & 306.20 &  -4.83 &  79.28 &  72.50 &  12.33 & 152.40 & 179.90 &  -1.02 \\ 
S2-50         &  74.75 &  78.10 &  -1.15 &  77.02 &  66.00 &   4.69 & -135.20 & -52.10 &  -0.71 \\ 
S2-17         & 346.96 & 354.30 & -15.22 &  -7.55 & -14.70 &  13.00 &  56.90 &  64.50 &  -0.17 \\ 
S2-19         & -308.83 & -319.90 &  20.12 &  29.45 &  26.00 &   5.50 &  54.50 &  41.20 &   0.50 \\ 
S2-58         & -30.58 & -28.10 &  -2.26 & 242.76 & 255.10 &  -8.70 &  77.30 &  61.90 &   0.44 \\ 
S2-74         & -327.33 & -341.60 &  22.78 &  41.53 &  36.60 &   8.96 &  43.80 &  35.20 &   0.29 \\ 
S3-3          & 138.94 & 134.50 &   1.19 & 155.17 & 152.60 &   0.48 &  23.30 &  45.40 &  -0.51 \\ 
S3-96         &  -0.76 & -33.90 &   5.40 & 196.32 & 207.40 &  -1.67 &  -3.00 &  41.80 &  -0.88 \\ 
S3-19         & 302.03 & 287.50 &   3.12 & -66.45 & -66.90 &   0.15 & -84.40 & -122.50 &   0.63 \\ 
S3-26         & 226.15 & 224.40 &   2.79 &  55.50 &  52.40 &   4.95 &  84.40 &  60.10 &   0.64 \\ 
S3-30         & -30.20 & -31.30 &   1.75 & 151.02 & 151.80 &  -1.62 &   6.50 &  31.90 &  -0.45 \\ 
S3-190        & -105.71 & -118.40 &  17.90 & -115.53 & -127.10 &  11.24 & -262.50 & -249.80 &  -0.14 \\ 
S3-331        & 225.02 & 227.50 &  -1.48 & 159.70 & 158.00 &   0.91 & -165.60 & -154.30 &  -0.24 \\ 
S3-374        & -10.57 & -19.40 &  12.45 & -165.74 & -171.50 &   8.12 &  33.80 &  17.80 &   0.60 \\ 
S4-71         &   3.02 &   3.40 &  -0.69 & -173.67 & -174.50 &   1.51 &  63.50 &  63.60 &  -0.00 \\ 
S4-169        & -95.52 & -106.70 &   6.08 & 149.88 & 150.60 &  -0.39 & 157.70 & -84.20 &   3.60 \\ 
S4-262        & -52.86 & -48.30 &  -3.62 & -205.38 & -196.00 &  -4.40 &  41.00 &  39.90 &   0.02 \\ 
S4-364        & 258.99 & 243.70 &   9.19 & -112.13 & -103.40 &  -3.73 & -109.30 & -150.90 &   0.84 \\ 
S5-237        & -49.46 & -59.40 &   1.60 & 233.70 & 244.60 &  -1.67 &  42.00 &  34.50 &   0.31 \\ 
S5-183        & -170.65 & -179.50 &   5.38 & -80.42 & -78.00 &  -1.19 & -146.30 & -187.00 &   0.96 \\ 
S5-191        & -57.39 & -55.20 &  -0.90 & -142.71 & -141.30 &  -0.73 & 127.00 & 107.00 &   0.29 \\ 
S6-89         & 108.73 &  98.10 &   2.52 & -243.52 & -246.20 &   0.91 & -128.20 & -128.50 &   0.00 \\ 
S6-96         & -30.20 & -24.90 &  -0.93 & 274.85 & 283.30 &  -1.13 &  21.80 & -20.10 &   0.64 \\ 
S6-81         & -101.56 & -104.90 &   0.45 & 191.04 & 187.70 &   0.70 & -21.80 &   9.90 &  -1.06 \\ 
S6-63         & 220.86 & 227.70 &  -3.06 &  62.67 &  60.00 &   0.72 & 143.50 & 140.80 &   0.06 \\ 
S7-30         & -113.64 & -101.10 &  -1.68 & -148.00 & -127.60 &  -3.48 & -35.40 & -27.40 &  -0.13 \\ 
S7-10         & -168.76 & -176.70 &   3.98 & -86.84 & -93.20 &   2.16 & -67.70 & -143.50 &   2.11 \\ 
S7-216        &  83.06 &  87.50 &  -1.28 & 215.58 & 202.20 &   2.85 & 128.10 &  71.00 &   1.42 \\ 
S7-236        & -95.90 & -92.70 &  -1.32 & -152.53 & -174.10 &  12.85 &  78.00 & -90.60 &   1.72 \\ 
S8-4          & -15.86 & -30.80 &   2.51 & 134.41 & 132.10 &   0.33 & -118.80 & -209.00 &   1.58 \\ 
S8-196        &  43.04 &  30.60 &   3.27 & -58.52 & -64.70 &   1.36 & 197.20 & 208.90 &  -0.17 \\ 
S9-143        &  22.28 &  24.50 &  -0.98 & -107.60 & -121.30 &   4.13 &  44.10 & 190.80 &  -1.31 \\ 
S10-50        & -29.45 & -19.00 &  -1.13 & -152.53 & -149.50 &  -0.43 &  69.90 &  88.50 &  -0.21 \\ 
S10-4         & -72.87 & -79.70 &   2.17 &  42.66 &  31.90 &   3.65 & -188.90 & -278.30 &   2.02 \\ 
S10-32        & 102.69 & 110.90 &  -2.61 & 131.01 & 150.50 &  -5.56 & 146.30 & 214.80 &  -0.92 \\ 
S10-48        &  72.49 &  55.00 &   2.71 &  30.58 &  18.90 &   2.16 & -285.30 & -208.10 &  -1.16 \\ 
S11-21        & -68.71 & -81.90 &   4.79 & -92.50 & -103.50 &   2.94 & -81.40 & -199.40 &   2.34 \\ 
IRS13E1       & -137.80 & -139.40 &   0.45 & -100.43 & -107.10 &   3.83 & -10.40 &  34.90 &  -1.51 \\ 
IRS16CC       & -57.01 & -75.60 &   5.06 & 246.54 & 244.40 &   0.75 & 253.20 & 246.20 &   0.22 \\ 
IRS33N        & 137.43 & 135.80 &   2.96 & -225.39 & -236.00 &  14.96 &  23.00 &  39.00 &  -0.35 \\ 
\enddata
\end{deluxetable*}

\clearpage

Even though the differences between each proper motion component are small in amplitude, we believe this inconsistency may lead to different membership probabilities to dynamical structures for stars (Appendix \ref{subsec:disk_comp}).
This emphasizes the importance of using position and proper motion data extracted from a common data set.
\edit1{Additionally, the $v_z$ differences are large in amplitude; but small in statistical significance. This will also impact the disk membership probabilities for individual stars, especially when $\Delta v_z \gtrsim 90$ km/s.
The fraction of stars with discrepant velocities is fairly consistent across the kinematic sub-structures detected in this paper and those from vF22 and does not depend on radius. This is likely due to the fact that $v_x$ and $v_y$ discrepancies dominate at small radii while large differences and uncertainties in $v_z$ dominate at large radii.}

\subsection{Significance and Structure Comparison}
\label{subsec:sig_comp}

In \citet{von_Fellenberg_2022}, they detect structures through an over-density of angular momentum vectors (their \S3), which is similar to what we use in \S\ref{sec:disk}. 
However, when quantifying the significance of each structure, they use a different definition (vF22 Eq 6) compared to ours (Eq \ref{eq:sig}). 
The main difference is that \citet{von_Fellenberg_2022} uses mean values of density while we use peak density in the ($i, \Omega$) map (for a detailed comparison, see their Appendix C).

They report ($i, \Omega$) \footnote{\citet{von_Fellenberg_2022} reports ($\theta, \phi$) and  we convert to ($i, \Omega$) for direct comparison.} locations of 5 structures in their Table 1. 
However, there is no reported uncertainty in the location of each structure. 
To calculate the significance of their structures using our definition, we assign an uncertainty of 20$\degree$ for the position of each structure (eg: CW1 will have $i = 124.5 \pm 20$, $\Omega = 106.9 \pm 20$).
We choose 20$\degree$ because this is the largest uncertainty our structures have and this covers regions of structures shown in \citet[Figure 7]{von_Fellenberg_2022}.
Then we search for the peak density of the normal vector in the assigned region of each structure and calculate the corresponding significance shown in Table \ref{tab:sig}.

We find that CW1 and F3 have comparable significance with our Disk1 and Plane2 while the significance of other structures do not exceed our criterion of 6$\sigma$. 
The conflicting results are likely due to sample differences (FOV size, stars included, reference frame\edit1{, velocity measurement differences}), different definitions of significance\edit1{, and perhaps our choice of uniform-acceleration prior (c.f. \S\ref{sec:mc})}.
For example, we do not detect the CW2 feature because for the total 13 CW2 stars, only 6 have completeness information and are thus included in our sample.
The locations of CW1 and F3 are within the uncertainties of our Disk1 and Plane2, which supports the definitive existence of these two structures. 
The eccentricity distributions within these two structures are also similar. 
For Disk1, we report the $<e> = 0.39 \pm 0.16$ (\S\ref{sec:ecc_dist}) while they find CW1 has a median eccentricity of 0.5, which is consistent within the uncertainty. 
For Plane2, we calculate that it has $<e> = 0.68 \pm 0.07$ while they report F3 has a median eccentricity of 0.7.

\edit1{We also notice that we detect an asymmetry in Plane2, which is not detected by vF22 in their F3 structure analysis. We note that vF22 has incomplete azimuthal coverage for stars outside of 10", where as both papers have relatively uniform coverage inside of 10". This, combined with variable completeness and reddening, which is not modeled in vF22, may explain why we detect asymmetry while vF22 does not. }

\startlongtable 
\begin{deluxetable}{lrrrrrr}
\tablecaption{Comparison of Significance\label{tab:sig}}
\tablehead{\colhead{Structure} & \colhead{$N_{stars}$} & \colhead{$i(\degree)$} & \colhead{$\Omega(\degree)$} & \colhead{$\sigma_{i}$ \tablenotemark{a}} & \colhead{$\sigma_{\Omega}$ \tablenotemark{a}} & \colhead{Significance}}
\startdata
CW1 & 33 & 124.5 & 106.9 & 20 & 20 & 12.40 \\
CW2 & 13 & 145.6 & 156.7 & 20 & 20 & 0.64 \\
CCW(F1) & 33 & 60.0 & 227.0 & 20 & 20 & 2.04 \\
F2 & 37 & 106.0 & 180.0 & 20 & 20 & 3.31 \\
F3 & 36 & 101.5 & 224.2 & 20 & 20 & 5.99 \\
\hline
Disk1 & 18 & 124.0 & 94.0 & 15 & 17 & 12.40 \\
Plane2 & 10 & 90.0 & 245.0 & 20 & 19 & 6.45 \\
\enddata
\tablenotetext{a}{For 5 structures reported in \cite{von_Fellenberg_2022}, we\\
assign an uncertainty of 20$\degree$ on their locations, which is similar\\
to our data.}
\end{deluxetable}

\subsection{Membership Comparison}
\label{subsec:disk_comp}

We compare the membership for 18 stars that we think belong to Disk1 and 10 stars to Plane2 to the structures assigned to these stars in \citet{von_Fellenberg_2022}. 
The results are shown in Table \ref{tab:disk1_comp} and Table \ref{tab:disk2_comp}, where 'Curr P$_\mathrm{Disk1}$' is the probability membership for a star being on Disk1 reported in our paper and 'F22 Structure' is the structure assigned to this star in \citet{von_Fellenberg_2022}.
We classify a star as belonging to a structure if its membership probability exceeds $P_\mathrm{threshold} = 0.2$.
Even though \citet{von_Fellenberg_2022} reports two CW features, only CW1 is consistent with $(i, \Omega)$ of our Disk1. 
There are 13 out of 18 stars that both of us agree to be on the CW disk, which enhances our conclusion of the same structure.

For Plane2 structure, we choose to compare all potential Plane2 stars shown in Table \ref{tab:disk2} to better match the data used by Fellenberg where they do not exclude stars without completeness. 
From Table \ref{tab:disk2_comp}, there are only 6 out of 17 stars that are belonging to Plane2 and F3.
We identify one particular star, S0-31, which they report belonging to CCW disk while we think it is on Plane2. 
This might be caused by the inconsistent $v_x$ and $v_y$ for this star (see Table \ref{tab:comp_vel}).
These comparisons show that Plane2 and F3 differ in their exact memberships, but other properties such as the orientation and eccentricity distribution support the conclusion that these are similar structure.

As we discussed above, inconsistent velocities may lead to differences in membership and conflicting conclusions for individual stars (Appendix \ref{subsec:vel_comp}). 
For the total 47 common stars that have reported structures in both paper, we identify 11 conflicted stars, as shown in Table \ref{tab:mis_class}.
Through checking Table \ref{tab:comp_vel}, we find that 8 of them have at least one inconsistent proper motion data. 
For the remaining 3 stars, S0-3 and S0-8 have full orbital solutions while S3-3 has consistent $v_x$, $v_y$, and $v_z$.
We also identify one particular star, S10-32, which is reported to be a CW1 star.
S10-32 has P$_{\mathrm{Disk1}} = 0.18$, which is slightly lower than our threshold probability of 0.2 and thus we classify it as an off-disk star.
However, we think these two classifications are consistent given this tiny difference in probability.

In summary, through comparisons and analysis of velocity data, significance calculation, disk properties, and disk membership, we think that our Disk1 is consistent with CW1 structure reported in von Fellenberg's work.
\edit1{Plane2 has $\sim 35\%$ members classified as belonging to F3, but the difference in $\Omega$ between these two structures is greater than $1-\sigma$ (34\degree compared to 27.6\degree).
Thus, we think these two structures may be the same; but further analysis is needed to confirm.}
For the other 3 structures (CW2, CCW, and F2), we do not find these features because they do not exceed \edit1{our} 6$\sigma$ criterion (Appendix \ref{subsec:sig_comp}).

\startlongtable 
\begin{deluxetable}{lrr}
\tablecaption{Comparison of Disk1 Star Membership\label{tab:disk1_comp}}
\tablehead{\colhead{Name} & \colhead{Curr P$_\mathrm{Disk1}$}  & \colhead{F22 Structure}}
\startdata
S1-3          &   0.49 & CW1 \\ 
S0-15         &   0.76 & CW1 \\ 
S1-2          &   0.50 & CW1 \\ 
S1-22         &   0.54 & CW1 \\ 
S1-19         &   0.43 & - \\ 
IRS16CC       &   0.25 & - \\ 
S2-4          &   0.22 & CW1 \\ 
S2-6          &   0.50 & CW1 \\ 
S2-17         &   0.37 & CW1 \\ 
S2-21         &   0.50 & CW1 \\ 
S2-19         &   0.24 & CW1 \\ 
S2-74         &   0.50 & CW1 \\ 
S3-19         &   0.36 & CW1 \\ 
S3-26         &   0.24 & CW1 \\ 
S3-190        &   0.35 & CW1/F2 \\ 
S4-169        &   0.40 & CW2/F2 \\ 
S4-314        &   0.47 & - \\ 
S6-63         &   0.27 & CW2 \\ 
\enddata
\end{deluxetable}

\startlongtable 
\begin{deluxetable}{lrr}
\tablecaption{Comparison of Plane2 Star Membership\label{tab:disk2_comp}}
\tablehead{\colhead{Name} & \colhead{Curr P$_\mathrm{Plane2}$}  & \colhead{F22 Structure}}
\startdata
S0-31         &   0.30 & CCW(F1) \\ 
IRS 16NW      &   0.38 & - \\ 
IRS 13E1      &   0.50 & - \\ 
IRS 13E3      &   0.34 & - \\ 
S4-258        &   0.40 & CCW(F1) \\ 
S5-34         &   0.36 & F3 \\ 
S5-106        &   0.34 & - \\ 
S5-236        &   0.71 & F3 \\ 
S9-114        &   0.52 & F3 \\ 
S9-221        &   0.77 & - \\ 
S10-185       &   0.73 & - \\ 
S10-238       &   0.78 & - \\ 
S12-5         &   1.00 & - \\ 
S13-3         &   0.27 & - \\ 
S0-16         &   0.42 & - \\ 
S3-26         &   0.24 & F3/CW1 \\ 
S8-196        &   0.22 & F3 \\ 
\enddata
\end{deluxetable}

\startlongtable 
\begin{deluxetable}{lrrrr}
\tablecaption{Conflicted Stars\label{tab:mis_class}}
\tablehead{\colhead{Name} & \colhead{P$_\mathrm{Disk1}$} & \colhead{P$_\mathrm{Plane2}$} & \colhead{P$_\mathrm{Non-disk}$} & \colhead{F22 Structure}}
\startdata
S0-8          &   0.00 &   0.00 &   1.00 & CW1 \\ 
S0-3          &   0.00 &   0.00 &   1.00 & F3 \\ 
S0-11         &   0.00 &   0.05 &   0.95 & CCW(F1)/F3 \\ 
S0-9          &   0.00 &   0.00 &   1.00 & CW1 \\ 
S0-31         &   0.00 &   0.30 &   0.70 & CCW(F1) \\ 
S2-50         &   0.00 &   0.00 &   1.00 & F3 \\ 
S3-3          &   0.00 &   0.00 &   1.00 & F3 \\ 
S4-169        &   0.40 &   0.00 &   0.60 & CW2/F2 \\ 
S6-63         &   0.27 &   0.00 &   0.73 & CW2 \\ 
S7-10         &   0.16 &   0.03 &   0.81 & CW2/F3 \\
S9-143        &   0.01 &   0.00 &   0.99 & F3 \\ 
S10-32 \tablenotemark{a}        &   0.18 &   0.00 &   0.82 & CW1 \\ 
\enddata
\tablenotetext{a}{See Appendix \ref{subsec:disk_comp} for a detailed discussion on S10-32.}
\end{deluxetable}

\section{Full Keplerian Orbit Analysis} % (fold)
\begin{figure*}[h]
    \centering
    \includegraphics[width=\textwidth]{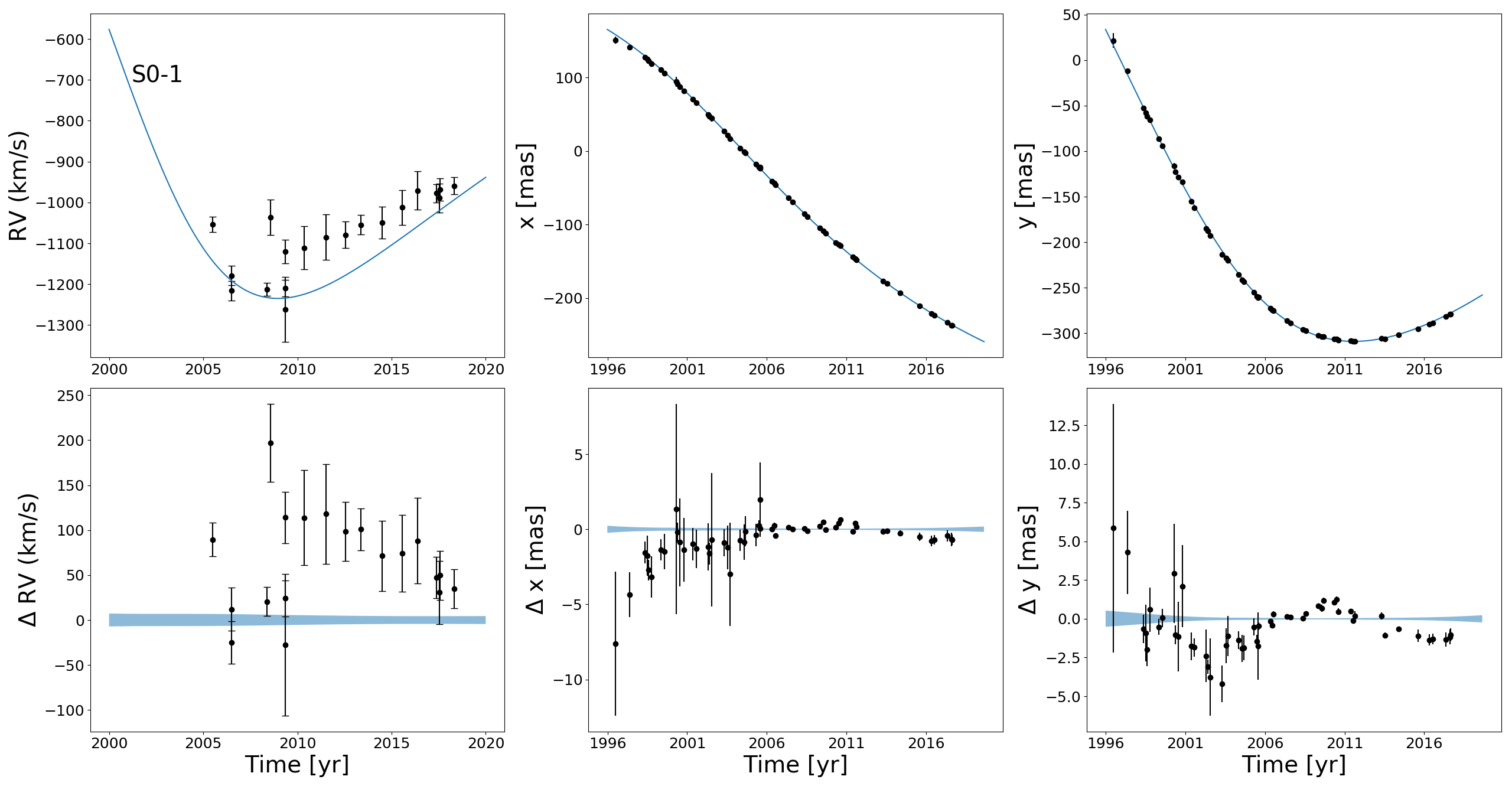}
    \caption{Examples of full Keplerian orbit analysis for S0-1.}
    \label{fig:e_fitS0-1}
\end{figure*}

\begin{figure*}[h]
    %\ContinuedFloat
    \includegraphics[width=\textwidth]{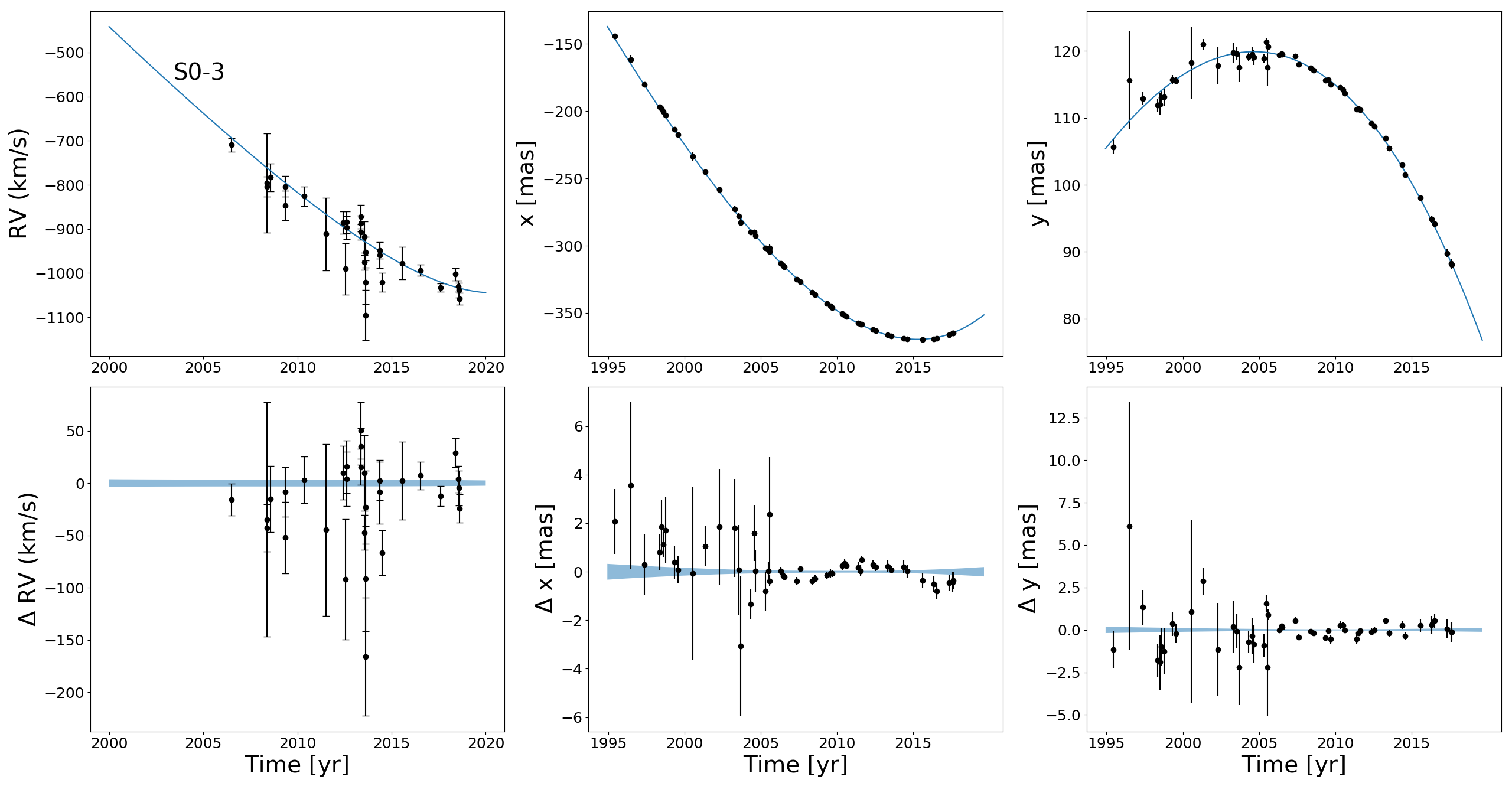}
    \caption{Examples of full Keplerian orbit analysis for S0-3.}
    \label{fig:e_fitS0-3}
\end{figure*}

\begin{figure*}[h]
    %\ContinuedFloat
    \includegraphics[width=\textwidth]{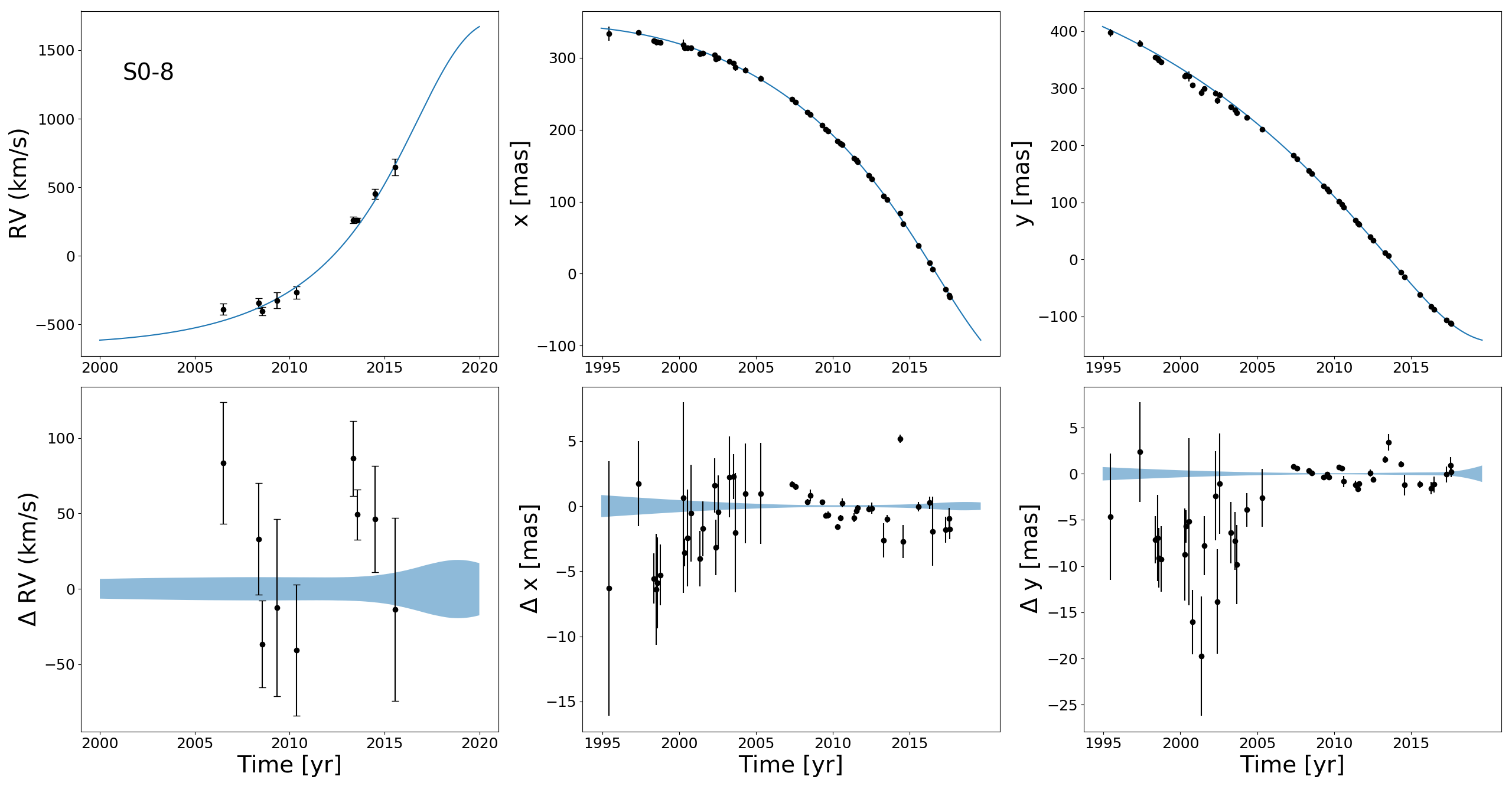}
    \caption{Examples of full Keplerian orbit analysis for S0-8.}
    \label{fig:e_fitS0-8}
\end{figure*}

\begin{figure*}[h]
    %\ContinuedFloat
    \includegraphics[width=\textwidth]{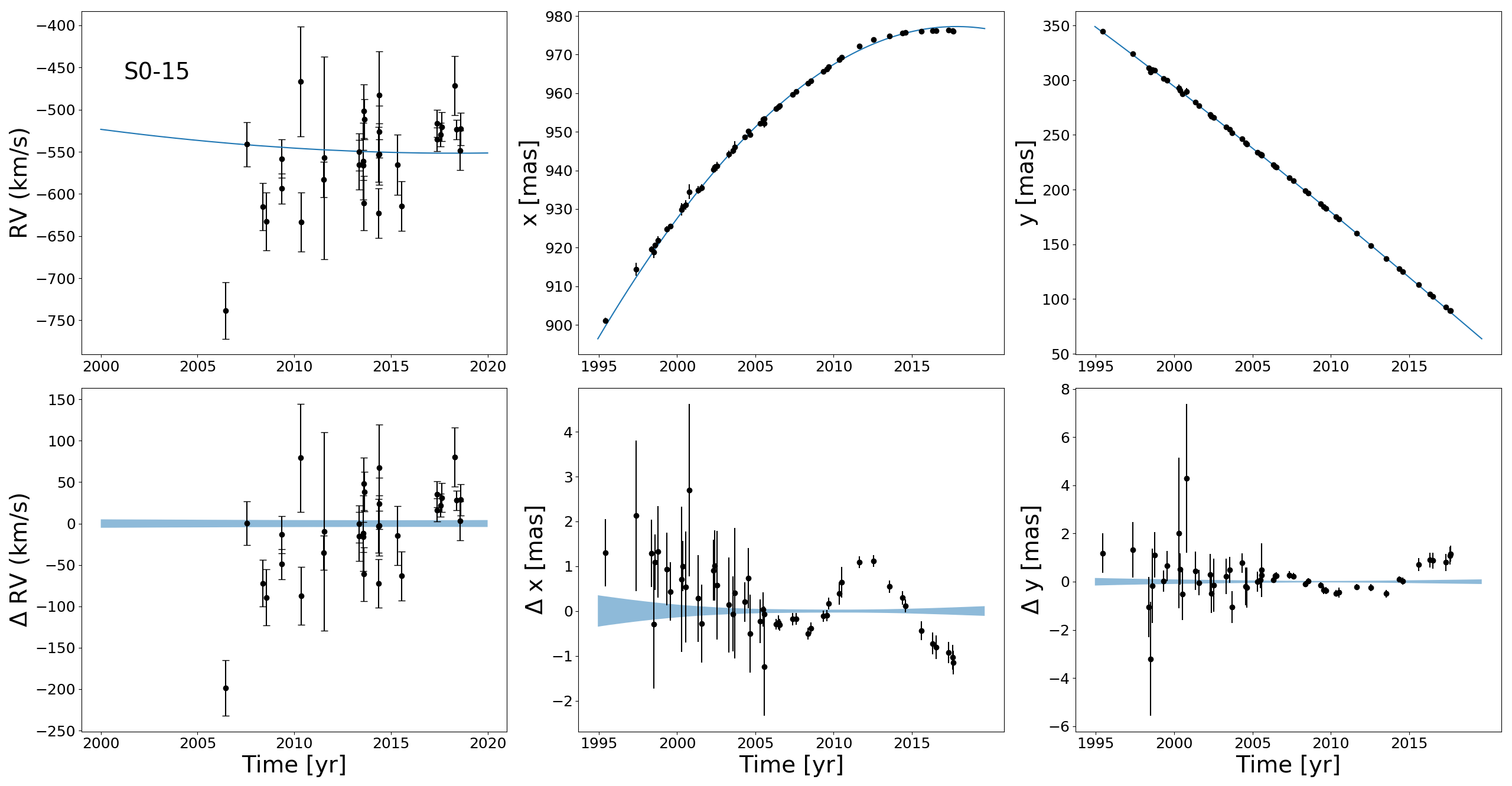}
    \caption{Examples of full Keplerian orbit analysis for S0-15.}
    \label{fig:e_fitS0-15}
\end{figure*}

\begin{figure*}[h]
    %\ContinuedFloat
    \includegraphics[width=\textwidth]{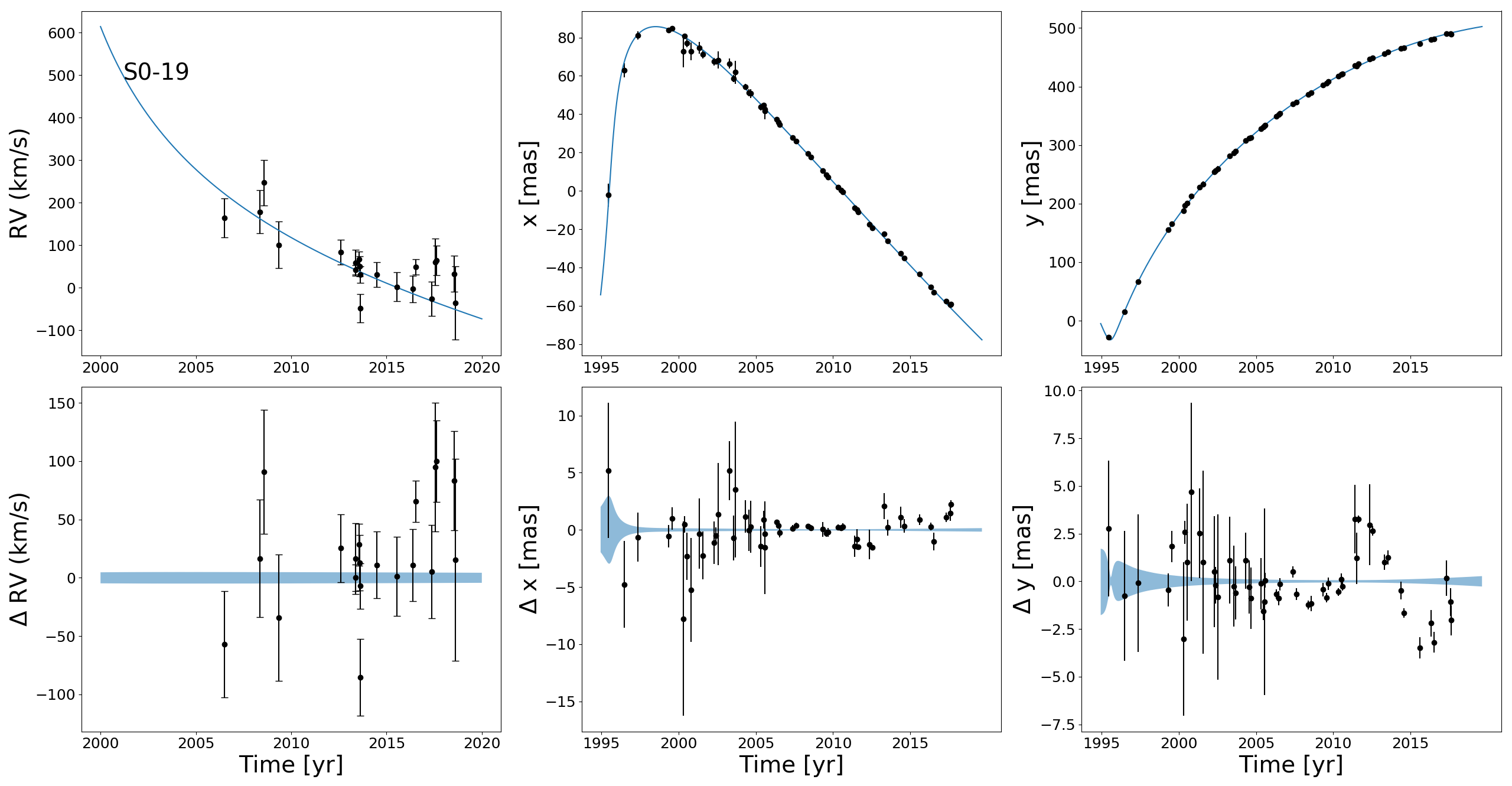}
    \caption{Examples of full Keplerian orbit analysis for S0-19.}
    \label{fig:e_fitS0-19}
\end{figure*}

\begin{figure*}[h]
    %\ContinuedFloat
    \includegraphics[width=\textwidth]{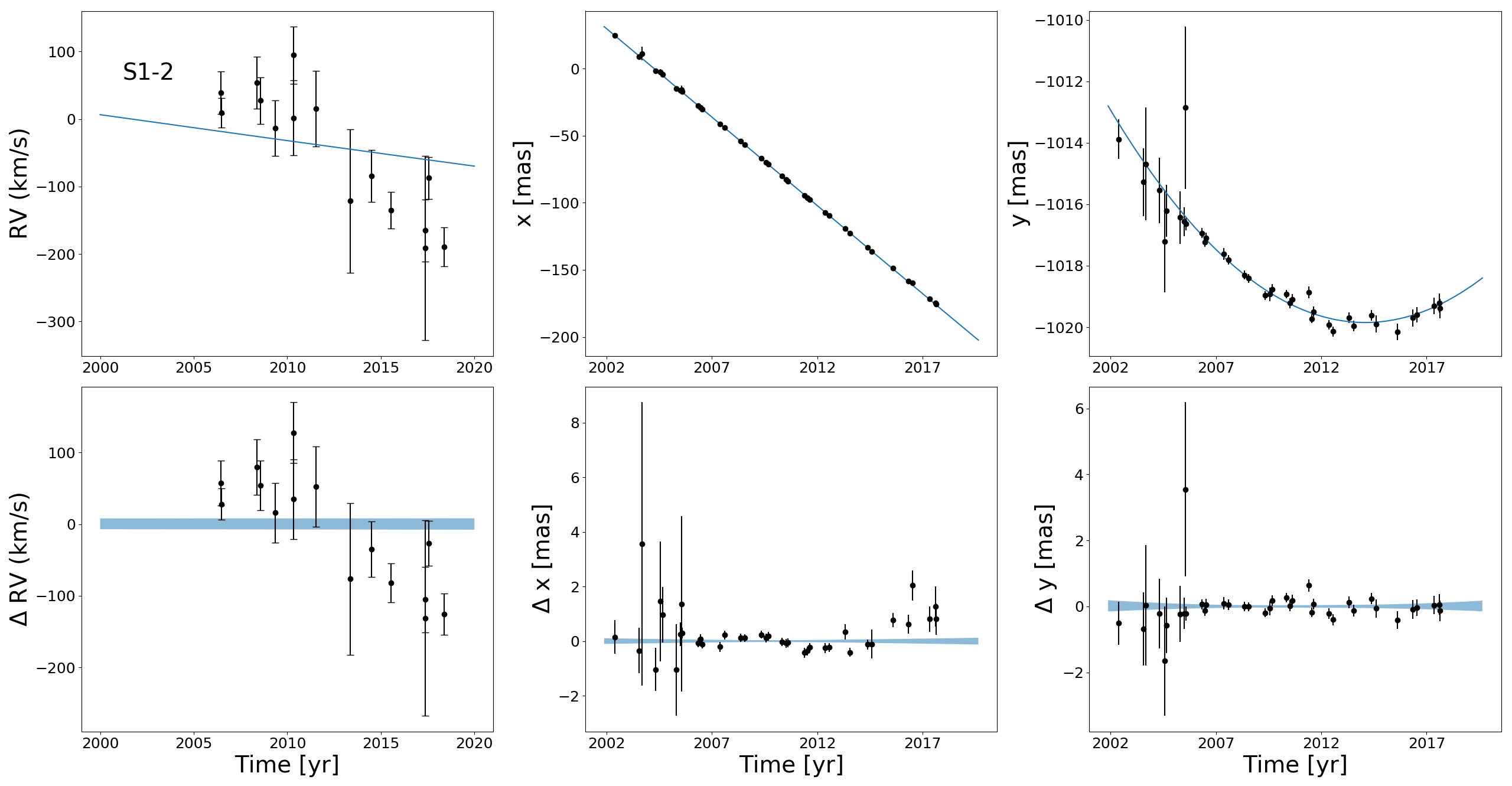}
    \caption{Examples of full Keplerian orbit analysis for S1-2.}
    \label{fig:e_fitS1-2}
\end{figure*}

\label{app:efit}
In Figure \Cref{fig:e_fitS0-1,fig:e_fitS0-3,fig:e_fitS0-8,fig:e_fitS0-15,fig:e_fitS0-19,fig:e_fitS1-2}, we plot the full Keplerian orbit fit for the other 6 orbit stars in our sample: S0-1, S0-3, S0-8, S0-15, S0-19, S1-2, in the same format as in Figure \ref{fig:efit_S0-2}.  


\clearpage

\bibliography{gc}

\end{document}